\documentclass[11pt]{article}

\usepackage{pstricks}
\usepackage{pst-plot}
\usepackage{amsmath}
\usepackage{amssymb}
\usepackage[T1]{fontenc}
\usepackage{ae}
\usepackage[ansinew]{inputenc}

\newrgbcolor{darkred}{0.8 0 0}
\newrgbcolor{darkerred}{0.7 0 0}
\newrgbcolor{darkestred}{0.5 0 0}
\newrgbcolor{darkgreen}{0 0.5 0.5}
\newrgbcolor{darkergreen}{0 0.6 0}
\newrgbcolor{darkestgreen}{0 0.5 0}
\newrgbcolor{darkblue}{0 0 0.6}
\newrgbcolor{darkestblue}{0 0 0.5}
\newrgbcolor{redgreen}{0.8 0.8 0}
\newrgbcolor{bluegreen}{0 0.4 0.7}
\newrgbcolor{bluewhite}{0 0.9 1}
\newrgbcolor{lightgreen}{0.6 1 0.6}
\newrgbcolor{lightred}{0.7 0.7 0.7}
\newrgbcolor{lightyred}{1 0.4 0.4}
\newrgbcolor{lightyblue}{0.5 0.5 1}
\newrgbcolor{lightblue}{1 1 1}
\newrgbcolor{lilas}{0.5 0.3 0.7}
\newrgbcolor{lilaswhite}{1 0.8 1}
\newrgbcolor{redy}{1 0.6 0.6}
\newrgbcolor{bluish}{0.6 0.6 1}
\newrgbcolor{sand}{0.9 0.9 0.7}
\newrgbcolor{brown}{0.4 0.4 0}
\newrgbcolor{blue1}{0 0.35 0.65}
\newrgbcolor{blue2}{0 0.4 0.6}
\newrgbcolor{blue3}{0 0.45 0.55}
\newrgbcolor{blue4}{0 0.5 0.5}
\newrgbcolor{blue5}{0 0.55 0.45}
\newrgbcolor{blue6}{0 0.6 0.4}
\newrgbcolor{blue7}{0 0.65 0.35}

\newcommand{\R }{\mathbb R}
\newcommand{\N }{\mathbb N}
\newcommand{\Z }{\mathbb Z}
\newcommand{\Q }{\mathbb Q}
\newcommand{\C }{\mathbb C}
\newcommand{\sen }{\ \! {\rm sen}\ \! }
\newcommand{\tg }{\ \! {\rm tg}\ \! }
\newcommand{\cosec }{\ \! {\rm cosec}\ \! }
\newcommand{\cotg }{\ \! {\rm cotg}\ \! }
\newcommand{\dx }{\ \! dx}
\newcommand{\e }{\ \! e}
\newcommand{\senh }{\ \! {\rm senh}\ \! }
\newcommand{\tgh }{\ \! {\rm tgh}\ \! }
\newcommand{\cotgh }{\ \! {\rm cotgh}\ \! }
\newcommand{\sech }{\ \! {\rm sech}\ \! }
\newcommand{\cosech }{\ \! {\rm cosech}\ \! }
\newcommand{\arcsen }{\ \! {\rm arcsen}\ \! }
\newcommand{\arctg }{\ \! {\rm arctg}\ \! }
\newcommand{\arccotg }{\ \! {\rm arccotg}\ \! }
\newcommand{\arcsenh }{\ \! {\rm arcsenh}\ \! }
\newcommand{\arccosh }{\ \! {\rm arccosh}\ \! }
\newcommand{\arctgh }{\ \! {\rm arctgh}\ \! }
\newcommand{\arccosech }{\ \! {\rm arccosech}\ \! }
\newcommand{\arcsech }{\ \! {\rm arcsech}\ \! }
\newcommand{\arcsec }{\ \! {\rm arcsec}\ \! }
\newcommand{\arccosec }{\ \! {\rm arccosec}\ \! }
\renewcommand{\arctgh }{\ \! {\rm arctgh}\ \! }
\newcommand{\arccotgh }{\ \! {\rm arccotgh}\ \! }
\newcommand{\nega }{\neg \ }
\newcommand{\eq }{\Leftrightarrow }

\begin{document}

\title{Building portfolios of stocks in the São Paulo Stock Exchange using Random Matrix Theory}

\author{Leonidas Sandoval Junior\thanks{E-mail: leonidassj@insper.edu.br (corresponding author)} \\ Maria Kelly Venezuela \thanks{E-mail: mariakv@insper.edu.br} \\ Adriana Bruscato \thanks{E-mail: adrianab@insper.edu.br} \\ \\ Insper, Instituto de Ensino e Pesquisa}

\maketitle

\vskip 0.4 cm

\noindent {\bf Abstract}

\vskip 0.2 cm

\noindent By using Random Matrix Theory, we build covariance matrices between stocks of the BM\&F-Bovespa (Bolsa de Valores, Mercadorias e Futuros de São Paulo) which are cleaned of some of the noise due to the complex interactions between the many stocks and the finiteness of available data. We also use a regression model in order to remove the market effect due to the common movement of all stocks. These two procedures are then used to build stock portfolios based on Markowitz's theory, trying to obtain better predictions of future risk based on past data.  This is done for years of both low and high volatility of the Brazilian stock market, from 2004 to 2010. The results show that the use of regression to subtract the market effect on returns greatly increases the accuracy of the prediction of risk, and that, although the cleaning of the correlation matrix often leads to portfolios that better predict risks, in periods of high volatility of the market this procedure may fail to do so.

\vskip 0.3 cm


\vskip 0.3 cm


\vskip 0.4 cm

\section{Introduction}

Modern portfolio theory is largely based on Markowitz's ideas, as described by Markowitz (1962), by Elton, Gruber, Brown, \& Goetzmann (2009), and by Bodie, Kane, \& Marcus (2009), where a portfolio of various equities is built on the principle of minimizing risk given an expected return. Risk is assessed as the volatility of each stock that comprises the portfolio, as well as their covariance. Preference is given to stocks that have negative or low covariance between each other, which leads to diversification of the equities held in one particular portfolio.

Both volatility and covariance are integrated into the covariance matrix, which is built using the stock returns of past data. This is used in order to predict the risk of a portfolio, and it is usualy different from the realized risk of the same portfolio.

Three problems arise from this approach. The first one is that past data reflect the market as it was, and not as it will be. So, the theory assumes the hypothesis that future events shall mimic past events, which is usually not true, since it does not incorporate news releases, or the current mood of the market. There is not much that can be done about this, but to minimize effects of events that might change the behavior of a market, one cannot use past data that is too old.

This leads us to the second problem, which is the noise associated with past data that arises purely from the fact that the available data are finite. Since one cannot go back in time indefinitely, and even if one could, it wouldn't be advisable given the discussion in the preceding paragraph, there is only a limited amount of data (in our case, price quotations) from which to build a covariance matrix. The problem gets even more severe if we think that an efficient portfolio should be built from many and diverse equities. A third source of noise comes from the complex interactions between the many elements of a stock market: traders, news, foreign markets, and the very prices of stocks interact in order to guide the price of a stock. Those interactions are usually too complex to be acommodated by any econometric model.

So, all this noise is incorporated into the covariance matrix that is used in the attempt to forecast the risk of a particular portfolio, and if one can remove some of that noise from the matrix, one is then able to make better risk predictions. Frankfurter, Phillips, \& Seagle (1971), Frankfurter, Phillips, \& Seagle (1972), Dickinson (1994), Jobson \& Korkie (1990), Michaud (1989), and Chopra \& Ziemba (1993) made studies on the influence of noise and other factors on the covariance matrix in the building of portfolios. Most of the approaches for solving them involve the reduction of the dimensionality of the covariance matrix by introducing some structure into it, obtained by principal component analysis, separation of stocks into economic sectors, among other means - see the works of Jorion (1986) and DeMiguel, Garlappi, Nogales, \& Uppal (2009) for two of these approaches.

A technique first developed for the study of the nuclei of the atoms of heavier elements, called Random Matrix Theory - see Mehta (2004), compares the eigenvalues of a correlation matrix with those of a correlation matrix built from a purely random matrix. From such a comparison, one may then discern elements which are clearly not random, and study them separately. Such technique has been applied to a number of complex systems, and, particularly, to financial markets. Of the many results that were obtained, the building of portfolios that most closely resemble the realized risk of the future market, based on past data, is one of them, as in Laloux, Cizeau, Bouchaud, \& Potters (1999), Laloux, Cizeau, Bouchaud, \& Potters (2000) and Rosenow, Plerou, Gopikrishnan, \& Stanley (2002), and it has been successfuly applied to stocks by Plerou, Gopikrishnan, Rosenow, Amaral, Guhr, \& Stanley (2002) and Sharifi, Crane, Shamaie, \& Ruskin (2004), and to mutual funds by Conlon, Ruskin, \& Crane (2007).

The building of robust correlation matrices using statistical methods has also been considered by other researchers. Damião \& Valls Pereira (2009) used robust statistical methods (influence function) in order to clean the covariance matrix of assets from the BM\&F-Bovespa from some of its noise, considering that the returns of such assets deviate from a normal distribution, specially for emerging markets, due to their ``heavier tails'' (an abnormal number of large positive and negative returns). Their method aimed at building efficient portfolios according to the mean-variance criterion, and succeeded in finding portfolios with better stability and variance.

Besides the cleaning of the correlation matrix, we used a regression model to remove the market effect on the asset returns. This procedure allows the estimation of the correlation matrix with greater precision, for there is just a part of the dependence which is due to the assets, which generates more reliable forecasts for the risk of a portfolio. This procedure is standard in many models in finance, most importantly in the CAPM (Capital Asset Pricing Model), and it is called {\sl Single Index Model}, based on the idea that the majority of the systemic risk is captured by a single market index. Other models, called {\sl factor models}, defend the hypothesis that the systemic risk is due to a number of factors, which may include statistical, macroeconomic, or fundamentalist influences. Ross (1976) presents a model, called {\sl Arbitrage Pricing Model}, which uses more than one factor to explain systemic risk. Schor, Bonomo, \& Valls Pereira (2002) studied the importance of macroeconomic factors on the returns of ten portfolios of assets of the Brazilian stock market, and showed that those factors were statistically significant for the majority of the portfolios. Vinha \& Chian (2007) built two factor models using data from 20 assets of the BM\&F-Bovespa, the first one based on statistical factors (eigenvectors of the correlation matrix of those assets), and the second one based on fundamentalist factors (market value of the company to which the asset is related, rate between profit and price of the asset, and financial leverage, as examples), and concluded that statistical factors were more significant.

Ledoit and Wolf (2003) used an optimally weighted average of the sample covariance matrix and the single-index covariance matrix in order to estimate the covariance matrix of stock returns, a process called {\sl shrinkage}. They obtained portfolios whith significantly lower out-of-sample variance than a set of existing estimators, including multifactor models. Reyna, Duarte, \& Mendes (2005) used a General Asset Management Model with a statistical utility function in order to eliminate some of the effects of extreme returns in estimating a robust covariance matrix. Their results showed a better ex-post performance when compared with the model based on the ordinary covariance matrices. Mendes \& Leal (2005) also used a new robust covariance matrix in order to account for both typical and atypical returns. Their results led to portfolios that yielded higher cumulative returns and weights for each of the components of such portfolio that were more stable in time.

The contribution of this article is the use of a method which is capable of ameliorating the risk forecasts of a portfolio built with Brazilian stock market assets, based on past data. This method involves three steps: (1) the removal of the market effect of the assets; (2) the cleaning of the correlation matrix, which encodes the structure of the dependence of the assets being considered, based on Random Matrix Theory, and (3) the construction of portfolios using Markowitz's theory and the cleaned correlation matrix. We calculate portfolios of stocks with and without the removal of the market effect so as to compare both results.

In order to analyze the suitability of the proposed method, we shall use the daily returns of BM\&F-Bovespa stocks with 100\% liquidity, what means that there was negotiation of those stocks every day the stock excahnged was open, whith pairs of years ranging from 2004 to 2010. For each year being analyzed, we build a portfolio using data from the previous year in order to make a forecast of the risk for a determined year, and that forecasted risk is then compared with the realized risk in that year. We use 61 stocks for 2004-2005, 72 stocks for 2005-2006, 86 stocks for 2006-2007, 105 stocks for 2007-2008, 148 stocks for 2008-2009, and 153 stocks for 2009-2010.

In order to analyze the evolution of the portfolios in time, we also make a study of 50 stocks that are 100\% liquid in the years ranging from 2004 to 2010 and study the differences betweeen predicted and realized portfolios in time. As data used in this article include periods of both low and high volatility in the BM\&F-Bovespa, in particular the data collected during the Subprime Mortgage Crisis of 2007 and 2008, we are able to study how this technique of cleaning the correlation matrix applies to times of high volatility.

The article is organized as follows: Section 2 introduces the basic concepts of Random Matrix Theory, and the characteristics of the eigenvalues of the correlation matrix. Section 3 is dedicated to the building of portfolios (according to Markowitz) with and without cleaning the correlation matrix for short selling allowed or not, as well as the regression for the removal of the market effect, with measures of how well predicted risk approximates realized risk for equal values of return. The analysis of how the proposed measures evolve in time is done in Section 4, and the article ends with a conclusion and comments on years of high volatility in Section 5.

\section{Methodology} \label{sec:meth}

In this section, we briefly describe the method proposed for the construction of portfolios by cleaning the correlation matrix and removing the market effect, aiming at a better forecasting of risk based on the previous behaviors of the assets. We use the year 2004 as an example of the application of such method in this section, and then apply the same methodology for the remaining years.

\subsection{Random matrix theory} \label{sec:rmt}

Random matrix theory had its origins in 1953, in the work of the Hungarian physicist Eugene Wigner (1955) (1958). He was studying the energy levels of complex atomic nuclei, such as uranium, and had no means of calculating the distance between those levels. He then assumed that those distances between energy levels should be similar to the ones obtained from a random matrix which expressed the connections between the many energy levels. Surprisingly, he could then be able to make sensible predictions about how the energy levels related to one another.

The theory was later developed, with many and surprising results arising. Today, random matrix theory is applied to quantum physics, nanotechnology, quantum gravity, the study of the structure of crystals, and may have applications in ecology, linguistics, and many other fields where a huge amount of apparently unrelated information may be understood as being somehow connected. The theory has also been applied to finance in a series of works dealing with the correlation matrices of stock prices, as well as with risk management in portfolios, as in Pafka \& Kondor (2002), Onnela, Chakraborti, \& Kaski (2003), Tola, Lillo, Gallegati, \& Mantegna (2008), and Pantaleo, Tumminello, Lillo, \& Mantegna (2011). For a recent review on the subject, see Bouchaud \& Potters (2011).

The first result of the theory that we shall mention is that, given an $L\times N$ matrix with random numbers from a Gaussian distribution with zero mean and standard deviation $\sigma $, then, in the limit $L\to \infty $ and $N\to \infty$ such that $Q=L/N$ remains finite and greater than one, the eigenvalues $\lambda$ of the correlation matrix built from this random matrix will have the following probability density function, called a Mar\v{c}enku-Pastur distribution, developed in Mar\v{e}nko \& Pastur (1967):
\begin{equation}
\label{dist}
\rho(\lambda )=\frac{Q}{2\pi \sigma ^2}\frac{\sqrt{(\lambda_+-\lambda )(\lambda -\lambda_-)}}{\lambda }\ ,
\end{equation}
where
\begin{equation}
\lambda_-=\sigma ^2\left( 1+\frac{1}{Q}-2\sqrt{\frac{1}{Q}}\right) \ \ ,\ \ \lambda_+=\sigma ^2\left( 1+\frac{1}{Q}+2\sqrt{\frac{1}{Q}}\right) \ ,
\end{equation}
and $\lambda $ is restricted to the interval $\left[ \lambda_-,\lambda_+\right] $.

Since the distribution (\ref{dist}) is only valid for the limit $L\to \infty $ and $N\to \infty $, finite distributions will present differences from this behavior. Another source of deviations is the fact that financial time series are better described by non-Gaussian distributions, such as t Student or Tsallis distribution. The eigenvalue distribution of correlation matrices derived from time series with t Student distributions was studied by Biroli, Bouchaud, \& Potters (2007). The authors derived a theoretical probability distribution for the eigenvalues which has no higher limit, but which decays exponentially.

Knowing that the returns of some assets have tails heavier than the tails of the normal distribution, we decided to use the Kolmogorov Smirnov test to check if the empirical distribution of eigenvalues of the Pearson´s correlation matrix is compliant with the Mar\v{c}enku-Pastur distribution or if we need to use the Student Ensemble distribution or a Maximum Likelihood estimator of correlations in place of the Pearson's estimator (Biroli, Bouchaud, \& Potters (2007)). The Kolmogorov Smirnov test did not reject the hypothesis that the eigenvalues follow the Mar\v{c}enku-Pastur distribution with a confidence level of 99\% for each one of the years 2004 to 2010. Thus, we consider the Mar\v{c}enku-Pastur distribution  valid for the eigenvalues calculated based on the Pearson correlation matrix in the case of assets traded in the São Paulo stock exchange in the period evaluated.

\subsection{Eigenvalues and eigenvectors of the correlation matrix}

We shall now explain why random matrix theory is useful for portfolio building, starting by clarifying how it can be used to remove part of the noise from the correlation matrix. In order to do that we shall consider the data concerning the year 2004, the first of the years considered here in our study. For this period we chose Bovespa stocks (by then, Bovespa had not yet joined with BM\&F) which were negotiated every trading day during the years 2004 and 2005 (2004 will be the past data that will be used to predict the risk in 2005), totalizing 61 stocks. Very similar results are obtained for the other pairs of years.

For each stock, we calculated the returns, more precisely the log-returns, given by
\begin{equation}
R_t=\ln (P_t)-\ln (P_{t-1})\approx \frac{P_t-P_{t-1}}{P_t}\ ,
\end{equation}
where $P_t$ is the closing price of one stock at the trading day $t$. The correlation matrix (a $61\times 61$ matrix) between the variables $X_t$ for the year 2004 was then calculated.

The distribution density of the eigenvalues of the correlation matrix thus obtained is shown in Figure 1 (left picture). Also, the eigenvalues are plotted in order of magnitude in the right picture of Figure 1. The shaded area (between $\lambda_-$ and $\lambda_+$) indicates the region predicted by the theory for the data related to a purely random behavior of the normalized returns, which is called the Wishart region.

We have $L=248$ days of data for each of the $N=61$ stocks, so that $Q=248/61\approx 4.06$. The probability distribution function for a random matrix with $L\to \infty $ and $N\to \infty $ with $Q\approx 4.06$ is also plotted in Figure 1, so that we may compare the result of pure noise with the one obtained for our data. The minimum ($\lambda_-$) and maximum ($\lambda_+$) values of the probability distribution function are given by
\[ \lambda_-=0.254\ \ \text{and}\ \ \ \lambda_+=2.238\ .\]

\begin{pspicture}(-1,0)(3.5,3.4)
\psset{xunit=1,yunit=200}
\psline(0,0)(0,0.00328)(0.1,0.00328)(0.1,0) \psline(0.1,0)(0.1,0.01148)(0.2,0.01148)(0.2,0) \psline(0.2,0)(0.2,0.00984)(0.3,0.00984)(0.3,0) \psline(0.3,0)(0.3,0.01148)(0.4,0.01148)(0.4,0) \psline(0.4,0)(0.4,0.00984)(0.5,0.00984)(0.5,0) \psline(0.5,0)(0.5,0.00656)(0.6,0.00656)(0.6,0) \psline(0.6,0)(0.6,0.00656)(0.7,0.00656)(0.7,0) \psline(0.7,0)(0.7,0.00984)(0.8,0.00984)(0.8,0) \psline(0.8,0)(0.8,0.00164)(0.9,0.00164)(0.9,0) \psline(0.9,0)(0.9,0.00492)(1,0.00492)(1,0) \psline(1,0)(1,0.00492)(1.1,0.00492)(1.1,0) \psline(1.1,0)(1.1,0.00328)(1.2,0.00328)(1.2,0) \psline(1.2,0)(1.2,0.00492)(1.3,0.00492)(1.3,0) \psline(1.3,0)(1.3,0.00164)(1.4,0.00164)(1.4,0) \psline(1.4,0)(1.4,0.00164)(1.5,0.00164)(1.5,0) \psline(1.5,0)(1.5,0.00164)(1.6,0.00164)(1.6,0) \psline(1.6,0)(1.6,0.00164)(1.7,0.00164)(1.7,0) \psline(1.8,0)(1.8,0.00164)(1.9,0.00164)(1.9,0) \psline(2.5,0)(2.5,0.00164)(2.6,0.00164)(2.6,0) \psline(4.5,0)(4.5,0.00164)(4.6,0.00164)(4.6,0)
\psplot[linecolor=gray,plotpoints=500,linewidth=1pt]{0.2541}{2.2378}{2.2378 x sub x 0.2541 sub mul 0.5 exp x -1 exp mul 0.64706 mul 0.01 mul}
\psline{->}(0,0)(6,0) \psline{->}(0,0)(0,0.015) \rput(6.25,0){$\lambda $} \rput(0.5,0.015){$\rho (\lambda )$} \psline[linecolor=white,linewidth=2pt](3.4,0)(3.6,0) \psline(3.3,-0.001)(3.5,0.001) \psline(3.5,-0.001)(3.7,0.001) \scriptsize \psline(1,-0.0005)(1,0.0005) \rput(1,-0.0015){1} \psline(2,-0.0005)(2,0.0005) \rput(2,-0.0015){2} \psline(3,-0.0005)(3,0.0005) \rput(3,-0.0015){3} \psline(-0.1,0.005)(0.1,0.005) \rput(-0.5,0.005){$0,5$} \psline(-0.1,0.01)(0.1,0.01) \rput(-0.3,0.01){1} \psline(4,-0.0005)(4,0.0005) \rput(4,-0.0015){23} \psline(5,-0.0005)(5,0.0005) \rput(5,-0.0015){24} \psline(-0.1,0.005)(0.1,0.005) \rput(-0.5,0.005){$0,5$}
\end{pspicture}
\begin{pspicture}(-4.5,0)(3.5,3.4)
\psset{xunit=1.2,yunit=2.2}
\pspolygon*[linecolor=lightgray](0.2739,0)(0.2739,0.8)(2.1366,0.8)(2.1366,0)
\psline{->}(0,0)(6,0)  \psline[linecolor=white,linewidth=2pt](3.4,0)(3.6,0) \psline(3.3,-0.1)(3.5,0.1) \psline(3.5,-0.1)(3.7,0.1) \rput(6.3,0){$\lambda $} \scriptsize \psline(1,-0.05)(1,0.05) \rput(1,-0.15){1} \psline(2,-0.05)(2,0.05) \rput(2,-0.15){2} \psline(3,-0.05)(3,0.05) \rput(3,-0.15){3}  \psline(4,-0.05)(4,0.05) \rput(4,-0.15){23} \psline(5,-0.05)(5,0.05) \rput(5,-0.15){24}
\psline[linewidth=1pt](4.505,0)(4.505,0.5) \psline[linewidth=1pt](2.540,0)(2.540,0.5) \psline[linewidth=1pt](1.779,0)(1.779,0.5) \psline[linewidth=1pt](1.610,0)(1.610,0.5)
\psline[linewidth=1pt](1.530,0)(1.530,0.5) \psline[linewidth=1pt](1.431,0)(1.431,0.5) \psline[linewidth=1pt](1.333,0)(1.333,0.5) \psline[linewidth=1pt](1.222,0)(1.222,0.5)
\psline[linewidth=1pt](1.174,0)(1.174,0.5) \psline[linewidth=1pt](1.159,0)(1.159,0.5) \psline[linewidth=1pt](1.103,0)(1.103,0.5) \psline[linewidth=1pt](1.070,0)(1.070,0.5)
\psline[linewidth=1pt](1.024,0)(1.024,0.5) \psline[linewidth=1pt](0.963,0)(0.963,0.5) \psline[linewidth=1pt](0.952,0)(0.952,0.5) \psline[linewidth=1pt](0.903,0)(0.903,0.5)
\psline[linewidth=1pt](0.878,0)(0.878,0.5) \psline[linewidth=1pt](0.853,0)(0.853,0.5) \psline[linewidth=1pt](0.783,0)(0.783,0.5) \psline[linewidth=1pt](0.749,0)(0.749,0.5)
\psline[linewidth=1pt](0.726,0)(0.726,0.5) \psline[linewidth=1pt](0.705,0)(0.705,0.5) \psline[linewidth=1pt](0.695,0)(0.695,0.5) \psline[linewidth=1pt](0.671,0)(0.671,0.5)
\psline[linewidth=1pt](0.655,0)(0.655,0.5) \psline[linewidth=1pt](0.629,0)(0.629,0.5) \psline[linewidth=1pt](0.591,0)(0.591,0.5) \psline[linewidth=1pt](0.579,0)(0.579,0.5)
\psline[linewidth=1pt](0.574,0)(0.574,0.5) \psline[linewidth=1pt](0.535,0)(0.535,0.5) \psline[linewidth=1pt](0.523,0)(0.523,0.5) \psline[linewidth=1pt](0.494,0)(0.494,0.5)
\psline[linewidth=1pt](0.463,0)(0.463,0.5) \psline[linewidth=1pt](0.444,0)(0.444,0.5) \psline[linewidth=1pt](0.434,0)(0.434,0.5) \psline[linewidth=1pt](0.429,0)(0.429,0.5)
\psline[linewidth=1pt](0.393,0)(0.393,0.5) \psline[linewidth=1pt](0.373,0)(0.373,0.5) \psline[linewidth=1pt](0.369,0)(0.369,0.5) \psline[linewidth=1pt](0.345,0)(0.345,0.5)
\psline[linewidth=1pt](0.337,0)(0.337,0.5) \psline[linewidth=1pt](0.319,0)(0.319,0.5) \psline[linewidth=1pt](0.306,0)(0.306,0.5) \psline[linewidth=1pt](0.279,0)(0.279,0.5)
\psline[linewidth=1pt](0.274,0)(0.274,0.5) \psline[linewidth=1pt](0.255,0)(0.255,0.5) \psline[linewidth=1pt](0.246,0)(0.246,0.5) \psline[linewidth=1pt](0.232,0)(0.232,0.5)
\psline[linewidth=1pt](0.216,0)(0.216,0.5) \psline[linewidth=1pt](0.190,0)(0.190,0.5) \psline[linewidth=1pt](0.178,0)(0.178,0.5) \psline[linewidth=1pt](0.163,0)(0.163,0.5)
\psline[linewidth=1pt](0.144,0)(0.144,0.5) \psline[linewidth=1pt](0.133,0)(0.133,0.5) \psline[linewidth=1pt](0.122,0)(0.122,0.5) \psline[linewidth=1pt](0.111,0)(0.111,0.5)
\psline[linewidth=1pt](0.085,0)(0.085,0.5) \psline[linewidth=1pt](0.078,0)(0.078,0.5) \psline[linewidth=1pt](0.061,0)(0.061,0.5) \psline[linewidth=1pt](0.046,0)(0.046,0.5) \psline[linewidth=1pt](0.039,0)(0.039,0.5)
\end{pspicture}

\vskip 0.7 cm

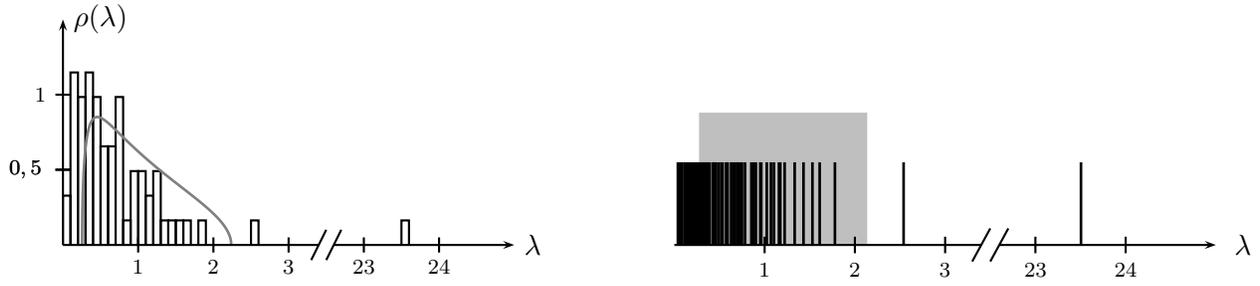
\noindent {\bf Fig. 1.} Left: histogram of eigenvalues for the correlation matrix of 61 stocks in 2004 and Mar\v{c}enku-Pastur theoretical distribution (solid line). Right: eigenvalues for the correlation matrix of 61 stocks in 2004 and purely random region.

\vskip 0.3 cm

The first striking feature is that the largest eigenvalue is more than ten times larger than the maximum value predicted for a purely random correlation matrix. About 72\% of the eigenvalues fall within the shaded region associated with pure noise, 15 of them fall below this region, and another one is above it.

The eigenvectors $e_1$ and $e_2$ for the two largest eigenvalues, $\lambda _1=23.505$, and $\lambda_2=2.540$, are represented in Figure 2 (first two graphs). The white bars represent positive values and the gray bars represent negative ones.



\vskip 0.4 cm

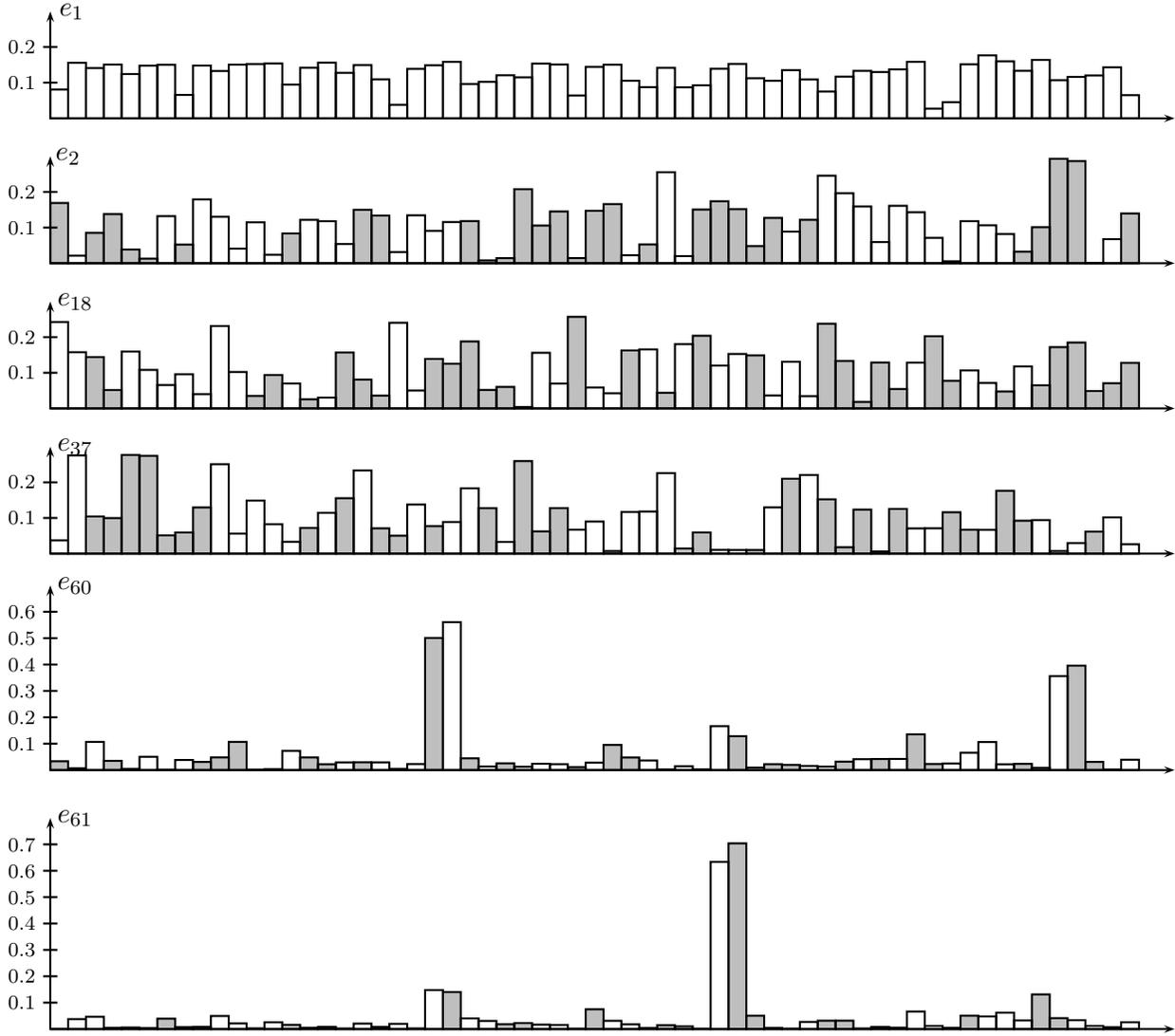
\noindent {\bf Fig. 2.} Eigenvectors of some fixed eigenvalues: $\lambda_{1}$, $\lambda_{2}$ (largest), $\lambda_{18}$, $\lambda_{37}$ (noise region), and $\lambda_{60}$, $\lambda_{61}$ (lowest eigenvalues).

\vskip 0.4 cm

The distribution of individual values of eigenvector $e_1$ is very similar for all the stocks considered, showing that all stocks contribute to this mode, which is considered ``the market mode''. For eigenvector $e_2$, one can see the prevalence of some stocks over others. In comparison, eigenvectors corresponding to the shaded region (Wishart region) do not show any preference for particular stocks.

The third and fourth graphs of Figure 2 show the distributions of the eigenvectors associated with the eigenvalues of two eigenvectors that are inside the Wishart region, $\lambda_{18}=0.853$, and $\lambda_{37}=0.393$. Note that there are no clearly defined stock structures.

We also show the eigenvectors corresponding to the two lowest eigenvalues of the correlation matrix, $\lambda_{60}=0.046$ and $\lambda_{62}=0.039$ (last two graphs in Figure 2). These eigenvectors corresponding to low eigenvalues represent ``portfolios'' of low risk, in opposition to the eigenvectors of the largest two eigenvetors, which represent the oscillations of the market and the common behavior of a cluster of stocks that behave in a similar way. Eigenvector $e_{61}$ represents a portfolio from which the investor buys PETR4 and short-sells PETR3, which are stocks belonging to the same company, Petrobras, and buys ELET3 and short-sells ELET6, which also belong to the same company, Eletrobras. Eigenvector $e_{60}$, in its turn, represents a portfolio from which the investor buys VALE3 and ELET6 and short-sells VALE5 and ELET3, which again are two pairs of stocks of the same companies, and also buys PETR3 and short-sells PETR4.

Considering the daily log-returns of portfolio $P_1$ built with eigenvector $e_{1}$ and the log-returns of the Ibovespa, which is an index that describes the general behavior of the São Paulo Stock Exchange, for the year 2004, the correlation between the two vectors is $0.9865$, which is a very strong indication that the portfolio $P_1$ corresponds to a combination of stocks that behave much like the market, although with a much larger volatility: the standard deviation of the returns of $P_1$ is $12.51\%$, and the standard deviation for the Ibovespa is $1.80\%$. The difference in volatilities is explained by the different weights that each stock has on the two indices.

The situation changes if we consider a portfolio built with eigenvector $e_{37}$, which corresponds to the noisy part of the eigenvalue spectrum: the correlation between this portfolio, $P_{37}$, and the Ibovespa is $0.1824$, and it has a standard deviation $1.72\%$, very close to the standard deviation of the Ibovespa. For the portfolio $P_{61}$, built with eigenvector $e_{61}$, which corresponds to the lowest eigenvalue, the correlation with the Ibovespa is $0.0932$, and its standard deviation is $0.44\%$. This portfolio presents the lowest correlation with the Ibovespa.

Previous works on the stock exchanges of emerging markets using Random Matrix Theory have been conducted by Wilcox and Gebbie (2004, 2007) for South Africa, Pan and Sinha (2007) for India, Nilantha, Ranasinghe and Malmini (2007) for Sri Lanka, and by Medina and Mansilla (2008) for Mexico. Their results show some differences between the stock exchanges of emerging markets and the stock exchanges of more developed ones, shuch as less liquidity for the stocks, and less integration of different sectors.

Eterovic and Eterovic (2013), following a similar methodology as ours, studied the stock market from Chile using Random Matrix Theory and performing an analysis of the eigenvalues and of the eigenvectors of the correlation matrix, and also studying the dynamics of those eigenvectors, revealing some structure based on some key industrial sectors of the Chilean economy. They also used Vector Autoregressive Analysis in order to pinpoint the main drivers of the Chilean stock market.

\subsection{Discussing the type of theoretical distribution to be used}

We said in Section 1 that the real probability distribution of the eigevalues of the correlation matrix, with the expception of the abnormally high eigenvalues, may be different from the Mar\v{c}enku-Pastur probability distribution and more similar to the one derived by Biroli, Bouchaud, and Potters (2007). Here we use three instruments to clarify this point. Figure 3 (left) shows the qq plot of the real eigenvalue distribution for 2004 and the probability distribution calculated using the Mar\v{c}enku-Pastur formula. A qq plot (quantile-quantile plot) consists on a graphical method for comparing two probability distributions by plotting their quantiles against each other. If the two distributions being compared are similar, then the points in the qq plot will approximately lie on the line correponding to the identity function, $y=x$. If the two distributions are linearly correlated, then the qq plot will correspond approximately to a straight line which is not necessarily the identity function. Figure 3 (right) shows the qq plot with the largest eigenvalue removed. Both figures show that the qq plot corresponds very strongly to a straight line, in particular for the eigenvalues considered to be inside the Wishart region. The same result can be obtained for the remaining years. From Figure 3, one may see that the highest eigenvalues are considered by the test as outliers, since most of the eigenvalues are in an approximately straight line.

\begin{pspicture}(-1,-0.5)(6,6)
\psline{->}(0,0)(5,0) \psline{->}(0,0)(0,5) \rput(5.5,0){MPq} \rput(0.5,5){Eq} \scriptsize \psline(1.667,-0.1)(1.667,0.1) \psline(3.333,-0.1)(3.333,0.1) \rput(-0.1,-0.2){0} \rput(1.667,-0.3){0.5} \rput(3.333,-0.3){1.0} \psline(-0.1,2)(0.1,2) \psline(-0.1,4)(0.1,4) \rput(-0.4,2){10} \rput(-0.4,4){20} %
\psdots[linewidth=0.2pt] (0.598,0.008) (0.648,0.009) (0.693,0.012) (0.735,0.016) (0.776,0.017) (0.816,0.022) (0.855,0.024) (0.894,0.027) (0.933,0.029) (0.972,0.033) (1.011,0.036) (1.050,0.038) (1.089,0.043) (1.129,0.046) (1.169,0.049) (1.209,0.051) (1.250,0.055) (1.291,0.056) (1.332,0.061) (1.374,0.064) (1.417,0.067) (1.460,0.069) (1.504,0.074) (1.549,0.075) (1.594,0.079) (1.640,0.086) (1.686,0.087) (1.733,0.089) (1.781,0.093) (1.830,0.099) (1.880,0.105) (1.930,0.107) (1.982,0.115) (2.034,0.116) (2.087,0.118) (2.142,0.126) (2.197,0.131) (2.254,0.134) (2.312,0.139) (2.371,0.141) (2.431,0.145) (2.493,0.150) (2.557,0.157) (2.622,0.171) (2.689,0.176) (2.757,0.181) (2.828,0.190) (2.901,0.193) (2.976,0.205) (3.054,0.214) (3.135,0.221) (3.219,0.232) (3.306,0.235) (3.398,0.244) (3.494,0.267) (3.596,0.286) (3.705,0.306) (3.823,0.322) (3.953,0.356) (4.099,0.508) (4.277,4.701)
\end{pspicture}
\begin{pspicture}(-3,-0.5)(6,6)
\psline{->}(0,0)(5,0) \psline{->}(0,0)(0,5) \rput(5.5,0){MPq} \rput(0.5,5){Eq} \scriptsize \psline(1.25,-0.1)(1.25,0.1) \psline(2.5,-0.1)(2.5,0.1) \psline(3.75,-0.1)(3.75,0.1) \rput(-0.1,-0.2){0} \rput(1.25,-0.3){0.5} \rput(2.5,-0.3){1.0} \rput(3.75,-0.3){1.5} \psline(-0.1,2)(0.1,2) \psline(-0.1,4)(0.1,4) \rput(-0.4,2){1} \rput(-0.4,4){2} %
\psdots[linewidth=0.2pt] (0.646,0.071) (0.700,0.082) (0.749,0.110) (0.794,0.140) (0.838,0.153) (0.881,0.200) (0.923,0.220) (0.966,0.239) (1.007,0.259) (1.049,0.294) (1.091,0.320) (1.134,0.343) (1.176,0.388) (1.219,0.418) (1.262,0.443) (1.305,0.459) (1.350,0.493) (1.394,0.502) (1.439,0.550) (1.484,0.575) (1.531,0.606) (1.577,0.622) (1.625,0.663) (1.673,0.671) (1.721,0.708) (1.771,0.771) (1.821,0.781) (1.872,0.799) (1.924,0.833) (1.976,0.889) (2.030,0.941) (2.085,0.962) (2.140,1.033) (2.197,1.042) (2.254,1.065) (2.313,1.132) (2.373,1.179) (2.434,1.208) (2.497,1.251) (2.561,1.268) (2.626,1.307) (2.693,1.347) (2.761,1.409) (2.832,1.536) (2.904,1.580) (2.978,1.625) (3.054,1.713) (3.133,1.733) (3.214,1.843) (3.298,1.927) (3.385,1.985) (3.476,2.086) (3.570,2.113) (3.669,2.200) (3.774,2.400) (3.884,2.577) (4.002,2.755) (4.129,2.898) (4.269,3.203) (4.427,4.571)
\end{pspicture}

\vskip 0.3 cm

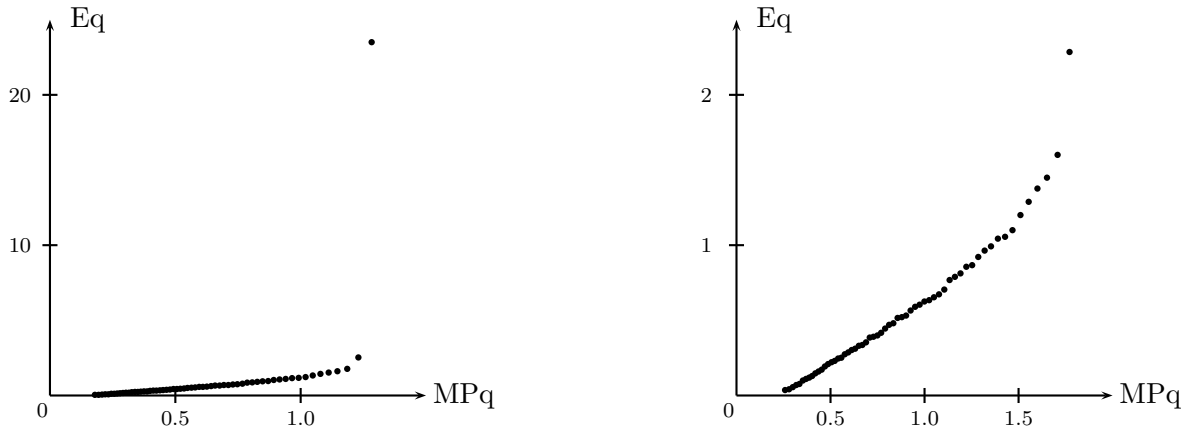
\noindent {\bf Fig. 3.} Qqplots using data for the year 2004, with all eigenvalues (left graph) and with the largest eigenvalue removed (right graph). The horizontal axis represents the quantiles of the Mar\v{c}enku-Pastur distribution relative to the data for the corresponding year, and the vertical axis represents que quantiles of the eigenvalue distribution of the correlation matrices obtained from the log-returns of the corresponding years.

\vskip 0.3 cm

The second instrument is to use the Kolmogorov-Smirnov test, which in its one-sample form compares a sample of a probability distribution with a reference probability distribution (in our case, the Mar\v{c}enku-Pastur distribution). The statistic quantifies a distance measure between the empirical distribution function of the sample and the cumulative distribution function of the reference distribution, and the null distribution of this statistic is calculated under the null hypothesis that the sample is drawn from the reference distribution.

Using the test for the eigenvalues obtained from the data from 2004, we obtained an average distance $D=0.1691$ and a p-value$=0.06108$, so that the test fails to reject the null hypothesis at the 1\% level. This means that, even considering the high eigenvalues, the probability distribution of the eigenvalues obtained from the correlation matrix correspond very closely to a Mar\v{c}enku-Pastur distribution. The results for 2004 and for the remaining years are summarized in Table 1.

\[ \begin{array}{c|c|c|c|c} \hline \text{Year} & \text{Number of stocks} & \text{Average distance} & \text{p-value} & \text{Rejects the null hypothesis?} \\ \hline 2004 & 61 & 0.1691 & 0.06108 & \text{No} \\ 2005 & 72 & 0.1508 & 0.07565 & \text{No} \\ 2006 & 86 & 0.1272 & 0.1239 & \text{No} \\ 2007 & 105 & 0.1058 & 0.1906 & \text{No} \\ 2008 & 148 & 0.1258 & 0.01851 & \text{No} \\ 2009 & 153 & 0.0846 & 0.2237 & \text{No} \\ \hline \end{array} \]

\noindent {\bf Table 1.} Results for the Kolmogorov-Smirnov test for the eigenvalue probability distribution of the correlation matrix for the log-returns of the years from 2004 to 2009 and the corresponding Mar\v{c}enku-Pastur distributions.

\vskip 0.3 cm

The third instrument is to calculate the correlation matrices based on randomized data, obtained by considering all time series and randomly changing their orders, individually. A total of 10,000 simulations of eigenvalues were made using such randomized data, obtaining a probability distribution for the eigenvalues which closely resembles the common Mar\v{c}enku-Pastur distribution, except for small differences in their borders, which now decay like a power law, in a similar way as the Wishart-Student distribution in Biroli, Bouchaud, \& Potters (2007), which behaves like $\lambda ^{-1-\mu /2}$ for larger values of the eigenvalue $\lambda $, where $\mu $ is the parameter that leads to the better fit of a t Student distribution for the time series of log-returns for each stock. 

In Figure 4, data relative to the year 2004 is presented. The real distributions are in block format, the Mar\v{c}enku-Pastur distributions are in gray, and the distributions resulting from the 10,000 simulations are in black lines. As it can be seen, the two distributions, the one obtained from randomized data and the one obtained from the Mar\v{c}enku-Pastur distribution, are nearly identical, except for a small difference at the latter's borders.

What we obtained by the 10,000 simulations was the distribution of time series that had the exact probability distributions as the original ones, be them closer to a Gaussian or a t Student distribution, but that were unrelated with one another. That is not a theoretical result, and it is independent of the type of distribution associated with the real data. Since what we wish to do is differentiate the real distribution from a distribution of uncorrelated time series, we thought it would be a fitting result to be used for deciding which eigenvalues were within the noise region. Due to the power law decaying character of the Wishart-Student distribution, the results are quite similar to the ones that would be obtained from that distribution.

\begin{pspicture}(-1,0)(3.5,3.8)
\psset{xunit=1,yunit=200}
\psline(0,0)(0,0.00328)(0.1,0.00328)(0.1,0) \psline(0.1,0)(0.1,0.01148)(0.2,0.01148)(0.2,0) \psline(0.2,0)(0.2,0.00984)(0.3,0.00984)(0.3,0) \psline(0.3,0)(0.3,0.01148)(0.4,0.01148)(0.4,0) \psline(0.4,0)(0.4,0.00984)(0.5,0.00984)(0.5,0) \psline(0.5,0)(0.5,0.00656)(0.6,0.00656)(0.6,0) \psline(0.6,0)(0.6,0.00656)(0.7,0.00656)(0.7,0) \psline(0.7,0)(0.7,0.00984)(0.8,0.00984)(0.8,0) \psline(0.8,0)(0.8,0.00164)(0.9,0.00164)(0.9,0) \psline(0.9,0)(0.9,0.00492)(1,0.00492)(1,0) \psline(1,0)(1,0.00492)(1.1,0.00492)(1.1,0) \psline(1.1,0)(1.1,0.00328)(1.2,0.00328)(1.2,0) \psline(1.2,0)(1.2,0.00492)(1.3,0.00492)(1.3,0) \psline(1.3,0)(1.3,0.00164)(1.4,0.00164)(1.4,0) \psline(1.4,0)(1.4,0.00164)(1.5,0.00164)(1.5,0) \psline(1.5,0)(1.5,0.00164)(1.6,0.00164)(1.6,0) \psline(1.6,0)(1.6,0.00164)(1.7,0.00164)(1.7,0) \psline(1.8,0)(1.8,0.00164)(1.9,0.00164)(1.9,0) \psline(2.5,0)(2.5,0.00164)(2.6,0.00164)(2.6,0) \psline(4.5,0)(4.5,0.00164)(4.6,0.00164)(4.6,0)
\psplot[linecolor=gray,plotpoints=500,linewidth=1pt]{0.2541}{2.2378}{2.2378 x sub x 0.2541 sub mul 0.5 exp x -1 exp mul 0.64706 mul 0.01 mul}
\psline (0,0.00000) (0.01,0.00000) (0.02,0.00000) (0.03,0.00000) (0.04,0.00000) (0.05,0.00000) (0.06,0.00000) (0.07,0.00000) (0.08,0.00000) (0.09,0.00000) (0.1,0.00000) (0.11,0.00000) (0.12,0.00000) (0.13,0.00000) (0.14,0.00000) (0.15,0.00000) (0.16,0.00000) (0.17,0.00000) (0.18,0.00000) (0.19,0.00000) (0.2,0.00000) (0.21,0.00001) (0.22,0.00004) (0.23,0.00017) (0.24,0.00049) (0.25,0.00128) (0.26,0.00251) (0.27,0.00402) (0.28,0.00497) (0.29,0.00573) (0.3,0.00615) (0.31,0.00657) (0.32,0.00686) (0.33,0.00716) (0.34,0.00743) (0.35,0.00778) (0.36,0.00784) (0.37,0.00795) (0.38,0.00808) (0.39,0.00819) (0.4,0.00814) (0.41,0.00812) (0.42,0.00830) (0.43,0.00848) (0.44,0.00839) (0.45,0.00850) (0.46,0.00841) (0.47,0.00860) (0.48,0.00832) (0.49,0.00831) (0.5,0.00836) (0.51,0.00831) (0.52,0.00835) (0.53,0.00843) (0.54,0.00823) (0.55,0.00822) (0.56,0.00829) (0.57,0.00819) (0.58,0.00820) (0.59,0.00822) (0.6,0.00805) (0.61,0.00814) (0.62,0.00784) (0.63,0.00804) (0.64,0.00798) (0.65,0.00780) (0.66,0.00785) (0.67,0.00793) (0.68,0.00758) (0.69,0.00773) (0.7,0.00756) (0.71,0.00739) (0.72,0.00770) (0.73,0.00763) (0.74,0.00748) (0.75,0.00743) (0.76,0.00760) (0.77,0.00721) (0.78,0.00713) (0.79,0.00730) (0.8,0.00722) (0.81,0.00716) (0.82,0.00720) (0.83,0.00702) (0.84,0.00721) (0.85,0.00672) (0.86,0.00695) (0.87,0.00679) (0.88,0.00676) (0.89,0.00694) (0.9,0.00673) (0.91,0.00646) (0.92,0.00673) (0.93,0.00661) (0.94,0.00669) (0.95,0.00656) (0.96,0.00644) (0.97,0.00639) (0.98,0.00627) (0.99,0.00635) (1,0.00621) (1.01,0.00620) (1.02,0.00616) (1.03,0.00604) (1.04,0.00628) (1.05,0.00607) (1.06,0.00603) (1.07,0.00594) (1.08,0.00588) (1.09,0.00580) (1.1,0.00602) (1.11,0.00572) (1.12,0.00574) (1.13,0.00570) (1.14,0.00567) (1.15,0.00554) (1.16,0.00563) (1.17,0.00546) (1.18,0.00547) (1.19,0.00571) (1.2,0.00536) (1.21,0.00541) (1.22,0.00527) (1.23,0.00530) (1.24,0.00519) (1.25,0.00523) (1.26,0.00508) (1.27,0.00499) (1.28,0.00514) (1.29,0.00505) (1.3,0.00498) (1.31,0.00505) (1.32,0.00491) (1.33,0.00488) (1.34,0.00490) (1.35,0.00474) (1.36,0.00480) (1.37,0.00471) (1.38,0.00472) (1.39,0.00467) (1.4,0.00463) (1.41,0.00451) (1.42,0.00442) (1.43,0.00448) (1.44,0.00443) (1.45,0.00433) (1.46,0.00445) (1.47,0.00433) (1.48,0.00433) (1.49,0.00420) (1.5,0.00431) (1.51,0.00415) (1.52,0.00410) (1.53,0.00396) (1.54,0.00403) (1.55,0.00401) (1.56,0.00396) (1.57,0.00396) (1.58,0.00388) (1.59,0.00383) (1.6,0.00379) (1.61,0.00370) (1.62,0.00376) (1.63,0.00367) (1.64,0.00361) (1.65,0.00352) (1.66,0.00359) (1.67,0.00356) (1.68,0.00349) (1.69,0.00335) (1.7,0.00347) (1.71,0.00342) (1.72,0.00329) (1.73,0.00324) (1.74,0.00315) (1.75,0.00321) (1.76,0.00318) (1.77,0.00306) (1.78,0.00317) (1.79,0.00299) (1.8,0.00292) (1.81,0.00290) (1.82,0.00282) (1.83,0.00288) (1.84,0.00273) (1.85,0.00270) (1.86,0.00262) (1.87,0.00273) (1.88,0.00259) (1.89,0.00253) (1.9,0.00258) (1.91,0.00239) (1.92,0.00245) (1.93,0.00225) (1.94,0.00235) (1.95,0.00222) (1.96,0.00232) (1.97,0.00223) (1.98,0.00216) (1.99,0.00205) (2,0.00195) (2.01,0.00200) (2.02,0.00197) (2.03,0.00185) (2.04,0.00180) (2.05,0.00180) (2.06,0.00173) (2.07,0.00157) (2.08,0.00157) (2.09,0.00147) (2.1,0.00151) (2.11,0.00133) (2.12,0.00128) (2.13,0.00124) (2.14,0.00117) (2.15,0.00107) (2.16,0.00098) (2.17,0.00083) (2.18,0.00075) (2.19,0.00068) (2.2,0.00052) (2.21,0.00049) (2.22,0.00037) (2.23,0.00030) (2.24,0.00025) (2.25,0.00017) (2.26,0.00017) (2.27,0.00012) (2.28,0.00011) (2.29,0.00007) (2.3,0.00005) (2.31,0.00003) (2.32,0.00003) (2.33,0.00002) (2.34,0.00002) (2.35,0.00001) (2.36,0.00000) (2.37,0.00000) (2.38,0.00000) (2.39,0.00000) (2.4,0.00000)
\psline{->}(0,0)(6,0) \psline{->}(0,0)(0,0.015) \rput(6.25,0){$\lambda $} \rput(0.5,0.015){$\rho (\lambda )$} \psline[linecolor=white,linewidth=2pt](3.4,0)(3.6,0) \psline(3.3,-0.001)(3.5,0.001) \psline(3.5,-0.001)(3.7,0.001) \scriptsize \psline(1,-0.0005)(1,0.0005) \rput(1,-0.0015){1} \psline(2,-0.0005)(2,0.0005) \rput(2,-0.0015){2} \psline(3,-0.0005)(3,0.0005) \rput(3,-0.0015){3} \psline(-0.1,0.005)(0.1,0.005) \rput(-0.5,0.005){$0,5$} \psline(-0.1,0.01)(0.1,0.01) \rput(-0.3,0.01){1} \psline(4,-0.0005)(4,0.0005) \rput(4,-0.0015){23} \psline(5,-0.0005)(5,0.0005) \rput(5,-0.0015){24} \psline(-0.1,0.005)(0.1,0.005) \rput(-0.5,0.005){$0,5$}
\end{pspicture}
\begin{pspicture}(-5,0)(3.5,3.8)
\psset{xunit=1,yunit=0.7}
\psline (0.2,3.34984) (0.3,3.13803) (0.4,3.05721) (0.5,3.06836) (0.6,2.93459) (0.7,2.66426) (0.8,2.66426) (0.9,2.49426) (1,2.57508) (1.1,2.25738) (1.2,2.17656) (1.3,2.10967) (1.4,1.98705) (1.5,1.81705) (1.6,1.66377) (1.7,1.41295) (1.8,1.27361) (1.9,1.16213) (2,0.88902) (2.1,0.83607) (2.2,0.63262) (2.3,0.50721) (2.4,0.41803) (2.5,0.28426) (2.6,0.28705) (2.7,0.20623) (2.8,0.18115) (2.9,0.11705) (3,0.07803) (3.1,0.05852) (3.2,0.04459) (3.3,0.03623) (3.4,0.03066) (3.5,0.01393) (3.6,0.00279) (3.7,0.00557) (3.8,0.00557) (3.9,0.00836) (4,0.00279)
\psline[linecolor=gray] (0.2,3.37774) (0.3,3.29234) (0.4,3.20541) (0.5,3.11682) (0.6,3.02640) (0.7,2.93398) (0.8,2.83936) (0.9,2.74229) (1,2.64250) (1.1,2.53966) (1.2,2.43336) (1.3,2.32312) (1.4,2.20834) (1.5,2.08826) (1.6,1.96190) (1.7,1.82794) (1.8,1.68455) (1.9,1.52906) (2,1.35732) (2.1,1.16211) (2.2,0.92873) (2.3,0.61514) (2.4,0.00000)
\psline{->}(0,0)(5,0) \psline{->}(0,0)(0,4.25) \rput(5.3,0){$\lambda $} \rput(0.5,4.25){$\rho (\lambda )$} \scriptsize \rput(0,-0.3){2} \psline(1,-0.1)(1,0.1) \rput(1,-0.3){2.1} \psline(2,-0.1)(2,0.1) \rput(2,-0.3){2.2} \psline(3,-0.1)(3,0.1) \rput(3,-0.3){2.3} \psline(4,-0.1)(4,0.1) \rput(4,-0.3){2.4} \psline(-0.1,0.85)(0.1,0.85) \rput(-0.6,0.85){0.0005} \psline(-0.1,1.7)(0.1,1.7) \rput(-0.5,1.7){0.001} \psline(-0.1,2.55)(0.1,2.55) \rput(-0.6,2.55){0.0015} \psline(-0.1,3.4)(0.1,3.4) \rput(-0.5,3.4){0.002}
\end{pspicture}

\vskip 0.7 cm

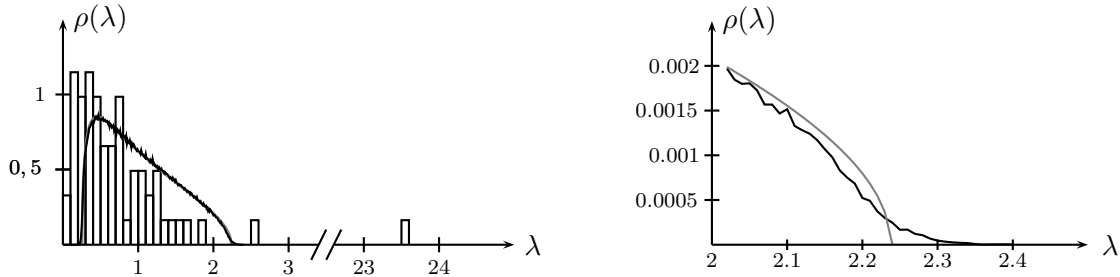
\noindent {\bf Fig. 4.} Left: histogram of eigenvalues for the correlation matrix of 61 stocks in 2004, the Mar\v{c}enku-Pastur theoretical distribution (gray line), and the distribution obtained from 10,000 simulations with randomized data (black line). Right: the same distribution, but centered around the border $\lambda_+$.

\vskip 0.3 cm

Using the Kolmogorov-Smirnov test in its two-sample form, which compares two samples of probability distributions, where the first one was the result of the simulations with randomized data and the second one was calculated from the
Mar\v{c}enku-Pastur distribution, the results were that the test fails to reject the null hypothesis that the two samples are from the same distribution. The test is sensitive to differences in both location and shape of the empirical cumulative distribution functions of the two samples, and the same results are obtained for the data obtained from the remaining years.

So, the conclusion is that we may be quite certain that the probability distribution of the eigenvalues correspond very closely to a Mar\v{c}enku-Pastur distribution, at least for the Wishart region.

\section{Building portfolios using Markowitz's Theory} \label{sec:marko}

In this section, we shall start by building portfolios using the $N=61$ stocks we are considering based on the correlation matrix of their returns (we shall reffer to log-returns as simply returns) in the year 2004, and then also for the remaining years. According to the usual portfolio theory, we can obtain $w$, the vector of weights of the portfolio due to each stock, by fixing the portfolio return (RE) and minimizing the risk (RI) of the portfolio, as in Markowitz (1952).

The return of the portfolio is given by
\begin{equation}
\label{return}
RE=w^{T}R\ ,
\end{equation}
where $R$ is the vector of average returns of each stock.

The risk is defined by the variance of the portfolio
\begin{equation}
\label{risk}
RI=w^{T}\Sigma_R w \ ,
\end{equation}
where $\Sigma_R$ is the estimated covariance matrix of the $N$ stocks.

The risk is then minimized with the constraint that the sum of all weights in the portfolio should be equal to one,
\begin{equation}
\label{sum1}
\sum_{i=1}^{N}w_i=1\ .
\end{equation}

One can do that for several values of the average return, leaving the coordinates of $w$ free to assume negative values, as well as positive ones, so that short selling is allowed. Short selling involves the borrowing and selling of stocks not owned by the seller. The stocks are then bought at the current market price and given back to the lender. This makes it possible to raise the returns of a portfolio, but at the cost of also raising its risk. In Finance, this is not always possible, or sometimes it is limited, and so we shall consider the case of no short selling for now.

In order to build a portfolio, the covariance matrix of a period of time (usually some months) prior to the period of investment is used together with a forecast of the expected returns. Those returns, which are unknown, may be approximated by many means, with relative degrees of success. There is a vast literature on the forecasting of returns, as in Elton,Gruber, Brown, \& Goetzmann (2009) and bibliography therein, but this does not concerns us in our study of how to improve the prediction of risk. So, in order to restrict ourselves to the analysis of the correlation matrix, we shall consider that our prediction of returns is the best one possible, which is a perfect forecast of expected returns. Of course, if we had a perfect forecast of expected returns, and we knew it was a perfect forecast of returns, we would not need to make any portfolio analysis. We use here the perfect forecast of expected returns in order to compare different ways of calculating risk in a fashion that is independent of the way one tries to forecast returns.

So, we first use the covariance matrix from 2004, together with the average returns of 2005 (perfect forecast of expected returns), in order to build minimum risk portfolios for 2005. Doing so, we build an {\sl efficient frontier}, which is a curve whose coordinates are the minimum risk for a given return. We also use the data from 2005, which means perfect forecasts of expected risk and expected return, in order to build an efficient frontier for the realized risk.

The covariance matrices for both the predicted and the realized risks are calculated using the formula
\begin{equation}
\label{cov}
\Sigma_R=\sigma^{T}\C \sigma \ ,
\end{equation}
where $\C$ is the correlation matrix obtained from the data and $\sigma $ is the vector of the standard deviations of the time series for the target year, in this case, 2005. So, both covariance matrices, the one for predicted risk (based on data from 2004), and the one for realized risk (based on data from 2005) are built on their respective correlation matrices, but with the standard deviations of 2005. This is done so as to isolate the effects of the correlation matrices alone, and not of the difference in volatility between the two years.

Figure 5 shows the predicted (gray lines) and realized (black lines) returns and risks of portfolios using the correlation matrices from 2004-2005, 2005-2006, 2006-2007, 2007-2008, 2008-2009, and 2009-2010. Note that the curves vary both in shape and in the values of risks and returns, and that the results of predicted and realized risks are particularly bad for years of high volatility.

Looking into the first graph of Figure 5, we have the predicted return and risk of portfolios using data from 2004 (gray line) and the realized return and risk using data from 2005 (black line). The graphs are made for positive values of returns, only, and go from the minimum possible average returns to the maximum possible average returns. Note that, for a given return, the predicted risk is sometimes smaller and sometimes higher than the realized one. This may lead to a false perception of how risky an investment truly is, and may cause wrong decisions by the portfolio manager.

One way to measure the agreement (AG) of the curves (predicted and realized risk) can be given by
\begin{equation}
\label{AG}
AG=\frac{1}{n}\sum_{i=1}^n\frac{RI_i^{real}-RI_i^{pred}}{RI_i^{pred}}\ ,
\end{equation}
where $RI_i^{real}$ is the realized risk and $Ri_i^{pred}$ is the predicted risk, both for $i=1,\cdots ,n$ values of fixed returns. In our case, this number is $AG=-0.073$ ($n=61$), what means that the predicted risk is, on average, $7\% $ smaller than the realized risk.

Since the Agreement measure consists on positive and negative values, which partially cancel one another, another measure, the Mean Squared Error (MSE), which is defined as
\begin{equation}
\label{MSNE}
MSE=\frac{1}{n}\sum_{i=1}^n\left( RI_i^{real}-RI_i^{pred}\right) ^2\ ,
\end{equation}
is also proposed. This measures the sum of the squared differences between two risk values (predicted and realized) with the same expected return. For the pair of years 2004-2005, the Mean Squared Error is $MSE=10.50 \times 10^{-11}$.

Columns 2 of Table 2 and of Table 3 show the results for the Agreement (AG) and the Mean Squared Error (MSE) for all pairs of years. The results for the Agreement (AG) mean that the forecasted risk for 2008 based on data from 2007, for example, was in average bellow the realized risk for 2008, and that the forecasted risk for 2009 based on data from 2008 was in average above the realized risk for 2008. The Minimum Squared Error (MSE) values show that the difference between predicted and realized risks was much larger for the pairs of years 2006-2007, 2007-2008, and 2008-2009. This was to be expected, since the MSE measures squared differences, which are larger if the risks are higher. Normalizing the MSE in order to account for this effect leads to other problems (when we wish to compare results with and without regression, as in the next session), but also points out that the results are worse for the pairs of years 2006-2007 and 2008-2009.



\vskip 0.3 cm

\noindent {\bf Fig. 5.} Predicted (gray lines) and realized (black lines) returns and risks of portfolios using the correlation matrices from 2004-2005, 2005-2006, 2006-2007, 2007-2008, 2008-2009, and 2009-2010.

\vskip 0.3 cm

In particular, one may notice that the returns and risks of the optimal portfolios for the data comprising the years 2006 and 2007 are much higher than for the other years. This was due to the stocks from Telebras (telecommunications), which had a very strong variation in price and risk during 2007. This was caused by rumors that the Federal Government would use this nearly deactivated company as the means to provide large bandwidth Internet services to 64\% of the households of the country. The rumors were not realized, and the stocks of Telebras fell again to low values.

Although we could have removed the time series of the stocks from Telebras from our sample, we decided to keep them, since nothing stoped a broker investing in those stocks, so they should be included in a possible portfolio, and since it makes an interesting case in our study of how volatility may affect portfolios, even the volatility of a single stock.

Another comment of importance is that the values of the AG and of the MSE depend on the number of points that are plotted in each graph. We decided to plot 100 points for each graph, starting with the minimum possible positive average return and finishing at the maximum possible average return for a single stock. So, the portfolio with the lowest risk is not necessarily made of a single stock. Since negative returns are undesirable for investors, we did not consider them as part of the efficient frontier.

\subsection{Building portfolios with a cleaned correlation matrix} \label{sec:clean}

The situation may be improved by trying to remove some of the noise of the correlation matrices of 2004 and 2005 returns. One way this can be done is by building a diagonal matrix $D$ where the elements of the diagonal are the eigenvalues of the original correlation matrix, but now with all eigenvalues corresponding to noise (those between $\lambda_-$ and $\lambda_+$) replaced by their average, as it is done in Laloux, Cizeau, Bouchaud, \& Potters (2000), Rosenow, Plerou, Gopikrishnan, \& Stanley (2002), Plerou, Gopikrishnan, Rosenow, Amaral, Guhr, \& Stanley (2002), Sharifi, Crane, Shamaie, \& Ruskin (2004), and Conlon, Ruskin, \& Crane (2007). In our present case, this average is $\bar \lambda =0.748$ for the eigenvalues based on data from 2004 and $\bar \lambda =0.790$ for the eigenvalues based on data from 2005. The cleaned correlation matrix is then built using the formula
\begin{equation}
\label{diag}
\C_{\text{clean}}=PDP^{-1}\ ,
\end{equation}
where $P$ are matrices whose columns are the eigenvectors of the original correlation matrix. The cleaned correlation matrix is then built using the average standard deviation of returns of the realized data.

Calculating now the efficient frontier built with the covariance matrix obtained from the cleaned correlation matrix of 2004, together with the average returns of 2005 (perfect forecast of expected returns), gray line, and comparing with the real curve calculated with the covariance matrix obtained from the cleaned correlation matrix of 2005, black line, we obtain the results represented in Figure 6.

The normalized difference between predicted and realized risks has now gone from $AG=-0.073$ (Figure 5) to $AG=0.033$ (Figure 6), what means that the predicted risk is, in average, $3\% $ larger than the realized risk. This is an improvement on the previous result and shows how the cleaning of the correlation matrix may help building portfolios which account best for the realized risk based on previous data. At the same time, the Mean Squared Error goes from $MSE=10.50 \times 10^{-11}$ to $MSE=24.99 \times 10^{-11}$, a worse result than the previous one. The difference is unexpected and is dificult to realize by visual inspection, only.

\begin{pspicture}(-6,-0.2)(5,4.5)
\psset{xunit=1,yunit=1}
\psline{->}(0,0)(5,0) \psline{->}(0,0)(0,4) \rput(5.3,0){$RI$} \rput(0.5,4){$RE$} \scriptsize \psline(1,-0.1)(1,0.1) \psline(2,-0.1)(2,0.1) \psline(3,-0.1)(3,0.1) \psline(4,-0.1)(4,0.1) \rput(-0.1,-0.2){0} \rput(1,-0.3){0.0001} \rput(2,-0.3){0.0002} \rput(3,-0.3){0.0003} \rput(4,-0.3){0.0004} \psline(-0.1,1)(0.1,1) \psline(-0.1,2)(0.1,2) \psline(-0.1,3)(0.1,3) \rput(-0.5,1){0.001} \rput(-0.5,2){0.002} \rput(-0.5,3){0.003}
\psline[linecolor=gray] (0.682,0.077) (0.682,0.111) (0.683,0.145) (0.684,0.180) (0.685,0.214) (0.687,0.248) (0.689,0.282) (0.692,0.316) (0.695,0.350) (0.699,0.384) (0.703,0.418) (0.708,0.453) (0.713,0.487) (0.719,0.521) (0.725,0.555) (0.731,0.589) (0.738,0.623) (0.745,0.657) (0.753,0.691) (0.761,0.726) (0.769,0.760) (0.777,0.794) (0.786,0.828) (0.796,0.862) (0.805,0.896) (0.815,0.930) (0.825,0.965) (0.836,0.999) (0.847,1.033) (0.858,1.067) (0.869,1.101) (0.881,1.135) (0.894,1.169) (0.906,1.203) (0.919,1.238) (0.933,1.272) (0.946,1.306) (0.960,1.340) (0.975,1.374) (0.990,1.408) (1.005,1.442) (1.020,1.476) (1.036,1.511) (1.053,1.545) (1.069,1.579) (1.087,1.613) (1.105,1.647) (1.123,1.681) (1.143,1.715) (1.163,1.749) (1.184,1.784) (1.205,1.818) (1.227,1.852) (1.250,1.886) (1.274,1.920) (1.299,1.954) (1.324,1.988) (1.351,2.022) (1.379,2.057) (1.408,2.091) (1.439,2.125) (1.471,2.159) (1.504,2.193) (1.538,2.227) (1.573,2.261) (1.611,2.295) (1.651,2.330) (1.692,2.364) (1.736,2.398) (1.782,2.432) (1.829,2.466) (1.879,2.500) (1.931,2.534) (1.984,2.568) (2.040,2.603) (2.098,2.637) (2.158,2.671) (2.219,2.705) (2.283,2.739) (2.349,2.773) (2.418,2.807) (2.489,2.841) (2.563,2.876) (2.640,2.910) (2.719,2.944) (2.802,2.978) (2.887,3.012) (2.975,3.046) (3.067,3.080) (3.166,3.115) (3.274,3.149) (3.391,3.183) (3.517,3.217) (3.652,3.251) (3.796,3.285) (3.949,3.319) (4.111,3.353) (4.282,3.388) (4.462,3.422) (4.651,3.456)
% Realized
\psline (0.567,0.077) (0.571,0.111) (0.576,0.145) (0.580,0.180) (0.585,0.214) (0.590,0.248) (0.596,0.282) (0.602,0.316) (0.608,0.350) (0.614,0.384) (0.621,0.418) (0.628,0.453) (0.636,0.487) (0.643,0.521) (0.652,0.555) (0.660,0.589) (0.669,0.623) (0.678,0.657) (0.688,0.691) (0.697,0.726) (0.708,0.760) (0.718,0.794) (0.729,0.828) (0.740,0.862) (0.752,0.896) (0.764,0.930) (0.776,0.965) (0.788,0.999) (0.801,1.033) (0.814,1.067) (0.828,1.101) (0.842,1.135) (0.856,1.169) (0.871,1.203) (0.886,1.238) (0.901,1.272) (0.917,1.306) (0.933,1.340) (0.949,1.374) (0.966,1.408) (0.983,1.442) (1.001,1.476) (1.019,1.511) (1.037,1.545) (1.056,1.579) (1.076,1.613) (1.095,1.647) (1.116,1.681) (1.136,1.715) (1.157,1.749) (1.179,1.784) (1.201,1.818) (1.223,1.852) (1.246,1.886) (1.269,1.920) (1.293,1.954) (1.317,1.988) (1.342,2.022) (1.368,2.057) (1.394,2.091) (1.420,2.125) (1.447,2.159) (1.475,2.193) (1.503,2.227) (1.532,2.261) (1.561,2.295) (1.591,2.330) (1.621,2.364) (1.654,2.398) (1.688,2.432) (1.724,2.466) (1.762,2.500) (1.802,2.534) (1.844,2.568) (1.887,2.603) (1.932,2.637) (1.979,2.671) (2.028,2.705) (2.078,2.739) (2.131,2.773) (2.187,2.807) (2.244,2.841) (2.305,2.876) (2.367,2.910) (2.432,2.944) (2.500,2.978) (2.572,3.012) (2.648,3.046) (2.730,3.080) (2.818,3.115) (2.918,3.149) (3.028,3.183) (3.150,3.217) (3.283,3.251) (3.426,3.285) (3.581,3.319) (3.751,3.353) (3.951,3.388) (4.201,3.422) (4.500,3.456)
\normalsize \rput(3.5,2){\boxed{2004-2005}}
\end{pspicture}

\vskip 0.3 cm

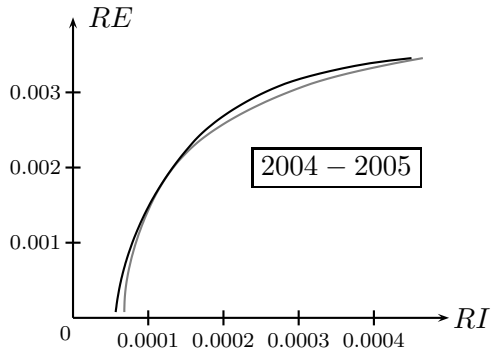
\noindent {\bf Fig. 6.} Predicted (gray lines) and realized (black lines) returns and risks of portfolios using the cleaned correlation matrices from 2004 to 2005.

\vskip 0.3 cm

Columns 2 and 4 of Table 2 show the results for the Agreement (AG) and the Mean Squared Error (MSE) for all pairs of years, without and with cleaning. In spite of the results for 2004-2005, all results for the remainnig years are consistently worse for the cleaned correlation matrices than the ones previously obtained, what is an unexpected result.

\subsection{Minimizing Systemic risk} \label{sec:risk}

When trying to predict the future expected risk of a portfolio, the volatility due to market movements may make it a difficult task, since one obtains a structure of dependence between the assets and the market, and not solely the dependence between assets. As an example, the prediction for 2008 using data from 2007 grossly underestimates the risk of 2008, since 2007 was a year with relatively low volatility while 2008 witnessed the height of the USA Subprime Mortgage Crisis. Similarly, risk prediction for 2009 using data from 2008 overestimates the risk for 2009.

The most common way to remove this so called systemic risk is to use a single index model, where all log returns $R_t$ are written in terms of a first degree function of a market index $I_t$, as, for example, the Ibovespa, plus an error $E_t$:
\begin{equation}
\label{singleindex}
R_t=a+bI_t+E_t\ .
\end{equation}
The coefficients $a$ and $b$ are estimated for each equity using simple linear regression.

As an alternative to the use of the Ibovespa as the market index, one may use the index obtained by the log returns of the portfolio of stocks that may be built using the eingenvector corresponding to the highest eigenvalue of the correlation matrix of those same stocks. As we showed for the data concerning the year 2004, both this index and the Ibovespa are very highly correlated, so the results should not be substantially altered by using any of those two indices.

We then calculated the residuals for all stocks being considered for each pair of years using the portfolios built from the eigenvector of the highest eigenvalue for each time period being studied as the market index. We then proceed into building portfolios using the correlation matrices between those residuals and also the cleaned correlation matrices. The resulting efficient frontiers for the pair 2004-2005 are drawn in Figure 7. Once again, the predicted results are in gray and the realized results are in black (almost indistinguishable in the graphs).

\begin{pspicture}(-1,-0.2)(5,4.5)
\psset{xunit=1,yunit=1}
\psline{->}(0,0)(5,0) \psline{->}(0,0)(0,4) \rput(5.3,0){$RI$} \rput(0.5,4){$RE$} \scriptsize \psline(1,-0.1)(1,0.1) \psline(2,-0.1)(2,0.1) \psline(3,-0.1)(3,0.1) \psline(4,-0.1)(4,0.1) \rput(-0.1,-0.2){0} \rput(1,-0.3){0.0001} \rput(2,-0.3){0.0002} \rput(3,-0.3){0.0003} \rput(4,-0.3){0.0004} \psline(-0.1,1)(0.1,1) \psline(-0.1,2)(0.1,2) \psline(-0.1,3)(0.1,3) \rput(-0.5,1){0.001} \rput(-0.5,2){0.002} \rput(-0.5,3){0.003}
\psline[linecolor=gray] (0.0081,0.0772) (0.0070,0.1113) (0.0059,0.1455) (0.0049,0.1796) (0.0040,0.2137) (0.0032,0.2478) (0.0025,0.2820) (0.0018,0.3161) (0.0013,0.3502) (0.0009,0.3843) (0.0005,0.4185) (0.0003,0.4526) (0.0001,0.4867) (0.0000,0.5209) (0.0000,0.5550) (0.0001,0.5891) (0.0003,0.6232) (0.0006,0.6574) (0.0010,0.6915) (0.0014,0.7256) (0.0020,0.7597) (0.0026,0.7939) (0.0034,0.8280) (0.0042,0.8621) (0.0051,0.8963) (0.0062,0.9304) (0.0073,0.9645) (0.0086,0.9986) (0.0100,1.0328) (0.0115,1.0669) (0.0131,1.1010) (0.0148,1.1351) (0.0167,1.1693) (0.0187,1.2034) (0.0209,1.2375) (0.0233,1.2716) (0.0258,1.3058) (0.0287,1.3399) (0.0317,1.3740) (0.0351,1.4082) (0.0387,1.4423) (0.0427,1.4764) (0.0469,1.5105) (0.0515,1.5447) (0.0565,1.5788) (0.0619,1.6129) (0.0678,1.6470) (0.0742,1.6812) (0.0811,1.7153) (0.0884,1.7494) (0.0963,1.7836) (0.1047,1.8177) (0.1137,1.8518) (0.1232,1.8859) (0.1333,1.9201) (0.1442,1.9542) (0.1558,1.9883) (0.1685,2.0224) (0.1825,2.0566) (0.1978,2.0907) (0.2146,2.1248) (0.2331,2.1589) (0.2532,2.1931) (0.2750,2.2272) (0.2986,2.2613) (0.3239,2.2955) (0.3512,2.3296) (0.3811,2.3637) (0.4135,2.3978) (0.4490,2.4320) (0.4877,2.4661) (0.5298,2.5002) (0.5754,2.5343) (0.6247,2.5685) (0.6776,2.6026) (0.7346,2.6367) (0.7957,2.6709) (0.8611,2.7050) (0.9312,2.7391) (1.0071,2.7732) (1.0894,2.8074) (1.1783,2.8415) (1.2744,2.8756) (1.3780,2.9097) (1.4890,2.9439) (1.6074,2.9780) (1.7332,3.0121) (1.8665,3.0462) (2.0073,3.0804) (2.1560,3.1145) (2.3144,3.1486) (2.4837,3.1828) (2.6650,3.2169) (2.8587,3.2510) (3.0649,3.2851) (3.2836,3.3193) (3.5202,3.3534) (3.7878,3.3875) (4.0981,3.4216) (4.4512,3.4558)
% Realized
\psline (0.0133,0.0772) (0.0115,0.1113) (0.0099,0.1455) (0.0085,0.1796) (0.0071,0.2137) (0.0059,0.2478) (0.0047,0.2820) (0.0037,0.3161) (0.0029,0.3502) (0.0021,0.3843) (0.0015,0.4185) (0.0009,0.4526) (0.0005,0.4867) (0.0002,0.5209) (0.0001,0.5550) (0.0000,0.5891) (0.0001,0.6232) (0.0002,0.6574) (0.0005,0.6915) (0.0010,0.7256) (0.0015,0.7597) (0.0021,0.7939) (0.0029,0.8280) (0.0038,0.8621) (0.0048,0.8963) (0.0060,0.9304) (0.0073,0.9645) (0.0088,0.9986) (0.0104,1.0328) (0.0122,1.0669) (0.0141,1.1010) (0.0162,1.1351) (0.0185,1.1693) (0.0209,1.2034) (0.0236,1.2375) (0.0264,1.2716) (0.0295,1.3058) (0.0329,1.3399) (0.0366,1.3740) (0.0406,1.4082) (0.0449,1.4423) (0.0495,1.4764) (0.0545,1.5105) (0.0598,1.5447) (0.0656,1.5788) (0.0717,1.6129) (0.0784,1.6470) (0.0855,1.6812) (0.0931,1.7153) (0.1012,1.7494) (0.1099,1.7836) (0.1192,1.8177) (0.1290,1.8518) (0.1395,1.8859) (0.1507,1.9201) (0.1625,1.9542) (0.1752,1.9883) (0.1886,2.0224) (0.2029,2.0566) (0.2181,2.0907) (0.2344,2.1248) (0.2520,2.1589) (0.2709,2.1931) (0.2911,2.2272) (0.3130,2.2613) (0.3368,2.2955) (0.3625,2.3296) (0.3902,2.3637) (0.4201,2.3978) (0.4524,2.4320) (0.4873,2.4661) (0.5253,2.5002) (0.5667,2.5343) (0.6120,2.5685) (0.6619,2.6026) (0.7165,2.6367) (0.7763,2.6709) (0.8419,2.7050) (0.9134,2.7391) (0.9910,2.7732) (1.0751,2.8074) (1.1663,2.8415) (1.2645,2.8756) (1.3698,2.9097) (1.4826,2.9439) (1.6032,2.9780) (1.7317,3.0121) (1.8687,3.0462) (2.0144,3.0804) (2.1686,3.1145) (2.3339,3.1486) (2.5127,3.1828) (2.7049,3.2169) (2.9107,3.2510) (3.1299,3.2851) (3.3626,3.3193) (3.6091,3.3534) (3.8751,3.3875) (4.1649,3.4216) (4.4865,3.4558)
\end{pspicture}
\begin{pspicture}(-3.8,-0.2)(5,4.5)
\psset{xunit=1,yunit=1}
\psline{->}(0,0)(5,0) \psline{->}(0,0)(0,4) \rput(5.3,0){$RI$} \rput(0.5,4){$RE$} \scriptsize \psline(1,-0.1)(1,0.1) \psline(2,-0.1)(2,0.1) \psline(3,-0.1)(3,0.1) \psline(4,-0.1)(4,0.1) \rput(-0.1,-0.2){0} \rput(1,-0.3){0.0001} \rput(2,-0.3){0.0002} \rput(3,-0.3){0.0003} \rput(4,-0.3){0.0004} \psline(-0.1,1)(0.1,1) \psline(-0.1,2)(0.1,2) \psline(-0.1,3)(0.1,3) \rput(-0.5,1){0.001} \rput(-0.5,2){0.002} \rput(-0.5,3){0.003}
\psline[linecolor=gray] (0.0087,0.0772) (0.0072,0.1113) (0.0057,0.1455) (0.0044,0.1796) (0.0033,0.2137) (0.0022,0.2478) (0.0013,0.2820) (0.0005,0.3161) (-0.0002,0.3502) (-0.0007,0.3843) (-0.0012,0.4185) (-0.0015,0.4526) (-0.0017,0.4867) (-0.0019,0.5209) (-0.0019,0.5550) (-0.0017,0.5891) (-0.0015,0.6232) (-0.0012,0.6574) (-0.0007,0.6915) (-0.0002,0.7256) (0.0005,0.7597) (0.0013,0.7939) (0.0022,0.8280) (0.0033,0.8621) (0.0044,0.8963) (0.0057,0.9304) (0.0072,0.9645) (0.0088,0.9986) (0.0105,1.0328) (0.0123,1.0669) (0.0143,1.1010) (0.0164,1.1351) (0.0187,1.1693) (0.0211,1.2034) (0.0236,1.2375) (0.0263,1.2716) (0.0292,1.3058) (0.0323,1.3399) (0.0356,1.3740) (0.0390,1.4082) (0.0426,1.4423) (0.0464,1.4764) (0.0505,1.5105) (0.0549,1.5447) (0.0595,1.5788) (0.0644,1.6129) (0.0697,1.6470) (0.0754,1.6812) (0.0814,1.7153) (0.0879,1.7494) (0.0949,1.7836) (0.1023,1.8177) (0.1103,1.8518) (0.1191,1.8859) (0.1286,1.9201) (0.1389,1.9542) (0.1502,1.9883) (0.1628,2.0224) (0.1766,2.0566) (0.1921,2.0907) (0.2091,2.1248) (0.2277,2.1589) (0.2481,2.1931) (0.2702,2.2272) (0.2941,2.2613) (0.3200,2.2955) (0.3483,2.3296) (0.3789,2.3637) (0.4119,2.3978) (0.4477,2.4320) (0.4863,2.4661) (0.5282,2.5002) (0.5734,2.5343) (0.6219,2.5685) (0.6740,2.6026) (0.7298,2.6367) (0.7894,2.6709) (0.8527,2.7050) (0.9197,2.7391) (0.9912,2.7732) (1.0682,2.8074) (1.1514,2.8415) (1.2410,2.8756) (1.3372,2.9097) (1.4405,2.9439) (1.5509,2.9780) (1.6685,3.0121) (1.7940,3.0462) (1.9305,3.0804) (2.0787,3.1145) (2.2385,3.1486) (2.4099,3.1828) (2.5934,3.2169) (2.7905,3.2510) (3.0013,3.2851) (3.2282,3.3193) (3.4734,3.3534) (3.7509,3.3875) (4.0723,3.4216) (4.4375,3.4558)
% Realized
\psline (0.0451,0.0772) (0.0433,0.1113) (0.0417,0.1455) (0.0401,0.1796) (0.0387,0.2137) (0.0374,0.2478) (0.0363,0.2820) (0.0352,0.3161) (0.0343,0.3502) (0.0335,0.3843) (0.0329,0.4185) (0.0323,0.4526) (0.0319,0.4867) (0.0316,0.5209) (0.0314,0.5550) (0.0313,0.5891) (0.0314,0.6232) (0.0316,0.6574) (0.0319,0.6915) (0.0323,0.7256) (0.0329,0.7597) (0.0336,0.7939) (0.0344,0.8280) (0.0353,0.8621) (0.0364,0.8963) (0.0375,0.9304) (0.0388,0.9645) (0.0403,0.9986) (0.0418,1.0328) (0.0435,1.0669) (0.0453,1.1010) (0.0472,1.1351) (0.0492,1.1693) (0.0514,1.2034) (0.0538,1.2375) (0.0564,1.2716) (0.0591,1.3058) (0.0620,1.3399) (0.0651,1.3740) (0.0685,1.4082) (0.0722,1.4423) (0.0761,1.4764) (0.0804,1.5105) (0.0851,1.5447) (0.0900,1.5788) (0.0954,1.6129) (0.1012,1.6470) (0.1074,1.6812) (0.1141,1.7153) (0.1212,1.7494) (0.1289,1.7836) (0.1372,1.8177) (0.1460,1.8518) (0.1555,1.8859) (0.1657,1.9201) (0.1766,1.9542) (0.1883,1.9883) (0.2009,2.0224) (0.2143,2.0566) (0.2288,2.0907) (0.2443,2.1248) (0.2610,2.1589) (0.2790,2.1931) (0.2985,2.2272) (0.3194,2.2613) (0.3418,2.2955) (0.3661,2.3296) (0.3925,2.3637) (0.4212,2.3978) (0.4527,2.4320) (0.4870,2.4661) (0.5245,2.5002) (0.5654,2.5343) (0.6097,2.5685) (0.6574,2.6026) (0.7091,2.6367) (0.7672,2.6709) (0.8319,2.7050) (0.9033,2.7391) (0.9815,2.7732) (1.0663,2.8074) (1.1583,2.8415) (1.2577,2.8756) (1.3645,2.9097) (1.4787,2.9439) (1.6006,2.9780) (1.7310,3.0121) (1.8700,3.0462) (2.0177,3.0804) (2.1740,3.1145) (2.3391,3.1486) (2.5154,3.1828) (2.7041,3.2169) (2.9053,3.2510) (3.1189,3.2851) (3.3468,3.3193) (3.5928,3.3534) (3.8575,3.3875) (4.1475,3.4216) (4.4773,3.4558)
\end{pspicture}

\vskip 0.3 cm

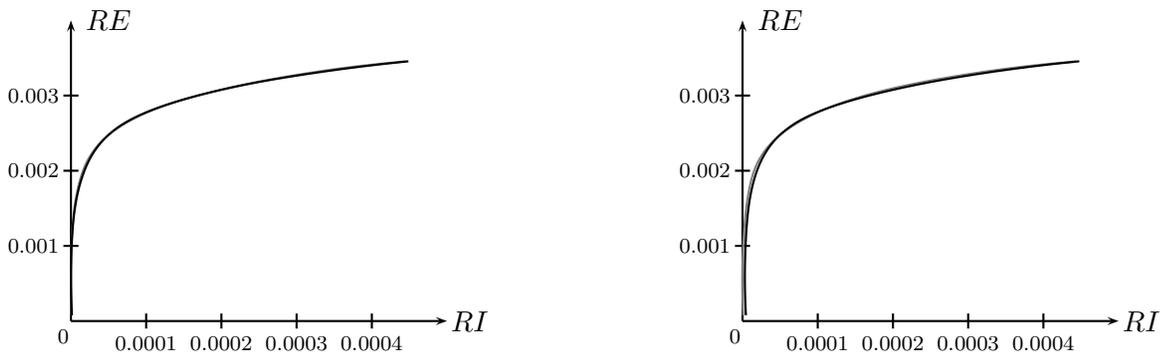
\noindent {\bf Fig. 7.} Predicted (gray lines) and realized (black lines) returns and risks of portfolios using the residues obtained from the single index regression with the original correlation matrices (left plots) and the cleaned correlation matrices (right plots) from 2004 to 2005.

\vskip 0.3 cm

The Agreement measure is now $AG=0.547$ for the original correlation matrix and $AG=-3.394$ for the cleaned correlation matrix. When compared with the results without the regression with the single index model, it seems like the results are much worse. For the Mean Square Error, we have $MSE=0.46 \times 10^{-11}$ for the original correlation matrix and $MSE=2.13 \times 10^{-11}$ for the cleaned correlation matrix, both much better results than for the original data. Nevertheless, the result with the cleaning procedure is not as good as the one without it.

A simple visual inspection will reveal that the results for the cleaned correlation matrix are much better than the results previously obtained. This shows that the Agreement measure (AG) is inadequate for comparing results obtained with and without a regression with a single index model. Since this measure, in order to normalize results, is divided by $R_i^{pred}$, it puts a lot of weigth in returns that are very close to zero. By looking at the graphs for the portfolios based on data from 2004, figures 5, 6 and 7, one may see that the efficient frontiers for the residues of the regression are much closer to zero, and that is the reason for the bad results in $AG$. Removing the normalization will lead to larger values when volatility and returns are higher.

An alternative which is not dependent on the volatilities of risks and returns is the {\sl angle between vectors}. Given equal returns, and considering the vectors $RI^{pred}$ and $RI^{real}$ given, respectivelly, by the values of $RI_i^{pred}$ and of $RI_i^{real}$, $i=1,\cdots ,n$, one may calculate the angle $\theta $ between them as
\begin{equation}
\label{angle}
\cos \theta =\frac{\left< RI^{pred},RI^{real}\right> }{\left| \left| RI^{pred} \right| \right| \cdot \left| \left| RI^{real} \right| \right| }=\frac{\sum_{i=1}^nRI_i^{pred}RI_i^{real}}{\sqrt{\sum_{i=1}^n\left( RI_i^{pred}\right) ^2}\sqrt{\sum_{i=1}^n\left( RI_i^{real}\right) ^2}}\ .
\end{equation}

\subsection{Results} \label{sec:results}

We shall now summarize the results obtained for each of the years considered in this article. We shall make a comparison of the predicted risk and the realized risk for the years 2004 to 2010 calculating the agreement $AG$, the mean squared error $MSE$, and the angle $\theta $.

Table 2 presents the results for the agreement measure (AG), the Mean Squared Error (MSE), and for the angles between vectors ($\theta $), calculated with and without the cleaning of the correlation matrix and with and without the regression for the removal of the market effect. The angles are shown in degrees so as to facilitate intuition.

\[ \begin{array}{c|ccccc} \hline \multicolumn{1}{c}{\boxed{\text{\bf Agreement (AG)}}} & \multicolumn{2}{c}{\text{Without Cleaning}} &	\hspace{2mm} & \multicolumn{2}{c}{\text{With Cleaning}} \\ \text{Previous-Predicted} & \text{No Regression} & \text{Regression} & \hspace{2mm} & \text{No Regression} & \text{Regression} \\ \hline \hline
2004-2005 &	-0.073 & 0.547 & \hspace{2mm} & \mathbf{-0.066} & -3.394 \\
2005-2006 &	\mathbf{-0.017} & 0.177 & \hspace{2mm} & -0.039 & -0.476 \\
2006-2007 &	0.186 & \mathbf{0.064} & \hspace{2mm} & 0.225 & -0.486 \\
2007-2008 &	\mathbf{0.119} & 0.314 & \hspace{2mm} & 0.143 & 0.330 \\
2008-2009 &	-0.429 &  \mathbf{0.082} & \hspace{2mm} & -0.439 & -0.872 \\
2009-2010 &	0.073 & 0.737 & \hspace{2mm} & \mathbf{0.033} & -3.876 \\
\hline \hline \multicolumn{1}{c}{\boxed{\text{\bf Mean Squared Error (MSE)}}} & \multicolumn{2}{c}{\text{Without Cleaning}} &	\hspace{2mm} & \multicolumn{2}{c}{\text{With Cleaning}} \\ \text{Previous-Predicted} & \text{No Regression} & \text{Regression} & \hspace{2mm} & \text{No Regression} & \text{Regression} \\ \hline \hline
2004-2005 &	10.50 \times 10^{-11} & \mathbf{0.46 \times 10^{-11}} & \hspace{2mm} & 24.99 \times 10^{-11} & 2.13 \times 10^{-11} \\
2005-2006 &	2.01 \times 10^{-11} & \mathbf{0.80 \times 10^{-11}} & \hspace{2mm} & 2.60 \times 10^{-11} & 2.85 \times 10^{-11} \\
2006-2007 &	10.87 \times 10^{-8} & 11.03 \times 10^{-8} & \hspace{2mm} & \mathbf{0.12 \times 10^{-8}} & 1.53 \times 10^{-8} \\
2007-2008 &	 10.56 \times 10^{-10} & \mathbf{5.13 \times 10^{-10}} & \hspace{2mm} & 15.19 \times 10^{-10} & 7.66 \times 10^{-10} \\
2008-2009 &	 39.79 \times 10^{-10} &  2.29 \times 10^{-10} & \hspace{2mm} & 35.94 \times 10^{-10} & \mathbf{0.59 \times 10^{-10}} \\
2009-2010 &	8.53 \times 10^{-11} & 11.18 \times 10^{-11} & \hspace{2mm} & \mathbf{6.79 \times 10^{-11}} & 4.77 \times 10^{-11} \\
\hline \hline \multicolumn{1}{c}{\boxed{\text{\bf Angle ($\mathbf \theta $)}}} & \multicolumn{2}{c}{\text{Without Cleaning}} &	\hspace{2mm} & \multicolumn{2}{c}{\text{With Cleaning}} \\ \text{Previous-Predicted} & \text{No Regression} & \text{Regression} & \hspace{2mm} & \text{No Regression} & \text{Regression} \\ \hline \hline
2004-2005 &	1.74 & \mathbf{0.71} & \hspace{2mm} & 2.10 & 1.49 \\
2005-2006 &	1.82 & \mathbf{1.38} & \hspace{2mm} & 2.10 & 1.64 \\
2006-2007 &	1.08 & \mathbf{0.29} & \hspace{2mm} & 1.12 & 0.66 \\
2007-2008 &	1.55 & 6.83 & \hspace{2mm} & \mathbf{1.31} & 6.28 \\
2008-2009 &	10.22 &  \mathbf{2.37} & \hspace{2mm} & 11.52 & 2.41 \\
2009-2010 &	3.42 & 3.91 & \hspace{2mm} & \mathbf{2.05} & 2.55 \\
\hline \end{array} \]

\noindent {\bf Table 2.} Agreement meausre (AG) and Mean Squared Error (MSE) of the curves, and angle ($\theta $) between vectors $RI^{pred}$ and $RI^{real}$. The best results for each line are shown in bold face.

\vskip 0.3 cm

In Table 2, the best results for the Agreement measure (AG) are scattered among the portfolios obtained using the original correlation matrices (cleaned or not cleaned) and the portfolios obtained with the uncleaned residues of the regression model. For the Mean Squared Error (MSE), many of the forecasted results were better with the use of regression in order to eliminate the effect of the market movements, and sometimes also with the cleaned correlation matrix without the regression. For the angle ($\theta $) between the returns, most of the best results are for the uncleaned correlation matrices obtained with the residues of the regression model.

\subsection{Results with Short Selling}

As mentioned before, short selling is not usually freely allowed in financial transactions, mainly due to the increase in risks it might bring to a portfolio, but it is an important instrument used by many practitioners. Short selling implies changing the constraint that the weight of each stock must be larger than zero and changing it to another limit. In our calculations, we established the constraints $-1\leq w_i\leq 2$.

The results with short selling allowed are summarized in Table 3. For the AG, the cleaned correlations based on the residues of the regression model clearly have the best results; for the MSE, the best results are with the cleaned correlation matrices, be them cleaned or not; for the angle $\theta $, the best results are obtained from the residues of the regression, be them with a cleaned or with an uncleaned correlation matrix.

\[ \begin{array}{c|ccccc} \hline \multicolumn{1}{c}{\boxed{\text{\bf Agreement (AG)}}} & \multicolumn{2}{c}{\text{Without Cleaning}} &	\hspace{2mm} & \multicolumn{2}{c}{\text{With Cleaning}} \\ \text{Previous-Predicted} & \text{No Regression} & \text{Regression} & \hspace{2mm} & \text{No Regression} & \text{Regression} \\ \hline \hline
2004-2005 &	0.588 & 0.296 & \hspace{2mm} & 0.528 & \mathbf{0.162} \\
2005-2006 &	0.302 & 0.291 & \hspace{2mm} & 0.608 & \mathbf{0.185} \\
2006-2007 &	\mathbf{-0.004} & 0.048 & \hspace{2mm} & 0.494 & 0.273 \\
2007-2008 &	0.537 & 0.568 & \hspace{2mm} & 0.690 & \mathbf{0.486} \\
2008-2009 &	0.125 &  \mathbf{0.024} & \hspace{2mm} & 0.655 & -0.128 \\
2009-2010 &	0.398 & 0.433 & \hspace{2mm} & 0.529 & \mathbf{0.239} \\
\hline \hline \multicolumn{1}{c}{\boxed{\text{\bf Mean Squared Error (MSE)}}} & \multicolumn{2}{c}{\text{Without Cleaning}} &	\hspace{2mm} & \multicolumn{2}{c}{\text{With Cleaning}} \\ \text{Previous-Predicted} & \text{No Regression} & \text{Regression} & \hspace{2mm} & \text{No Regression} & \text{Regression} \\ \hline \hline
2004-2005 &	31.82 \times 10^{-6} & 30.43 \times 10^{-6} & \hspace{2mm} & 6.24 \times 10^{-6} & \mathbf{4.68 \times 10^{-6}} \\
2005-2006 &	16.10 \times 10^{-6} & 31.55 \times 10^{-6} & \hspace{2mm} & 11.52 \times 10^{-6} & \mathbf{2.62 \times 10^{-6}} \\
2006-2007 &	2.11 \times 10^{-5} & 7.30 \times 10^{-5} & \hspace{2mm} & \mathbf{1.57 \times 10^{-5}} & 4.05 \times 10^{-5} \\
2007-2008 &	 6.98 \times 10^{-3} & 10.36 \times 10^{-3} & \hspace{2mm} & \mathbf{5.16 \times 10^{-3}} & 11.21 \times 10^{-3} \\
2008-2009 &	 \mathbf{0.60 \times 10^{-4}} &  5.01 \times 10^{-4} & \hspace{2mm} & 11.80 \times 10^{-4} & 1.00 \times 10^{-4} \\
2009-2010 &	3.50 \times 10^{-4} & 4.42 \times 10^{-4} & \hspace{2mm} & 4.68 \times 10^{-4} & \mathbf{2.93 \times 10^{-4}} \\
\hline \hline \multicolumn{1}{c}{\boxed{\text{\bf Angle ($\mathbf \theta $)}}} & \multicolumn{2}{c}{\text{Without Cleaning}} &	\hspace{2mm} & \multicolumn{2}{c}{\text{With Cleaning}} \\ \text{Previous-Predicted} & \text{No Regression} & \text{Regression} & \hspace{2mm} & \text{No Regression} & \text{Regression} \\ \hline \hline
2004-2005 &	4.70 & 1.96 & \hspace{2mm} & 4.71 & \mathbf{1.29} \\
2005-2006 &	2.37 & \mathbf{2.16} & \hspace{2mm} & 6.91 & 3.36 \\
2006-2007 &	0.69 & \mathbf{0.46} & \hspace{2mm} & 2.15 & 2.92 \\
2007-2008 &	4.50 & \mathbf{3.54} & \hspace{2mm} & 11.28 & 8.09 \\
2008-2009 &	2.69 &  2.33 & \hspace{2mm} & 2.51 & \mathbf{1.65} \\
2009-2010 &	1.75 & 2.36 & \hspace{2mm} & 4.37 & \mathbf{1.66} \\
\hline \end{array} \]

\noindent {\bf Table 3.} Agreement meausre (AG) and Mean Squared Error (MSE) of the curves, and angle ($\theta $) between vectors $RI^{pred}$ and $RI^{real}$, all with short selling allowed. The best results for each line are shown in bold face.

\vskip 0.3 cm

\subsection{Measures of distance of the correlation matrices}

Other measures of how well a portfolio is related with another portfolio obtained by some cleaning procedure can be devised, some of them based directly on the distances between the correlation matrices. Such distances avoid the actual building of the portfolios and the issue of using or not short selling. One of theses measures is the simple distance between matrices, which can be defined by (as in Anton \& Rorres (2005))
\begin{equation}
\label{distance}
Dist={\rm Tr}\left( \left( \C -\C _{clean}\right) ^t\left( \C -\C _{clean}\right) \right) \ .
\end{equation}
Another measure that can be used as a distance between matrices is the Kullback Leibler distance, established in Kullback \& Leibler (1951), and used by Tumminello, Lillo, \& Mantegna (2007a), Tumminello, Lillo, \& Mantegna (2007b), Biroli, Bouchaud, \& Potters (2007), and Tumminello, Lillo, \& Mantegna (2010) in order to compare some cleaning procedures. The discrete version of this measure is based on the probability distributions $P$ and $P_{clean}$ of, respectively, the correlation matrix and the cleaned correlation matrix. It is given by
\begin{equation}
\label{KL}
D_{KL}=\sum_{i=1}^NP_i\ln \frac{P_i}{Q_i}\ ,
\end{equation}
where $P_i$ and $Q_i$ are the probabilities of bin $i$, $i=1,\cdots ,N$, occuring in state $P$ of the correlation matrix and in state $Q$ of the cleaned correlation matrix, respectively, and the element of the sum is considered as zero if $P_i$ or $Q_i$ are zero.

By applying these two measures to the data, we obtain the results given in Table 4 for the simple distance measure ($Dist$) and for the Kullback-Leibler distance ($D_{KL}$). The results depend directly on the correlation matrices (cleaned or not cleaned, obtained from the original data or from the residues of the regression), and so do not change for short selling allowed or not. The results are again different for each type of measure. For the simple distance ($Dist$), the cleaning of the correlation matrix obtained from the residues of the regression is definitely the best result; for the Kullback-Leibler distance, the best result is for the cleaning of the correlation matrix obtained from the original data. Both measures depend on the size of the correlation matrices, so that one period of time cannot be truly compared with another, and the Kullback-Leibler distance also depends on the choice and number of bins used to derive the probability distributions of the correlation matrices. A brief sutdy with other choices for bins reveals that the results are not significantly altered with the number of bins used, under certain reasonable limits.

\[ \begin{array}{c|ccccc} \hline \multicolumn{1}{c}{\boxed{\text{\bf Distance ($\mathbf Dist$)}}} & \multicolumn{2}{c}{\text{Without Cleaning}} &	\hspace{2mm} & \multicolumn{2}{c}{\text{With Cleaning}} \\ \text{Previous-Predicted} & \text{No Regression} & \text{Regression} & \hspace{2mm} & \text{No Regression} & \text{Regression} \\ \hline \hline
2004-2005 &	50 & 37 & \hspace{2mm} & 38 & \mathbf{25} \\
2005-2006 &	68 & 52 & \hspace{2mm} & 49 & \mathbf{36} \\
2006-2007 &	112 & 78 & \hspace{2mm} & 92 & \mathbf{52} \\
2007-2008 &	222 & 149 & \hspace{2mm} & 202 & \mathbf{112} \\
2008-2009 &	794 &  277 & \hspace{2mm} & 695 & \mathbf{172} \\
2009-2010 &	268 & 226 & \hspace{2mm} & 182 & \mathbf{133} \\
\hline \hline \multicolumn{1}{c}{\boxed{\text{\bf Kullback-Leibler distance ($\mathbf D_{KL}$)}}} & \multicolumn{2}{c}{\text{Without Cleaning}} &	\hspace{2mm} & \multicolumn{2}{c}{\text{With Cleaning}} \\ \text{Previous-Predicted} & \text{No Regression} & \text{Regression} & \hspace{2mm} & \text{No Regression} & \text{Regression} \\ \hline \hline
2004-2005 &	0.1014 & \mathbf{0.0102} & \hspace{2mm} & 0.0925 & 0.0284 \\
2005-2006 &	0.0215 & \mathbf{0.0092} & \hspace{2mm} & 0.0145 & 0.0184 \\
2006-2007 &	0.2360 & \mathbf{0.0049} & \hspace{2mm} & 0.2858 & 0.0216 \\
2007-2008 &	0.2982 & \mathbf{0.0524} & \hspace{2mm} & 0.3427 & 0.1423 \\
2008-2009 &	0.8547 &  0.0518 & \hspace{2mm} & 0.8984 & \mathbf{0.0363} \\
2009-2010 &	0.0456 & \mathbf{0.0032} & \hspace{2mm} & 0.0570 & 0.0103 \\
\hline \end{array} \]

\noindent {\bf Table 4.} Distance ($Dist$) and Kullback-Leibler distance ($D_{KL}$) between predicted and realized correlation matrices. The best results for each line are shown in bold face.

\vskip 0.3 cm

\section{Evolution in Time}

Our analysis so far is based on large windows, with a varying number of stocks for each window, and large jumps from one window to the other. In order to perform a temporal analysis of the evolution of the porfolios, we now consider the 50 stocks that were 100\% liquid in the period from 2004 to 2010 in moving windows of 100 days each, with a lag of 5 days between each window. For each of these windows, portfolios are built on efficient frontiers with and without regression, with and without cleaning, and with and without short selling.

\subsection{Mean Squared Error}

We start by studying the effects of the window size of the sample in the results. In order to do this, we calculated the MSE (Mean Squared Error) for the 50 stocks in sliding windows of 100 and 55 days, both with a lag of 5 days between each window. By doing this, we study the cases $Q=100/50=2$ and $Q=55/50=1.1$. Figure 8 shows the MSE for each window for the case of original data (no regression), no short selling, and no cleaning of the correlation matrices.

Both graphs are very similar, with the graph for windows of 55 days less smooth, as it was to be expected. Both graphs show strong peaks by window 230. Figure 9 shows the average volatility of the Ibovespa for windows of 100 days (left) and of 55 days (right). Comparing both figures, one may see that the peaks in difference between predicted and realized risks occur in times of high volatility of the BM\&F-Bovespa. This is partially expected from the definition of the Mean Squared Error, which makes this measure larger when considering times of higher risks.

\begin{pspicture}(-0.6,-0.6)(8,5.5)
\psset{xunit=0.02,yunit=1}
\psline{->}(0,0)(320,0) \rput(330,0){t} \psline{->}(0,0)(0,5) \rput(65,5){MSE ($\times 10^{-8}$)} \scriptsize  \rput(-20,-0.2){0} \psline(100,-0.1)(100,0.1) \rput(100,-0.3){100} \psline(200,-0.1)(200,0.1) \rput(200,-0.3){200} \psline(300,-0.1)(300,0.1) \rput(300,-0.3){300} \psline(-5,1)(5,1) \rput(-20,1){$1$} \psline(-5,2)(5,2) \rput(-20,2){$2$} \psline(-5,3)(5,3) \rput(-20,3){$3$} \psline(-5,4)(5,4) \rput(-20,4){$4$}
\psline (0,0.2183) (1,0.1965) (2,0.2576) (3,0.2142) (4,0.2238) (5,0.1717) (6,0.1205) (7,0.1292) (8,0.1624) (9,0.1525) (10,0.1433) (11,0.1341) (12,0.0806) (13,0.0954) (14,0.0740) (15,0.0916) (16,0.0751) (17,0.1031) (18,0.0122) (19,0.0089) (20,0.0055) (21,0.0071) (22,0.0247) (23,0.0216) (24,0.0452) (25,0.0611) (26,0.1115) (27,0.1006) (28,0.0435) (29,0.0968) (30,0.1230) (31,0.1273) (32,0.3292) (33,0.2990) (34,0.0816) (35,0.0094) (36,0.0179) (37,0.0041) (38,0.0107) (39,0.0252) (40,0.0205) (41,0.0129) (42,0.0043) (43,0.0079) (44,0.0128) (45,0.0205) (46,0.0177) (47,0.0263) (48,0.0139) (49,0.0069) (50,0.0164) (51,0.0423) (52,0.0582) (53,0.0039) (54,0.0092) (55,0.0055) (56,0.0132) (57,0.0197) (58,0.0212) (59,0.0164) (60,0.0574) (61,0.0798) (62,0.0483) (63,0.0237) (64,0.0163) (65,0.0423) (66,0.0320) (67,0.0195) (68,0.0592) (69,0.0522) (70,0.0787) (71,0.0559) (72,0.0510) (73,0.0425) (74,0.0392) (75,0.0345) (76,0.0336) (77,0.0475) (78,0.0457) (79,0.0375) (80,0.0266) (81,0.0318) (82,0.0403) (83,0.0643) (84,0.0807) (85,0.0395) (86,0.0451) (87,0.0153) (88,0.0599) (89,0.0328) (90,0.0486) (91,0.1264) (92,0.0175) (93,0.0341) (94,0.0293) (95,0.0184) (96,0.0391) (97,0.0462) (98,0.0481) (99,0.0242) (100,0.0148) (101,0.0089) (102,0.0274) (103,0.0387) (104,0.0314) (105,0.0290) (106,0.0433) (107,0.0725) (108,0.0530) (109,0.0723) (110,0.0413) (111,0.0255) (112,0.0510) (113,0.0244) (114,0.0613) (115,0.0387) (116,0.0516) (117,0.0094) (118,0.0054) (119,0.0050) (120,0.0063) (121,0.0105) (122,0.0739) (123,0.0425) (124,0.0198) (125,0.0274) (126,0.0375) (127,0.0688) (128,0.0568) (129,0.0708) (130,0.0175) (131,0.0385) (132,0.0323) (133,0.0171) (134,0.0246) (135,0.0367) (136,0.0660) (137,0.0024) (138,0.0079) (139,0.0121) (140,0.0188) (141,0.0190) (142,0.0061) (143,0.0044) (144,0.0022) (145,0.0084) (146,0.0081) (147,0.0056) (148,0.0086) (149,0.0042) (150,0.0218) (151,0.0092) (152,0.0131) (153,0.0097) (154,0.0209) (155,0.0216) (156,0.0125) (157,0.0613) (158,0.0327) (159,0.0291) (160,0.0328) (161,0.0683) (162,0.0671) (163,0.1430) (164,0.0332) (165,0.0153) (166,0.0217) (167,0.0051) (168,0.0147) (169,0.0281) (170,0.0246) (171,0.0454) (172,0.0491) (173,0.0301) (174,0.0302) (175,0.0409) (176,0.0694) (177,0.0366) (178,0.0357) (179,0.0475) (180,0.0200) (181,0.1113) (182,0.0372) (183,0.0036) (184,0.0039) (185,0.0097) (186,0.0068) (187,0.0048) (188,0.0327) (189,0.0232) (190,0.0829) (191,0.0206) (192,0.0298) (193,0.0370) (194,0.0079) (195,0.0137) (196,0.0000) (197,0.0886) (198,0.3037) (199,0.0000) (200,0.0000) (201,0.5461) (202,0.4927) (203,0.4000) (204,0.1499) (205,0.2424) (206,0.0430) (207,0.2433) (208,0.0559) (209,0.2706) (210,0.4748) (211,0.2508) (212,0.5049) (213,0.3894) (214,0.0990) (215,0.8440) (216,0.1032) (217,1.0055) (218,1.2393) (219,4.5842) (220,2.2037) (221,3.0005) (222,2.1249) (223,1.5516) (224,2.3006) (225,1.3824) (226,0.9301) (227,0.5746) (228,1.3530) (229,0.9487) (230,0.4752) (231,0.3051) (232,0.6309) (233,0.4570) (234,0.4783) (235,0.2984) (236,0.2557) (237,0.1902) (238,0.1026) (239,0.0193) (240,0.0249) (241,0.0099) (242,0.0508) (243,0.0525) (244,0.0218) (245,0.0042) (246,0.0092) (247,0.0191) (248,0.0140) (249,0.0138) (250,0.0104) (251,0.0220) (252,0.0172) (253,0.0320) (254,0.0229) (255,0.0572) (256,0.0255) (257,0.0135) (258,0.0190) (259,0.0254) (260,0.0049) (261,0.0084) (262,0.0085) (263,0.0041) (264,0.0460) (265,0.0603) (266,0.0432) (267,0.1503) (268,0.0947) (269,0.0068) (270,0.0213) (271,0.0217) (272,0.0072) (273,0.0112) (274,0.0063) (275,0.0021) (276,0.0007) (277,0.0036) (278,0.0232) (279,0.0087) (280,0.0666) (281,0.0596) (282,0.0071) (283,0.0265) (284,0.0138) (285,0.0143) (286,0.0146) (287,0.0097) (288,0.0061) (289,0.0020) (290,0.0024) (291,0.0058) (292,0.0130) (293,0.0290) (294,0.0207) (295,0.0232) (296,0.0473) (297,0.0547) (298,0.0735) (299,0.0665) (300,0.0702) (301,0.0638) (302,0.0318) (303,0.0390) (304,0.0351) (305,0.0210) (306,0.0165)
\end{pspicture}
\begin{pspicture}(-1,-0.5)(4,6)
\psset{xunit=0.02,yunit=0.9}
\psline{->}(0,0)(340,0) \rput(350,0){t} \psline{->}(0,0)(0,6) \rput(65,6){MSE ($\times 10^{-8}$)} \scriptsize  \rput(-20,-0.2){0} \psline(100,-0.1)(100,0.1) \rput(100,-0.3){100} \psline(200,-0.1)(200,0.1) \rput(200,-0.3){200} \psline(300,-0.1)(300,0.1) \rput(300,-0.3){300} \psline(-5,1)(5,1) \rput(-20,1){$2$} \psline(-5,2)(5,2) \rput(-20,2){$4$} \psline(-5,3)(5,3) \rput(-20,3){$6$} \psline(-5,4)(5,4) \rput(-20,4){$8$} \psline(-5,5)(5,5) \rput(-20,5){$10$}
\psline (0,0.0147) (1,0.0082) (2,0.0033) (3,0.0353) (4,0.0952) (5,0.0130) (6,0.0116) (7,0.2290) (8,0.1273) (9,0.0937) (10,0.0846) (11,0.1171) (12,0.1708) (13,0.1341) (14,0.1197) (15,0.0634) (16,0.0824) (17,0.0669) (18,0.0093) (19,0.0252) (20,0.0476) (21,0.0071) (22,0.0306) (23,0.0518) (24,0.0190) (25,0.0163) (26,0.0409) (27,0.0396) (28,0.0161) (29,0.0268) (30,0.0143) (31,0.0122) (32,0.0103) (33,0.0036) (34,0.0088) (35,0.0106) (36,0.0182) (37,0.0320) (38,0.0201) (39,0.0788) (40,0.0890) (41,0.5265) (42,0.5802) (43,0.0796) (44,0.0246) (45,0.0152) (46,0.0035) (47,0.0010) (48,0.0015) (49,0.0451) (50,0.0298) (51,0.0302) (52,0.0337) (53,0.0095) (54,0.0319) (55,0.0194) (56,0.0078) (57,0.0050) (58,0.0220) (59,0.0285) (60,0.1092) (61,0.0678) (62,0.0375) (63,0.0710) (64,0.0400) (65,0.0149) (66,0.0086) (67,0.0696) (68,0.0265) (69,0.0714) (70,0.1205) (71,0.0101) (72,0.0023) (73,0.0028) (74,0.0186) (75,0.0466) (76,0.0121) (77,0.0668) (78,0.0645) (79,0.0800) (80,0.0189) (81,0.0127) (82,0.0071) (83,0.0064) (84,0.0090) (85,0.0126) (86,0.0269) (87,0.0109) (88,0.0189) (89,0.0330) (90,0.0091) (91,0.0363) (92,0.0181) (93,0.3890) (94,0.9481) (95,0.1971) (96,0.0179) (97,0.0062) (98,0.0043) (99,0.0186) (100,0.6065) (101,0.0629) (102,0.0498) (103,0.0376) (104,0.0135) (105,0.0329) (106,0.0782) (107,0.1121) (108,0.0467) (109,0.0176) (110,0.0118) (111,0.0429) (112,0.0173) (113,0.0162) (114,0.0579) (115,0.0941) (116,0.0550) (117,0.0549) (118,0.1075) (119,0.0769) (120,0.1031) (121,0.0468) (122,0.0037) (123,0.0064) (124,0.0077) (125,0.0126) (126,0.0060) (127,0.0030) (128,0.0032) (129,0.0028) (130,0.0013) (131,0.0083) (132,0.0140) (133,0.0091) (134,0.0138) (135,0.0155) (136,0.0692) (137,0.0727) (138,0.0794) (139,0.0135) (140,0.0658) (141,0.0316) (142,0.0141) (143,0.0138) (144,0.0633) (145,0.1277) (146,0.0908) (147,0.0977) (148,0.0819) (149,0.1228) (150,0.0234) (151,0.0164) (152,0.0190) (153,0.0153) (154,0.0145) (155,0.0101) (156,0.0250) (157,0.0054) (158,0.0344) (159,0.0389) (160,0.0536) (161,0.0865) (162,0.0726) (163,0.0797) (164,0.1245) (165,0.1180) (166,0.0407) (167,0.0669) (168,0.0368) (169,0.0103) (170,0.0828) (171,0.1654) (172,0.5755) (173,0.4157) (174,0.2362) (175,0.2043) (176,0.2924) (177,0.1529) (178,0.0112) (179,0.0850) (180,0.1038) (181,0.0315) (182,0.0237) (183,0.0524) (184,0.0191) (185,0.0629) (186,0.0826) (187,0.1452) (188,0.1683) (189,0.1287) (190,0.0301) (191,0.0138) (192,0.0042) (193,0.0031) (194,0.0067) (195,0.0111) (196,0.1003) (197,0.1687) (198,0.2177) (199,0.2361) (200,0.0748) (201,0.0341) (202,0.1279) (203,0.0898) (204,0.0303) (205,0.0416) (206,0.0174) (207,0.0454) (208,0.0029) (209,0.0113) (210,0.0000) (211,0.1324) (212,0.3485) (213,0.4671) (214,0.2009) (215,1.6386) (216,2.0046) (217,0.5376) (218,0.2099) (219,0.4125) (220,0.8781) (221,0.4820) (222,0.5874) (223,0.1090) (224,0.3170) (225,0.0731) (226,1.9914) (227,2.4403) (228,5.1954) (229,3.4731) (230,4.3942) (231,4.2333) (232,1.5994) (233,1.5822) (234,0.3542) (235,0.1297) (236,0.0996) (237,0.1062) (238,0.0767) (239,0.0290) (240,0.0336) (241,0.0191) (242,0.0256) (243,0.0460) (244,0.0044) (245,0.1831) (246,0.0085) (247,0.1192) (248,0.2693) (249,0.0248) (250,0.0032) (251,0.0062) (252,0.0036) (253,0.0380) (254,0.0043) (255,0.0324) (256,0.0045) (257,0.0174) (258,0.0150) (259,0.0789) (260,0.1800) (261,0.2186) (262,0.1311) (263,0.0506) (264,0.1876) (265,0.1315) (266,0.0261) (267,0.0015) (268,0.0037) (269,0.0598) (270,0.0649) (271,0.0341) (272,0.0187) (273,0.0441) (274,0.3684) (275,0.1398) (276,0.0854) (277,0.0082) (278,0.1918) (279,0.0781) (280,0.0249) (281,0.0201) (282,0.0222) (283,0.0131) (284,0.0091) (285,0.0107) (286,0.0054) (287,0.0036) (288,0.0042) (289,0.0053) (290,0.0145) (291,0.0069) (292,0.0055) (293,0.0009) (294,0.0028) (295,0.0008) (296,0.0064) (297,0.0175) (298,0.0021) (299,0.0033) (300,0.0183) (301,0.0406) (302,0.0053) (303,0.0003) (304,0.0111) (305,0.0302) (306,0.0310) (307,0.0376) (308,0.0521) (309,0.0319) (310,0.0433) (311,0.0559) (312,0.0702) (313,0.0488) (314,0.0249) (315,0.0182) (316,0.0138) (317,0.0060) (318,0.0030) (319,0.0121) (320,0.0156) (321,0.0180) (322,0.0152) (323,0.0422) (324,0.0229)
\end{pspicture}

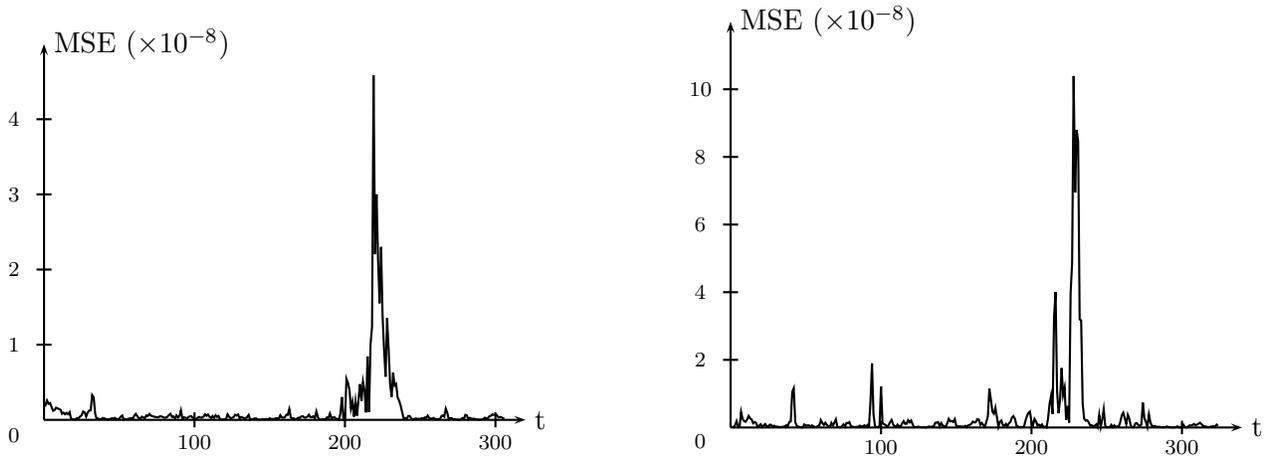
\noindent {\bf Fig. 8.} MSE for windows of 100 days (left) and for windows of 50 days (right) with sliding windows of 5 days, for original data, without short-selling, and no cleaning.

\vskip 0.3 cm

\begin{pspicture}(-0.6,-0.6)(8,5.3)
\psset{xunit=0.02,yunit=1.2}
\psline{->}(0,0)(320,0) \rput(330,0){t} \psline{->}(0,0)(0,4) \rput(30,4){Vol} \scriptsize  \rput(-20,-0.2){0} \psline(100,-0.1)(100,0.1) \rput(100,-0.3){100} \psline(200,-0.1)(200,0.1) \rput(200,-0.3){200} \psline(300,-0.1)(300,0.1) \rput(300,-0.3){300} \psline(-5,1)(5,1) \rput(-20,1){$0.01$} \psline(-5,2)(5,2) \rput(-20,2){$0.02$} \psline(-5,3)(5,3) \rput(-20,3){$0.03$}
% Volatility of the Ibovespa for 100 days windows and 5 days of sliding
\psline (0,1.8321) (1,1.7813) (2,1.7886) (3,1.7866) (4,1.6875) (5,1.6441) (6,1.6317) (7,1.5770) (8,1.5660) (9,1.5654) (10,1.4712) (11,1.4379) (12,1.4559) (13,1.4462) (14,1.4224) (15,1.4167) (16,1.3475) (17,1.3192) (18,1.2088) (19,1.1630) (20,1.1413) (21,1.1595) (22,1.1277) (23,1.0968) (24,1.1057) (25,1.0760) (26,1.0368) (27,1.0464) (28,1.0290) (29,1.0105) (30,0.9890) (31,0.9913) (32,0.9693) (33,1.0003) (34,1.0268) (35,1.0506) (36,1.0710) (37,1.0922) (38,1.1255) (39,1.1388) (40,1.1366) (41,1.1676) (42,1.1816) (43,1.2076) (44,1.2345) (45,1.2673) (46,1.2917) (47,1.2953) (48,1.3110) (49,1.3389) (50,1.3543) (51,1.3992) (52,1.4159) (53,1.3821) (54,1.3735) (55,1.3382) (56,1.3435) (57,1.3265) (58,1.3400) (59,1.3399) (60,1.3314) (61,1.3161) (62,1.3228) (63,1.3264) (64,1.2799) (65,1.2477) (66,1.2457) (67,1.2261) (68,1.2751) (69,1.2883) (70,1.3352) (71,1.3015) (72,1.2961) (73,1.2854) (74,1.2692) (75,1.2662) (76,1.2618) (77,1.2602) (78,1.2031) (79,1.1650) (80,1.1804) (81,1.1665) (82,1.1405) (83,1.1416) (84,1.1599) (85,1.1470) (86,1.1755) (87,1.1478) (88,1.0881) (89,1.0869) (90,1.0096) (91,0.9893) (92,1.0012) (93,1.0076) (94,1.0426) (95,1.0313) (96,1.0264) (97,1.0268) (98,1.0788) (99,1.1597) (100,1.2210) (101,1.2495) (102,1.3257) (103,1.3195) (104,1.3378) (105,1.3652) (106,1.3274) (107,1.4066) (108,1.4157) (109,1.3841) (110,1.3657) (111,1.3620) (112,1.3505) (113,1.3609) (114,1.3468) (115,1.3572) (116,1.3580) (117,1.3664) (118,1.3531) (119,1.2968) (120,1.2143) (121,1.1630) (122,1.1020) (123,1.0737) (124,1.0375) (125,1.0360) (126,1.0407) (127,0.9726) (128,0.9258) (129,0.9531) (130,1.0154) (131,1.0026) (132,0.9911) (133,0.9986) (134,0.9723) (135,0.9991) (136,0.9758) (137,1.0654) (138,1.0916) (139,1.1274) (140,1.1391) (141,1.1667) (142,1.1397) (143,1.1518) (144,1.1381) (145,1.1207) (146,1.1190) (147,1.1505) (148,1.1702) (149,1.1808) (150,1.1419) (151,1.1794) (152,1.1659) (153,1.1329) (154,1.1298) (155,1.1116) (156,1.1157) (157,1.0761) (158,1.0601) (159,1.0699) (160,1.1208) (161,1.1488) (162,1.2000) (163,1.2155) (164,1.2472) (165,1.2934) (166,1.3235) (167,1.3360) (168,1.3667) (169,1.3792) (170,1.3848) (171,1.4069) (172,1.4702) (173,1.5210) (174,1.5846) (175,1.5801) (176,1.6039) (177,1.5902) (178,1.5552) (179,1.5411) (180,1.5534) (181,1.6775) (182,1.6800) (183,1.6970) (184,1.7099) (185,1.6652) (186,1.6524) (187,1.7023) (188,1.7159) (189,1.7184) (190,1.7229) (191,1.6752) (192,1.6230) (193,1.5842) (194,1.6168) (195,1.6113) (196,1.6118) (197,1.5568) (198,1.5885) (199,1.6197) (200,1.5515) (201,1.4148) (202,1.4063) (203,1.4228) (204,1.3917) (205,1.4288) (206,1.4305) (207,1.4268) (208,1.4265) (209,1.3829) (210,1.3905) (211,1.4194) (212,1.5019) (213,1.6436) (214,1.7969) (215,1.9931) (216,2.1565) (217,2.3898) (218,2.5347) (219,2.8336) (220,2.9836) (221,3.0349) (222,3.1365) (223,3.2794) (224,3.3293) (225,3.3729) (226,3.4102) (227,3.3876) (228,3.4524) (229,3.5328) (230,3.5919) (231,3.5911) (232,3.5602) (233,3.4959) (234,3.3127) (235,3.1219) (236,3.0390) (237,2.8449) (238,2.6824) (239,2.3900) (240,2.3123) (241,2.2522) (242,2.1187) (243,1.9788) (244,1.9940) (245,1.9599) (246,1.9635) (247,1.9209) (248,1.8437) (249,1.7911) (250,1.7171) (251,1.7042) (252,1.6623) (253,1.5995) (254,1.5692) (255,1.5736) (256,1.4700) (257,1.4352) (258,1.4354) (259,1.3807) (260,1.2908) (261,1.2715) (262,1.2765) (263,1.2198) (264,1.1611) (265,1.1198) (266,1.0582) (267,1.0624) (268,1.0604) (269,1.1702) (270,1.1822) (271,1.2008) (272,1.1766) (273,1.1864) (274,1.1619) (275,1.1419) (276,1.1726) (277,1.1655) (278,1.1257) (279,1.1023) (280,1.1444) (281,1.1357) (282,1.1861) (283,1.1948) (284,1.1907) (285,1.1786) (286,1.1647) (287,1.1351) (288,1.1018) (289,0.9505) (290,0.9203) (291,0.8767) (292,0.8869) (293,0.8988) (294,0.9437) (295,0.9955) (296,1.0279) (297,1.0767) (298,1.1166) (299,1.1328) (300,1.0776) (301,1.1340) (302,1.0875) (303,1.0598) (304,1.0936) (305,1.0731) (306,1.0765)
\end{pspicture}
\begin{pspicture}(-0.5,-0.6)(4,5.3)
\psset{xunit=0.02,yunit=1}
\psline{->}(0,0)(330,0) \rput(340,0){t} \psline{->}(0,0)(0,5) \rput(30,5){Vol} \scriptsize  \rput(-20,-0.2){0} \psline(100,-0.1)(100,0.1) \rput(100,-0.3){100} \psline(200,-0.1)(200,0.1) \rput(200,-0.3){200} \psline(300,-0.1)(300,0.1) \rput(300,-0.3){300} \psline(-5,1)(5,1) \rput(-20,1){$0.01$} \psline(-5,2)(5,2) \rput(-20,2){$0.02$} \psline(-5,3)(5,3) \rput(-20,3){$0.03$} \psline(-5,4)(5,4) \rput(-20,4){$0.04$}
% Volatility of the Ibovespa for 55 days windows and 5 days of sliding
\psline (0,1.9985) (1,2.0036) (2,1.9735) (3,1.9288) (4,1.7653) (5,1.6485) (6,1.6885) (7,1.6177) (8,1.7838) (9,1.8651) (10,1.6656) (11,1.5590) (12,1.6037) (13,1.6444) (14,1.6097) (15,1.6397) (16,1.5750) (17,1.5362) (18,1.3483) (19,1.2657) (20,1.2769) (21,1.3167) (22,1.3080) (23,1.2480) (24,1.2352) (25,1.1937) (26,1.1199) (27,1.1021) (28,1.0692) (29,1.0603) (30,1.0057) (31,1.0023) (32,0.9473) (33,0.9456) (34,0.9763) (35,0.9583) (36,0.9538) (37,0.9906) (38,0.9887) (39,0.9607) (40,0.9723) (41,0.9804) (42,0.9913) (43,1.0550) (44,1.0773) (45,1.1429) (46,1.1883) (47,1.1937) (48,1.2624) (49,1.3170) (50,1.3009) (51,1.3548) (52,1.3719) (53,1.3602) (54,1.3918) (55,1.3916) (56,1.3950) (57,1.3969) (58,1.3596) (59,1.3608) (60,1.4077) (61,1.4436) (62,1.4599) (63,1.4041) (64,1.3553) (65,1.2847) (66,1.2919) (67,1.2561) (68,1.3204) (69,1.3190) (70,1.2550) (71,1.1886) (72,1.1857) (73,1.2487) (74,1.2044) (75,1.2106) (76,1.1994) (77,1.1961) (78,1.2298) (79,1.2576) (80,1.4154) (81,1.4144) (82,1.4066) (83,1.3221) (84,1.3340) (85,1.3218) (86,1.3243) (87,1.3243) (88,1.1764) (89,1.0724) (90,0.9455) (91,0.9187) (92,0.8745) (93,0.9610) (94,0.9858) (95,0.9722) (96,1.0268) (97,0.9714) (98,0.9998) (99,1.1015) (100,1.0737) (101,1.0598) (102,1.1278) (103,1.0542) (104,1.0995) (105,1.0904) (106,1.0259) (107,1.0822) (108,1.1577) (109,1.2179) (110,1.3683) (111,1.4392) (112,1.5236) (113,1.5848) (114,1.5761) (115,1.6400) (116,1.6288) (117,1.7310) (118,1.6736) (119,1.5503) (120,1.3632) (121,1.2849) (122,1.1774) (123,1.1370) (124,1.1174) (125,1.0745) (126,1.0871) (127,1.0018) (128,1.0325) (129,1.0432) (130,1.0655) (131,1.0411) (132,1.0266) (133,1.0103) (134,0.9576) (135,0.9974) (136,0.9942) (137,0.9433) (138,0.8190) (139,0.8630) (140,0.9653) (141,0.9642) (142,0.9556) (143,0.9869) (144,0.9869) (145,1.0008) (146,0.9574) (147,1.1875) (148,1.3642) (149,1.3917) (150,1.3129) (151,1.3693) (152,1.3238) (153,1.3166) (154,1.2894) (155,1.2406) (156,1.2806) (157,1.1135) (158,0.9763) (159,0.9698) (160,0.9708) (161,0.9896) (162,1.0080) (163,0.9491) (164,0.9701) (165,0.9826) (166,0.9508) (167,1.0388) (168,1.1438) (169,1.1701) (170,1.2708) (171,1.3079) (172,1.3920) (173,1.4819) (174,1.5243) (175,1.6041) (176,1.6963) (177,1.6331) (178,1.5895) (179,1.5884) (180,1.4988) (181,1.5059) (182,1.5484) (183,1.5601) (184,1.6450) (185,1.5560) (186,1.5116) (187,1.5472) (188,1.5210) (189,1.4938) (190,1.6080) (191,1.8491) (192,1.8116) (193,1.8340) (194,1.7749) (195,1.7744) (196,1.7932) (197,1.8574) (198,1.9108) (199,1.9430) (200,1.8377) (201,1.5012) (202,1.4344) (203,1.3345) (204,1.4587) (205,1.4483) (206,1.4303) (207,1.2563) (208,1.2661) (209,1.2963) (210,1.2653) (211,1.3283) (212,1.3783) (213,1.5111) (214,1.3246) (215,1.4094) (216,1.4308) (217,1.5972) (218,1.5869) (219,1.4694) (220,1.5158) (221,1.5105) (222,1.6255) (223,1.7762) (224,2.2691) (225,2.5768) (226,2.8822) (227,3.1823) (228,3.4826) (229,4.1978) (230,4.4515) (231,4.5593) (232,4.6476) (233,4.7825) (234,4.3895) (235,4.1690) (236,3.9383) (237,3.5929) (238,3.4221) (239,2.8679) (240,2.7323) (241,2.6229) (242,2.4727) (243,2.2094) (244,2.2359) (245,2.0748) (246,2.1397) (247,2.0970) (248,1.9427) (249,1.9122) (250,1.8923) (251,1.8815) (252,1.7647) (253,1.7483) (254,1.7521) (255,1.8451) (256,1.7873) (257,1.7448) (258,1.7447) (259,1.6700) (260,1.5420) (261,1.5269) (262,1.5600) (263,1.4506) (264,1.3863) (265,1.3022) (266,1.1528) (267,1.1256) (268,1.1261) (269,1.0914) (270,1.0396) (271,1.0162) (272,0.9929) (273,0.9890) (274,0.9359) (275,0.9375) (276,0.9636) (277,0.9991) (278,0.9947) (279,1.2490) (280,1.3248) (281,1.3855) (282,1.3603) (283,1.3838) (284,1.3878) (285,1.3463) (286,1.3816) (287,1.3318) (288,1.2566) (289,0.9556) (290,0.9639) (291,0.8860) (292,1.0119) (293,1.0058) (294,0.9936) (295,1.0109) (296,0.9477) (297,0.9385) (298,0.9469) (299,0.9454) (300,0.8766) (301,0.8674) (302,0.7619) (303,0.7917) (304,0.8937) (305,0.9801) (306,1.1082) (307,1.2148) (308,1.2864) (309,1.3201) (310,1.2786) (311,1.4007) (312,1.4131) (313,1.3280) (314,1.2935) (315,1.1660) (316,1.0448) (317,0.9538) (318,0.9041) (319,0.8882) (320,1.0425) (321,0.8814) (322,0.8575) (323,0.9008) (324,0.8165)
\end{pspicture}

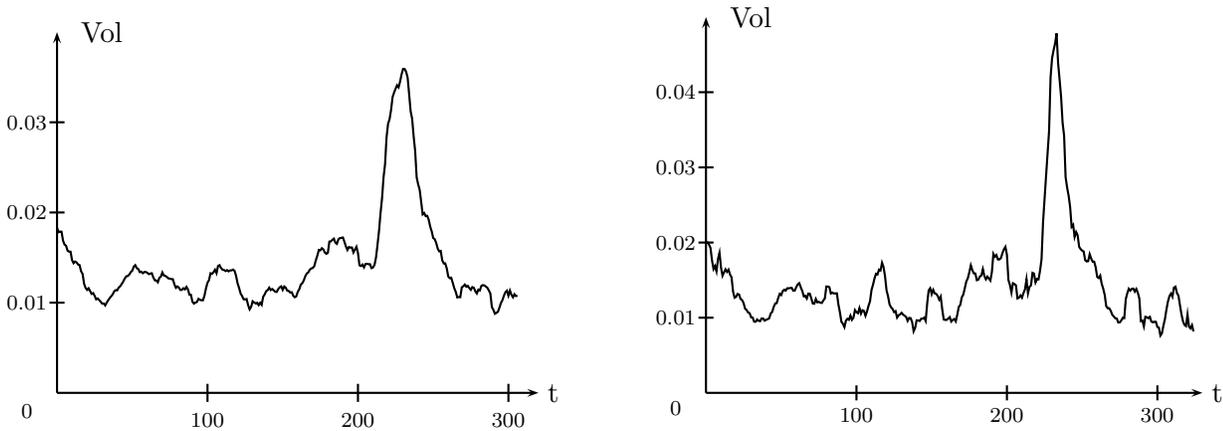
\noindent {\bf Fig. 9.} Volatility of the Ibovespa measured as an average over 100 days (left figure) and as an average over 55 days (right figure), both sliding windows with 5 days.

\vskip 0.3 cm

Now, we compare the results obtained with and without cleaning, with and without regression. Figure 10 shows the MSE for sliding windows of 100 days, with steps of 5 days, for the original data, with and without cleaning (left figure), and for the residues of the regression, with and without cleaning (right figure). The results without cleaning are in black lines, and the results with cleaning are in gray lines. The MSE for the residues of the regression are, in average, almost two orders of magnitude lower. The cleaning or not of the correlation matrices, though, offer very little difference in both cases.

\begin{pspicture}(-0.6,-0.6)(8,5.5)
\psset{xunit=0.02,yunit=1}
\psline{->}(0,0)(320,0) \rput(330,0){t} \psline{->}(0,0)(0,5) \rput(65,5){MSE ($\times 10^{-8}$)} \scriptsize  \rput(-20,-0.2){0} \psline(100,-0.1)(100,0.1) \rput(100,-0.3){100} \psline(200,-0.1)(200,0.1) \rput(200,-0.3){200} \psline(300,-0.1)(300,0.1) \rput(300,-0.3){300} \psline(-5,1)(5,1) \rput(-20,1){$1$} \psline(-5,2)(5,2) \rput(-20,2){$2$} \psline(-5,3)(5,3) \rput(-20,3){$3$} \psline(-5,4)(5,4) \rput(-20,4){$4$}
\psline (0,0.2183) (1,0.1965) (2,0.2576) (3,0.2142) (4,0.2238) (5,0.1717) (6,0.1205) (7,0.1292) (8,0.1624) (9,0.1525) (10,0.1433) (11,0.1341) (12,0.0806) (13,0.0954) (14,0.0740) (15,0.0916) (16,0.0751) (17,0.1031) (18,0.0122) (19,0.0089) (20,0.0055) (21,0.0071) (22,0.0247) (23,0.0216) (24,0.0452) (25,0.0611) (26,0.1115) (27,0.1006) (28,0.0435) (29,0.0968) (30,0.1230) (31,0.1273) (32,0.3292) (33,0.2990) (34,0.0816) (35,0.0094) (36,0.0179) (37,0.0041) (38,0.0107) (39,0.0252) (40,0.0205) (41,0.0129) (42,0.0043) (43,0.0079) (44,0.0128) (45,0.0205) (46,0.0177) (47,0.0263) (48,0.0139) (49,0.0069) (50,0.0164) (51,0.0423) (52,0.0582) (53,0.0039) (54,0.0092) (55,0.0055) (56,0.0132) (57,0.0197) (58,0.0212) (59,0.0164) (60,0.0574) (61,0.0798) (62,0.0483) (63,0.0237) (64,0.0163) (65,0.0423) (66,0.0320) (67,0.0195) (68,0.0592) (69,0.0522) (70,0.0787) (71,0.0559) (72,0.0510) (73,0.0425) (74,0.0392) (75,0.0345) (76,0.0336) (77,0.0475) (78,0.0457) (79,0.0375) (80,0.0266) (81,0.0318) (82,0.0403) (83,0.0643) (84,0.0807) (85,0.0395) (86,0.0451) (87,0.0153) (88,0.0599) (89,0.0328) (90,0.0486) (91,0.1264) (92,0.0175) (93,0.0341) (94,0.0293) (95,0.0184) (96,0.0391) (97,0.0462) (98,0.0481) (99,0.0242) (100,0.0148) (101,0.0089) (102,0.0274) (103,0.0387) (104,0.0314) (105,0.0290) (106,0.0433) (107,0.0725) (108,0.0530) (109,0.0723) (110,0.0413) (111,0.0255) (112,0.0510) (113,0.0244) (114,0.0613) (115,0.0387) (116,0.0516) (117,0.0094) (118,0.0054) (119,0.0050) (120,0.0063) (121,0.0105) (122,0.0739) (123,0.0425) (124,0.0198) (125,0.0274) (126,0.0375) (127,0.0688) (128,0.0568) (129,0.0708) (130,0.0175) (131,0.0385) (132,0.0323) (133,0.0171) (134,0.0246) (135,0.0367) (136,0.0660) (137,0.0024) (138,0.0079) (139,0.0121) (140,0.0188) (141,0.0190) (142,0.0061) (143,0.0044) (144,0.0022) (145,0.0084) (146,0.0081) (147,0.0056) (148,0.0086) (149,0.0042) (150,0.0218) (151,0.0092) (152,0.0131) (153,0.0097) (154,0.0209) (155,0.0216) (156,0.0125) (157,0.0613) (158,0.0327) (159,0.0291) (160,0.0328) (161,0.0683) (162,0.0671) (163,0.1430) (164,0.0332) (165,0.0153) (166,0.0217) (167,0.0051) (168,0.0147) (169,0.0281) (170,0.0246) (171,0.0454) (172,0.0491) (173,0.0301) (174,0.0302) (175,0.0409) (176,0.0694) (177,0.0366) (178,0.0357) (179,0.0475) (180,0.0200) (181,0.1113) (182,0.0372) (183,0.0036) (184,0.0039) (185,0.0097) (186,0.0068) (187,0.0048) (188,0.0327) (189,0.0232) (190,0.0829) (191,0.0206) (192,0.0298) (193,0.0370) (194,0.0079) (195,0.0137) (196,0.0000) (197,0.0886) (198,0.3037) (199,0.0000) (200,0.0000) (201,0.5461) (202,0.4927) (203,0.4000) (204,0.1499) (205,0.2424) (206,0.0430) (207,0.2433) (208,0.0559) (209,0.2706) (210,0.4748) (211,0.2508) (212,0.5049) (213,0.3894) (214,0.0990) (215,0.8440) (216,0.1032) (217,1.0055) (218,1.2393) (219,4.5842) (220,2.2037) (221,3.0005) (222,2.1249) (223,1.5516) (224,2.3006) (225,1.3824) (226,0.9301) (227,0.5746) (228,1.3530) (229,0.9487) (230,0.4752) (231,0.3051) (232,0.6309) (233,0.4570) (234,0.4783) (235,0.2984) (236,0.2557) (237,0.1902) (238,0.1026) (239,0.0193) (240,0.0249) (241,0.0099) (242,0.0508) (243,0.0525) (244,0.0218) (245,0.0042) (246,0.0092) (247,0.0191) (248,0.0140) (249,0.0138) (250,0.0104) (251,0.0220) (252,0.0172) (253,0.0320) (254,0.0229) (255,0.0572) (256,0.0255) (257,0.0135) (258,0.0190) (259,0.0254) (260,0.0049) (261,0.0084) (262,0.0085) (263,0.0041) (264,0.0460) (265,0.0603) (266,0.0432) (267,0.1503) (268,0.0947) (269,0.0068) (270,0.0213) (271,0.0217) (272,0.0072) (273,0.0112) (274,0.0063) (275,0.0021) (276,0.0007) (277,0.0036) (278,0.0232) (279,0.0087) (280,0.0666) (281,0.0596) (282,0.0071) (283,0.0265) (284,0.0138) (285,0.0143) (286,0.0146) (287,0.0097) (288,0.0061) (289,0.0020) (290,0.0024) (291,0.0058) (292,0.0130) (293,0.0290) (294,0.0207) (295,0.0232) (296,0.0473) (297,0.0547) (298,0.0735) (299,0.0665) (300,0.0702) (301,0.0638) (302,0.0318) (303,0.0390) (304,0.0351) (305,0.0210) (306,0.0165)
\psline[linecolor=gray] (0,0.1671) (1,0.1787) (2,0.2239) (3,0.1558) (4,0.1483) (5,0.1545) (6,0.0959) (7,0.1064) (8,0.1034) (9,0.1259) (10,0.1208) (11,0.1245) (12,0.0852) (13,0.0667) (14,0.0529) (15,0.1431) (16,0.0602) (17,0.1515) (18,0.0181) (19,0.0122) (20,0.0098) (21,0.0067) (22,0.0389) (23,0.0528) (24,0.0784) (25,0.0879) (26,0.1085) (27,0.0551) (28,0.0507) (29,0.0654) (30,0.2856) (31,0.0614) (32,0.2913) (33,0.2876) (34,0.0337) (35,0.0155) (36,0.0494) (37,0.0303) (38,0.0553) (39,0.0823) (40,0.0117) (41,0.0116) (42,0.0052) (43,0.0190) (44,0.0241) (45,0.0416) (46,0.0596) (47,0.1038) (48,0.0119) (49,0.0065) (50,0.0065) (51,0.0196) (52,0.0299) (53,0.0096) (54,0.0214) (55,0.0203) (56,0.0228) (57,0.0489) (58,0.0149) (59,0.0248) (60,0.1058) (61,0.0346) (62,0.0362) (63,0.0220) (64,0.0120) (65,0.0167) (66,0.0179) (67,0.0088) (68,0.0400) (69,0.0287) (70,0.0382) (71,0.0211) (72,0.0236) (73,0.0214) (74,0.0178) (75,0.0199) (76,0.0381) (77,0.0163) (78,0.0464) (79,0.0647) (80,0.0785) (81,0.1384) (82,0.0935) (83,0.0908) (84,0.1120) (85,0.0782) (86,0.0545) (87,0.0057) (88,0.0396) (89,0.0191) (90,0.0624) (91,0.1295) (92,0.0152) (93,0.0501) (94,0.0372) (95,0.0297) (96,0.0551) (97,0.0521) (98,0.0397) (99,0.0175) (100,0.0087) (101,0.0051) (102,0.0136) (103,0.0090) (104,0.0061) (105,0.0086) (106,0.0186) (107,0.0452) (108,0.0531) (109,0.0538) (110,0.0359) (111,0.0259) (112,0.0583) (113,0.0326) (114,0.0465) (115,0.0418) (116,0.0710) (117,0.0082) (118,0.0052) (119,0.0083) (120,0.0102) (121,0.0167) (122,0.0562) (123,0.0571) (124,0.0621) (125,0.0585) (126,0.0581) (127,0.1038) (128,0.1119) (129,0.1299) (130,0.0498) (131,0.0777) (132,0.0808) (133,0.0471) (134,0.0615) (135,0.0622) (136,0.0843) (137,0.0023) (138,0.0221) (139,0.0165) (140,0.0237) (141,0.0176) (142,0.0047) (143,0.0016) (144,0.0032) (145,0.0073) (146,0.0117) (147,0.0106) (148,0.0066) (149,0.0008) (150,0.0109) (151,0.0054) (152,0.0286) (153,0.0514) (154,0.0378) (155,0.2457) (156,0.3981) (157,0.5522) (158,0.5378) (159,0.8865) (160,0.3594) (161,0.7254) (162,0.3963) (163,0.6217) (164,0.1873) (165,0.0190) (166,0.0259) (167,0.0137) (168,0.0210) (169,0.0445) (170,0.0384) (171,0.0726) (172,0.0691) (173,0.0503) (174,0.0367) (175,0.0441) (176,0.0555) (177,0.0239) (178,0.0144) (179,0.0249) (180,0.0089) (181,0.0876) (182,0.0786) (183,0.0139) (184,0.0185) (185,0.0223) (186,0.0249) (187,0.0103) (188,0.0402) (189,0.0520) (190,0.0428) (191,0.0438) (192,0.0350) (193,0.0220) (194,0.0072) (195,0.0088) (196,0.0000) (197,0.1562) (198,0.1018) (199,0.0000) (200,0.0000) (201,0.0695) (202,0.0920) (203,0.0411) (204,0.0094) (205,0.4486) (206,0.3007) (207,3.4359) (208,0.1135) (209,0.5566) (210,0.6176) (211,1.0457) (212,0.1375) (213,0.3508) (214,0.0442) (215,0.5789) (216,0.0317) (217,0.8390) (218,0.8049) (219,3.4317) (220,0.9164) (221,2.0734) (222,1.7144) (223,0.6919) (224,1.2917) (225,0.8438) (226,0.7209) (227,0.3542) (228,0.8776) (229,0.7195) (230,0.2847) (231,0.2498) (232,0.4065) (233,0.3226) (234,0.3050) (235,0.2418) (236,0.5810) (237,0.2967) (238,0.1086) (239,0.0493) (240,0.0954) (241,0.0489) (242,0.0189) (243,0.0357) (244,0.0113) (245,0.0113) (246,0.0163) (247,0.0196) (248,0.0081) (249,0.0069) (250,0.0112) (251,0.0256) (252,0.0339) (253,0.0195) (254,0.0142) (255,0.0407) (256,0.0238) (257,0.0125) (258,0.0237) (259,0.0169) (260,0.0083) (261,0.0070) (262,0.0120) (263,0.0064) (264,0.0680) (265,0.0644) (266,0.0694) (267,0.1152) (268,0.0683) (269,0.0136) (270,0.0099) (271,0.0254) (272,0.0150) (273,0.0133) (274,0.0100) (275,0.0145) (276,0.0054) (277,0.0060) (278,0.0084) (279,0.0027) (280,0.0166) (281,0.0069) (282,0.0042) (283,0.0267) (284,0.0113) (285,0.0163) (286,0.0245) (287,0.0304) (288,0.0258) (289,0.0124) (290,0.0060) (291,0.0028) (292,0.0152) (293,0.0167) (294,0.0205) (295,0.0179) (296,0.0411) (297,0.0517) (298,0.0748) (299,0.0627) (300,0.0567) (301,0.0555) (302,0.0237) (303,0.0303) (304,0.0208) (305,0.0158) (306,0.0131)
\end{pspicture}
\begin{pspicture}(-1,-0.6)(4,5.5)
\psset{xunit=0.02,yunit=1}
\psline{->}(0,0)(320,0) \rput(330,0){t} \psline{->}(0,0)(0,5) \rput(70,5){MSE ($\times 10^{-10}$)} \scriptsize  \rput(-20,-0.2){0} \psline(100,-0.1)(100,0.1) \rput(100,-0.3){100} \psline(200,-0.1)(200,0.1) \rput(200,-0.3){200} \psline(300,-0.1)(300,0.1) \rput(300,-0.3){300} \psline(-5,1)(5,1) \rput(-20,1){$4$} \psline(-5,2)(5,2) \rput(-20,2){$8$} \psline(-5,3)(5,3) \rput(-20,3){$10$} \psline(-5,4)(5,4) \rput(-20,4){$12$}
\psline (0,0.0312) (1,0.0049) (2,0.0005) (3,0.0043) (4,0.0097) (5,0.0024) (6,0.0012) (7,0.0113) (8,0.0113) (9,0.0053) (10,0.0112) (11,0.0050) (12,0.0055) (13,0.0118) (14,0.0045) (15,0.0185) (16,0.0410) (17,0.0031) (18,0.0140) (19,0.0074) (20,0.0149) (21,0.0035) (22,0.0040) (23,0.0032) (24,0.0032) (25,0.0021) (26,0.0353) (27,0.0889) (28,0.0026) (29,0.0272) (30,0.0370) (31,0.0188) (32,0.0084) (33,0.0859) (34,0.0827) (35,0.0222) (36,0.0586) (37,0.0284) (38,0.0096) (39,0.1165) (40,0.1151) (41,0.0172) (42,0.0062) (43,0.0021) (44,0.0004) (45,0.0125) (46,0.0074) (47,0.0807) (48,0.0976) (49,0.0642) (50,0.0337) (51,0.1276) (52,0.0837) (53,0.0135) (54,0.0011) (55,0.0021) (56,0.0027) (57,0.0106) (58,0.0052) (59,0.0098) (60,0.0081) (61,0.0106) (62,0.0047) (63,0.0044) (64,0.0044) (65,0.0426) (66,0.0181) (67,0.0249) (68,0.0367) (69,0.0165) (70,0.0110) (71,0.0093) (72,0.0055) (73,0.0085) (74,0.0340) (75,0.0085) (76,0.0052) (77,0.0065) (78,0.0230) (79,0.1204) (80,0.0479) (81,0.4356) (82,0.2060) (83,0.3210) (84,0.0036) (85,0.0519) (86,0.0315) (87,0.0200) (88,0.0023) (89,0.0052) (90,0.0186) (91,0.0217) (92,0.0049) (93,0.0094) (94,0.0069) (95,0.0110) (96,0.0126) (97,0.0233) (98,0.0347) (99,0.0256) (100,0.0119) (101,0.0022) (102,0.0895) (103,0.0179) (104,0.0501) (105,0.0224) (106,0.0200) (107,0.0272) (108,0.0099) (109,0.0297) (110,0.0237) (111,0.0060) (112,0.0098) (113,0.0148) (114,0.0636) (115,0.0079) (116,0.0016) (117,0.0013) (118,0.0003) (119,0.0005) (120,0.0008) (121,0.0007) (122,0.0187) (123,0.0048) (124,0.0251) (125,0.0108) (126,0.0051) (127,0.0039) (128,0.0134) (129,0.0113) (130,0.0256) (131,0.0196) (132,0.0299) (133,0.0150) (134,0.0113) (135,0.0030) (136,0.0154) (137,0.0046) (138,0.0049) (139,0.0025) (140,0.0039) (141,0.0108) (142,0.0079) (143,0.0633) (144,0.0169) (145,0.0007) (146,0.0245) (147,0.0105) (148,0.0037) (149,0.0011) (150,0.0031) (151,0.0010) (152,0.0248) (153,0.0489) (154,0.1375) (155,0.2100) (156,0.2912) (157,0.2904) (158,0.2597) (159,0.3338) (160,0.0836) (161,0.1561) (162,0.1530) (163,0.2006) (164,0.1010) (165,0.0657) (166,0.0571) (167,0.0904) (168,0.0907) (169,0.0309) (170,0.0367) (171,0.0053) (172,0.0499) (173,0.0404) (174,0.0292) (175,0.0117) (176,0.0035) (177,0.0035) (178,0.0310) (179,0.1653) (180,0.0520) (181,0.0167) (182,0.0549) (183,0.0199) (184,0.0120) (185,0.0050) (186,0.0664) (187,0.0578) (188,0.0371) (189,0.0005) (190,0.0651) (191,0.0294) (192,0.0802) (193,0.1383) (194,0.0049) (195,0.1863) (196,0.0000) (197,0.0332) (198,0.9053) (199,0.0000) (200,0.0000) (201,1.8321) (202,1.0268) (203,3.3702) (204,0.0531) (205,0.0612) (206,0.1267) (207,0.0483) (208,0.4631) (209,0.7344) (210,2.2533) (211,4.6361) (212,3.6387) (213,0.5281) (214,0.3689) (215,0.4890) (216,0.1309) (217,1.4464) (218,0.1041) (219,0.0688) (220,0.0660) (221,1.5909) (222,0.1464) (223,0.1902) (224,0.3360) (225,0.0302) (226,0.0084) (227,0.0171) (228,0.4618) (229,0.0244) (230,0.0809) (231,0.0511) (232,0.0778) (233,0.0568) (234,0.0524) (235,0.0557) (236,0.8162) (237,0.2465) (238,0.1846) (239,0.1419) (240,0.0772) (241,0.0262) (242,0.0270) (243,0.0276) (244,0.0143) (245,0.0014) (246,0.0103) (247,0.0450) (248,0.0240) (249,0.0264) (250,0.0345) (251,0.0136) (252,0.0427) (253,0.0067) (254,0.0144) (255,0.0682) (256,0.0891) (257,0.0441) (258,0.0466) (259,0.0261) (260,0.0004) (261,0.0162) (262,0.0149) (263,0.0673) (264,0.0039) (265,0.0822) (266,0.0063) (267,0.0418) (268,0.0580) (269,0.0627) (270,0.0544) (271,0.1067) (272,0.0008) (273,0.0064) (274,0.0035) (275,0.0094) (276,0.0125) (277,0.0054) (278,0.0155) (279,0.0140) (280,0.1113) (281,0.2012) (282,0.0004) (283,0.0044) (284,0.0017) (285,0.0013) (286,0.0051) (287,0.0260) (288,0.0124) (289,0.0153) (290,0.0060) (291,0.0025) (292,0.0013) (293,0.0087) (294,0.0061) (295,0.0055) (296,0.0025) (297,0.0062) (298,0.0025) (299,0.0145) (300,0.0035) (301,0.0024) (302,0.0100) (303,0.0261) (304,0.0019) (305,0.0029) (306,0.0003)
\psline[linecolor=gray] (0,0.0051) (1,0.0054) (2,0.0136) (3,0.0096) (4,0.0097) (5,0.0087) (6,0.0037) (7,0.0168) (8,0.0033) (9,0.0051) (10,0.0064) (11,0.0078) (12,0.0117) (13,0.0243) (14,0.0059) (15,0.0661) (16,0.0775) (17,0.0311) (18,0.0090) (19,0.0186) (20,0.0129) (21,0.0066) (22,0.0047) (23,0.0113) (24,0.0465) (25,0.0871) (26,0.1271) (27,0.0712) (28,0.0171) (29,0.1924) (30,0.2467) (31,0.1045) (32,0.1267) (33,0.1655) (34,0.0871) (35,0.0409) (36,0.2309) (37,0.0483) (38,0.0245) (39,0.0354) (40,0.0083) (41,0.0145) (42,0.0157) (43,0.0117) (44,0.0401) (45,0.0559) (46,0.0172) (47,0.0266) (48,0.0559) (49,0.0328) (50,0.0410) (51,0.0795) (52,0.0621) (53,0.0458) (54,0.0084) (55,0.0149) (56,0.0050) (57,0.0070) (58,0.0117) (59,0.0124) (60,0.0153) (61,0.0138) (62,0.0172) (63,0.0335) (64,0.0410) (65,0.0102) (66,0.0102) (67,0.0106) (68,0.0057) (69,0.0113) (70,0.0121) (71,0.0165) (72,0.0138) (73,0.0050) (74,0.0068) (75,0.0288) (76,0.0130) (77,0.0231) (78,0.0203) (79,0.0361) (80,0.0294) (81,0.0144) (82,0.2884) (83,0.1002) (84,0.0052) (85,0.0146) (86,0.0104) (87,0.0036) (88,0.0135) (89,0.0145) (90,0.0109) (91,0.0078) (92,0.0050) (93,0.0217) (94,0.0190) (95,0.0258) (96,0.0290) (97,0.0034) (98,0.0029) (99,0.0022) (100,0.0029) (101,0.0030) (102,0.0426) (103,0.0145) (104,0.0165) (105,0.0064) (106,0.0044) (107,0.0019) (108,0.0027) (109,0.0015) (110,0.0023) (111,0.0027) (112,0.0097) (113,0.0030) (114,0.0061) (115,0.0231) (116,0.0067) (117,0.0034) (118,0.0021) (119,0.0036) (120,0.0046) (121,0.0028) (122,0.0035) (123,0.0060) (124,0.0019) (125,0.0021) (126,0.0087) (127,0.0071) (128,0.0056) (129,0.0055) (130,0.0144) (131,0.0128) (132,0.0145) (133,0.0065) (134,0.0045) (135,0.0029) (136,0.0023) (137,0.0018) (138,0.0029) (139,0.0024) (140,0.0181) (141,0.0028) (142,0.0029) (143,0.0031) (144,0.0061) (145,0.0065) (146,0.0079) (147,0.0075) (148,0.0066) (149,0.0103) (150,0.0205) (151,0.0101) (152,0.0466) (153,0.0373) (154,0.2337) (155,0.2578) (156,0.4024) (157,0.2280) (158,0.2018) (159,0.2797) (160,0.1646) (161,0.3393) (162,0.3842) (163,0.0946) (164,0.0837) (165,0.0349) (166,0.0189) (167,0.0315) (168,0.0202) (169,0.0041) (170,0.0141) (171,0.0459) (172,0.0816) (173,0.0721) (174,0.0499) (175,0.0672) (176,0.0717) (177,0.0374) (178,0.0497) (179,0.0090) (180,0.0297) (181,0.0122) (182,0.0058) (183,0.0049) (184,0.0062) (185,0.0042) (186,0.0115) (187,0.0207) (188,0.0074) (189,0.0135) (190,0.0059) (191,0.0106) (192,0.0547) (193,0.0567) (194,0.0025) (195,0.2292) (196,0.0000) (197,0.1295) (198,0.9224) (199,0.0000) (200,0.0000) (201,0.3365) (202,1.1595) (203,0.5932) (204,0.4225) (205,1.4982) (206,2.0330) (207,4.7449) (208,0.2781) (209,0.5726) (210,0.3661) (211,0.2808) (212,0.0509) (213,0.1273) (214,1.1919) (215,1.3174) (216,0.1026) (217,0.5053) (218,0.0622) (219,0.0560) (220,0.1361) (221,1.8702) (222,0.1756) (223,0.1300) (224,0.4269) (225,0.0203) (226,0.0078) (227,0.0124) (228,0.1620) (229,0.0126) (230,0.0175) (231,0.0197) (232,0.0171) (233,0.0719) (234,0.0749) (235,0.0865) (236,0.4408) (237,0.3297) (238,0.0590) (239,0.3073) (240,0.3632) (241,0.0293) (242,0.0275) (243,0.0242) (244,0.0185) (245,0.0039) (246,0.0032) (247,0.0056) (248,0.0217) (249,0.0100) (250,0.0048) (251,0.0023) (252,0.0074) (253,0.0077) (254,0.0037) (255,0.0818) (256,0.0113) (257,0.0198) (258,0.0091) (259,0.0043) (260,0.0070) (261,0.0110) (262,0.0199) (263,0.0371) (264,0.0063) (265,0.0367) (266,0.0043) (267,0.0163) (268,0.0219) (269,0.1793) (270,0.1151) (271,0.2194) (272,0.0099) (273,0.0081) (274,0.0011) (275,0.0070) (276,0.0197) (277,0.0063) (278,0.0151) (279,0.0203) (280,0.0070) (281,0.0373) (282,0.0182) (283,0.0146) (284,0.0119) (285,0.0142) (286,0.0120) (287,0.0351) (288,0.0145) (289,0.0095) (290,0.0114) (291,0.0083) (292,0.0116) (293,0.0040) (294,0.0031) (295,0.0030) (296,0.0073) (297,0.0058) (298,0.0119) (299,0.0092) (300,0.0061) (301,0.0050) (302,0.0043) (303,0.0060) (304,0.0026) (305,0.0045) (306,0.0034)
\end{pspicture}

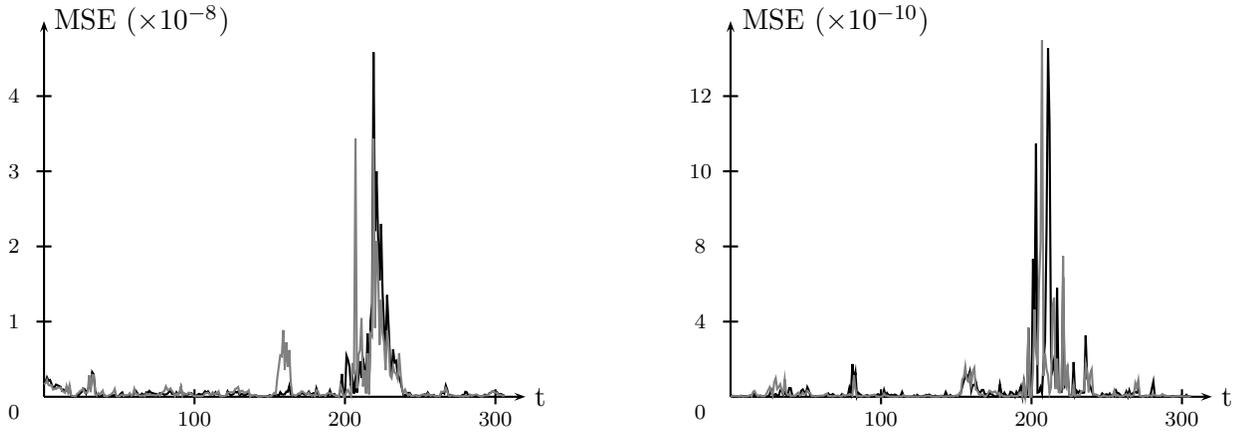
\noindent Fig. 10. MSE for windows of 100 days with sliding windows of 5 days, for the original data (left) and for the residues of single index model (right), without short-selling, without cleaning (black lines) and with cleaning (gray lines).

\vskip 0.3 cm

Very similar results are obtained for the case when short selling is allowed. So, times of high volatility coincide with higher values of the MSE.

\subsection{Agreement}

The results for the AG are not so well established. As shown in figure 11, they vary almost erratically, and the results for the cleaning procedure (gray line) are actually slightly worse than without the cleaning (black line). As we said before, the agreement measure is very sensitive to mutual cancelations of the factors $RI_i^{pred}$ and $RI_i^{real}$, and also to the values of the $RI_i^{pred}$ in the denominator of its definition.

\begin{pspicture}(-5.5,-2)(8,3.8)
\psset{xunit=0.02,yunit=0.8}
\psline{->}(0,0)(320,0) \rput(330,0){t} \psline{->}(0,-2)(0,4) \rput(30,4){AG} \scriptsize \rput(-20,0){0} \psline(100,-0.1)(100,0.1) \rput(100,-0.3){100} \psline(200,-0.1)(200,0.1) \rput(200,-0.3){200} \psline(300,-0.1)(300,0.1) \rput(300,-0.3){300} \psline(-5,-1)(5,-1) \rput(-25,-1){$-0.2$} \psline(-5,1)(5,1) \rput(-25,1){$0.2$} \psline(-5,2)(5,2) \rput(-25,2){$0.4$} \psline(-5,3)(5,3) \rput(-25,3){$0.6$}
\psline (0,-1.7469) (1,-1.7438) (2,-1.8610) (3,-1.7167) (4,-1.6916) (5,-1.6520) (6,-1.4252) (7,-1.4403) (8,-1.5414) (9,-1.5932) (10,-1.6938) (11,-1.6529) (12,-1.5157) (13,-1.4448) (14,-1.4041) (15,-1.1770) (16,-1.2386) (17,-1.3787) (18,-0.6662) (19,-0.6056) (20,-0.2855) (21,0.1131) (22,0.3564) (23,0.3045) (24,0.6694) (25,0.8437) (26,0.8672) (27,0.2838) (28,0.6185) (29,0.8681) (30,1.3174) (31,1.2950) (32,2.1114) (33,1.8083) (34,0.6637) (35,0.2062) (36,0.1045) (37,0.0783) (38,0.2108) (39,0.3120) (40,0.1073) (41,0.0758) (42,0.0524) (43,0.0855) (44,-0.3909) (45,-0.5045) (46,-0.5699) (47,-0.5960) (48,-0.1143) (49,0.1015) (50,0.2862) (51,0.0766) (52,0.2694) (53,-0.0248) (54,0.1841) (55,-0.2493) (56,-0.0793) (57,0.0720) (58,-0.7174) (59,-0.7144) (60,-0.9557) (61,-1.0796) (62,-0.9169) (63,-0.6831) (64,-0.5520) (65,-0.4469) (66,-0.1660) (67,-0.4560) (68,-1.2276) (69,-1.1585) (70,-1.3696) (71,-1.2298) (72,-1.2311) (73,-1.1879) (74,-0.9853) (75,-1.0705) (76,-0.8386) (77,-1.1331) (78,-0.0741) (79,0.1953) (80,0.0404) (81,0.1006) (82,0.4514) (83,0.3586) (84,0.8673) (85,0.5690) (86,0.7062) (87,0.4204) (88,1.1670) (89,0.8782) (90,1.2204) (91,0.8683) (92,0.9214) (93,1.1178) (94,1.1007) (95,0.8258) (96,1.2252) (97,1.8745) (98,2.0484) (99,1.1304) (100,0.8013) (101,0.6826) (102,-0.8953) (103,-0.6560) (104,-0.5875) (105,-0.5811) (106,-0.8070) (107,-1.1945) (108,-1.3042) (109,-1.2613) (110,-0.7902) (111,-0.7051) (112,-0.9165) (113,-0.9048) (114,-1.1216) (115,-1.0824) (116,-1.3272) (117,-0.3293) (118,0.0150) (119,0.1157) (120,0.1464) (121,0.3522) (122,1.6894) (123,1.3353) (124,1.0171) (125,1.1109) (126,1.3168) (127,2.0000) (128,1.8618) (129,1.8863) (130,0.7061) (131,0.9017) (132,0.9536) (133,0.9611) (134,1.3415) (135,1.6864) (136,2.4073) (137,0.1365) (138,-0.4858) (139,-0.5366) (140,-0.4586) (141,-0.4865) (142,-0.2052) (143,-0.0929) (144,0.2128) (145,0.4273) (146,0.3448) (147,0.3116) (148,0.4293) (149,0.1191) (150,0.6149) (151,0.2295) (152,0.0500) (153,0.1656) (154,0.3483) (155,0.4123) (156,0.1597) (157,0.8568) (158,0.6292) (159,-0.0298) (160,-0.1447) (161,-0.1318) (162,0.0131) (163,-0.4120) (164,-0.1781) (165,-0.1199) (166,-0.2449) (167,-0.0505) (168,0.0950) (169,0.5057) (170,0.4637) (171,0.6372) (172,0.5787) (173,0.4849) (174,0.2908) (175,0.3185) (176,0.4811) (177,0.2173) (178,0.2693) (179,0.3546) (180,0.2266) (181,-0.8312) (182,-0.5288) (183,-0.0447) (184,-0.1079) (185,0.0376) (186,0.1242) (187,-0.0553) (188,-0.1430) (189,-0.2817) (190,-0.0317) (191,-0.3313) (192,-0.1294) (193,0.0714) (194,-0.1099) (195,0.0833) (196,-0.0001) (197,0.2831) (198,0.6166) (199,-0.0001) (200,0.0000) (201,0.6386) (202,0.6844) (203,0.1916) (204,0.3467) (205,0.4550) (206,0.1190) (207,0.2785) (208,0.1223) (209,0.2954) (210,0.2324) (211,0.0448) (212,-0.0771) (213,0.4463) (214,-0.1967) (215,-0.2333) (216,0.0353) (217,-0.7185) (218,-0.9266) (219,-2.0911) (220,-1.7803) (221,-1.5421) (222,-1.5618) (223,-2.0141) (224,-2.0952) (225,-2.2070) (226,-1.9142) (227,-1.6813) (228,-2.0980) (229,-2.1781) (230,-1.8304) (231,-1.6198) (232,-1.8323) (233,-1.5564) (234,-1.5968) (235,-1.3080) (236,-1.0305) (237,-0.9230) (238,-0.7048) (239,-0.1340) (240,-0.2849) (241,-0.2508) (242,0.5346) (243,0.2805) (244,0.3203) (245,-0.2596) (246,-0.4732) (247,-0.3812) (248,0.1005) (249,0.6007) (250,0.5286) (251,0.5559) (252,0.4999) (253,0.4956) (254,0.1366) (255,-0.1545) (256,0.0980) (257,0.2565) (258,-0.3740) (259,-0.4747) (260,-0.0924) (261,-0.1261) (262,0.0313) (263,0.3363) (264,1.2574) (265,1.2105) (266,1.0236) (267,1.3757) (268,0.7997) (269,-0.2044) (270,-0.2777) (271,-0.4190) (272,-0.4650) (273,-0.4680) (274,-0.1663) (275,-0.1577) (276,0.0858) (277,0.1334) (278,0.7065) (279,0.5362) (280,0.6504) (281,0.9204) (282,-0.2265) (283,-0.8684) (284,-0.7006) (285,-0.7435) (286,-0.7769) (287,-0.5893) (288,-0.4850) (289,-0.2507) (290,-0.0609) (291,0.3059) (292,-0.2899) (293,-0.1751) (294,-0.9047) (295,-0.9004) (296,-1.5004) (297,-1.8717) (298,-1.9204) (299,-1.8560) (300,-1.9325) (301,-1.8275) (302,-1.5005) (303,-1.4832) (304,-1.4457) (305,-1.1136) (306,-0.9847)
\psline[linecolor=gray] (0,-1.5975) (1,-1.6358) (2,-1.7386) (3,-1.5473) (4,-1.5024) (5,-1.6223) (6,-1.3702) (7,-1.4351) (8,-1.4150) (9,-1.5607) (10,-1.7302) (11,-1.6952) (12,-1.6108) (13,-1.3410) (14,-1.2600) (15,-1.4987) (16,-1.0872) (17,-1.5926) (18,-0.7447) (19,-0.5573) (20,-0.3642) (21,0.0681) (22,0.2275) (23,-0.1026) (24,0.2527) (25,0.4943) (26,0.6586) (27,0.1743) (28,0.3453) (29,0.6587) (30,1.7270) (31,1.0685) (32,1.9644) (33,1.7685) (34,0.4948) (35,0.1602) (36,0.1167) (37,0.0042) (38,0.2613) (39,0.5772) (40,0.4416) (41,0.4408) (42,0.2371) (43,-0.1346) (44,-0.5443) (45,-0.7440) (46,-0.9632) (47,-1.0610) (48,-0.3363) (49,-0.1670) (50,0.0519) (51,0.1002) (52,0.2697) (53,0.1519) (54,0.3744) (55,0.0560) (56,0.1208) (57,0.2934) (58,-0.6153) (59,-0.6660) (60,-0.9703) (61,-0.9168) (62,-0.9457) (63,-0.7003) (64,-0.5172) (65,-0.4434) (66,-0.1284) (67,-0.3604) (68,-1.0719) (69,-0.9152) (70,-0.9976) (71,-0.8489) (72,-0.9331) (73,-0.8355) (74,-0.6076) (75,-0.7805) (76,-0.7326) (77,-0.7402) (78,0.3834) (79,0.7584) (80,0.6862) (81,0.9552) (82,0.7588) (83,0.8519) (84,1.2152) (85,0.8615) (86,0.8401) (87,0.3542) (88,0.9073) (89,0.7374) (90,1.2345) (91,1.0298) (92,0.8491) (93,1.4369) (94,1.2401) (95,1.1105) (96,1.5278) (97,2.0621) (98,1.8442) (99,1.1640) (100,0.7960) (101,0.5841) (102,-0.5679) (103,-0.4769) (104,-0.4342) (105,-0.4822) (106,-0.6825) (107,-1.1688) (108,-1.4172) (109,-1.3624) (110,-1.0599) (111,-1.0551) (112,-1.2516) (113,-1.2118) (114,-1.3498) (115,-1.2089) (116,-1.5273) (117,-0.4190) (118,0.0520) (119,0.2848) (120,0.3860) (121,0.6519) (122,1.6637) (123,1.6921) (124,1.9091) (125,1.8423) (126,1.9188) (127,2.8597) (128,3.0733) (129,2.9935) (130,1.5123) (131,1.5941) (132,1.6499) (133,1.7343) (134,2.3957) (135,2.3730) (136,2.8039) (137,0.1954) (138,-0.7723) (139,-0.6274) (140,-0.4838) (141,-0.4331) (142,-0.1218) (143,0.1381) (144,0.2482) (145,0.3858) (146,0.4702) (147,0.4546) (148,0.3462) (149,0.0625) (150,0.4055) (151,0.1434) (152,0.0471) (153,0.0587) (154,0.3380) (155,0.0476) (156,-0.2205) (157,0.1761) (158,0.1144) (159,-0.5186) (160,-0.1641) (161,-0.3086) (162,-0.1010) (163,-0.4743) (164,-0.3232) (165,0.0649) (166,-0.1080) (167,0.0309) (168,0.2622) (169,0.6367) (170,0.5469) (171,0.7631) (172,0.6710) (173,0.6590) (174,0.4473) (175,0.4971) (176,0.5793) (177,0.2422) (178,0.2980) (179,0.4000) (180,0.2388) (181,-0.8025) (182,-0.7723) (183,-0.0402) (184,-0.1337) (185,0.0227) (186,0.1844) (187,-0.2222) (188,-0.5056) (189,-0.6402) (190,-0.4366) (191,-0.5211) (192,-0.1448) (193,-0.1103) (194,-0.1158) (195,0.0382) (196,-0.0004) (197,0.4026) (198,0.4224) (199,0.0000) (200,-0.0001) (201,0.2200) (202,0.3118) (203,0.1701) (204,0.0958) (205,0.7401) (206,0.4496) (207,1.1021) (208,0.2101) (209,0.5414) (210,0.6358) (211,0.5897) (212,0.2212) (213,0.5008) (214,0.1256) (215,0.0319) (216,-0.0512) (217,-0.6525) (218,-0.7668) (219,-1.8896) (220,-1.2758) (221,-1.3032) (222,-1.3739) (223,-1.4797) (224,-1.6562) (225,-1.8259) (226,-1.7071) (227,-1.3938) (228,-1.8311) (229,-1.9483) (230,-1.5611) (231,-1.5165) (232,-1.6173) (233,-1.3977) (234,-1.3269) (235,-1.0953) (236,-1.1311) (237,-0.9508) (238,-0.6122) (239,-0.2596) (240,-0.5938) (241,-0.3030) (242,0.1123) (243,0.2255) (244,-0.1436) (245,-0.7142) (246,-0.8355) (247,-0.7305) (248,-0.1580) (249,0.4005) (250,0.3681) (251,0.5188) (252,0.3189) (253,0.4587) (254,0.1080) (255,-0.1734) (256,0.2705) (257,0.6719) (258,0.3094) (259,0.2285) (260,0.7439) (261,0.0063) (262,0.7596) (263,0.4044) (264,1.5981) (265,1.5390) (266,1.5376) (267,1.6288) (268,1.1143) (269,-0.2563) (270,-0.2943) (271,-0.6450) (272,-0.5646) (273,-0.6090) (274,-0.4351) (275,-0.5419) (276,-0.2950) (277,0.1682) (278,0.2701) (279,-0.0335) (280,-0.0419) (281,-0.0279) (282,-0.4284) (283,-0.9390) (284,-0.7596) (285,-0.9381) (286,-1.0939) (287,-1.0640) (288,-1.1242) (289,-0.9094) (290,-0.5992) (291,-0.2668) (292,-0.6755) (293,-0.5319) (294,-0.8925) (295,-0.8194) (296,-1.3990) (297,-1.7832) (298,-1.9225) (299,-1.6756) (300,-1.6618) (301,-1.6805) (302,-1.2517) (303,-1.2441) (304,-1.1406) (305,-0.9042) (306,-0.7783)
\end{pspicture}

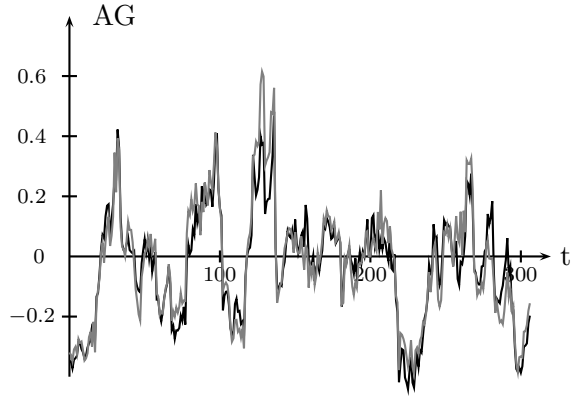
\noindent Fig. 11. AG for windows of 100 days with sliding windows of 5 days, for original data, without short-selling, without cleaning (black line) and with cleaning (gray line).

\vskip 0.3 cm

\subsection{Angle between risks}

Now, Figure 12 shows the results for the angle $\theta $ between vectors $RI^{pred}$ and $RI^{real}$ for windows of 100 days with sliding windows of 5 days, for the original data (left) and for the residues of the regression (right), both without short-selling. The results without cleaning are in black lines, and the results with cleaning are in gray lines. Again, there is an almost erratic behavior, showing that the removal of the higher risk effect also removes the peaks when volatility is high.

\begin{pspicture}(-1,-0.6)(8,5.6)
\psset{xunit=0.02,yunit=1}
\psline{->}(0,0)(320,0) \rput(330,0){t} \psline{->}(0,0)(0,5) \rput(15,5){$\theta $} \scriptsize  \rput(-20,-0.2){0} \psline(100,-0.1)(100,0.1) \rput(100,-0.3){100} \psline(200,-0.1)(200,0.1) \rput(200,-0.3){200} \psline(300,-0.1)(300,0.1) \rput(300,-0.3){300} \psline(-5,1.667)(5,1.667) \rput(-20,1.667){$5$} \psline(-5,3.333)(5,3.333) \rput(-20,3.333){$10$}
\psline (0,2.2822) (1,3.1074) (2,3.3811) (3,3.3839) (4,3.0749) (5,2.7962) (6,2.6583) (7,2.4200) (8,2.6569) (9,2.7490) (10,2.2252) (11,2.8273) (12,2.2073) (13,2.4301) (14,1.4123) (15,1.6916) (16,2.0471) (17,2.0256) (18,1.1038) (19,1.0050) (20,1.3489) (21,1.1559) (22,1.8906) (23,1.3271) (24,2.3570) (25,2.3743) (26,2.6216) (27,1.4134) (28,1.1669) (29,1.2015) (30,1.4289) (31,1.5951) (32,2.9265) (33,1.6026) (34,1.1023) (35,1.3134) (36,0.6151) (37,0.3700) (38,0.3467) (39,0.5336) (40,0.8251) (41,0.8893) (42,0.8176) (43,1.3825) (44,1.0821) (45,2.1707) (46,0.9839) (47,1.9219) (48,0.9394) (49,0.4823) (50,0.4576) (51,0.8696) (52,0.8925) (53,0.5230) (54,0.4784) (55,1.0464) (56,1.1473) (57,0.7420) (58,1.3960) (59,1.5497) (60,1.2125) (61,1.5971) (62,1.8091) (63,1.6894) (64,1.4602) (65,2.1613) (66,1.4640) (67,1.5054) (68,2.7613) (69,2.3907) (70,2.4375) (71,2.1483) (72,1.8385) (73,2.1514) (74,1.6575) (75,1.6284) (76,1.3623) (77,1.8012) (78,2.2603) (79,1.5177) (80,1.1384) (81,1.2198) (82,0.6152) (83,0.9411) (84,1.5936) (85,1.6855) (86,1.8492) (87,0.9325) (88,1.3091) (89,1.0438) (90,1.1999) (91,0.7225) (92,1.4270) (93,1.1066) (94,1.2609) (95,0.9996) (96,1.3584) (97,3.9067) (98,4.0121) (99,3.6937) (100,2.9957) (101,1.7551) (102,2.1859) (103,2.1354) (104,1.7894) (105,1.8939) (106,2.4244) (107,1.6158) (108,1.3113) (109,1.7644) (110,1.3137) (111,2.8751) (112,3.2168) (113,1.2619) (114,1.9041) (115,1.9553) (116,2.1506) (117,1.8572) (118,1.2933) (119,1.1182) (120,1.2163) (121,1.1721) (122,1.9726) (123,1.6775) (124,1.4511) (125,1.7457) (126,0.8299) (127,1.5603) (128,2.2054) (129,2.3414) (130,1.1251) (131,1.3287) (132,1.2285) (133,1.6627) (134,2.1101) (135,2.2316) (136,3.9014) (137,1.1921) (138,1.1956) (139,1.4583) (140,1.3650) (141,1.0008) (142,0.7573) (143,0.5514) (144,0.3939) (145,0.7005) (146,1.2006) (147,0.8557) (148,0.8904) (149,1.0510) (150,1.8175) (151,1.4801) (152,1.1135) (153,0.9701) (154,1.8009) (155,1.2243) (156,0.7680) (157,1.7122) (158,1.1772) (159,0.5771) (160,0.6829) (161,0.9000) (162,0.9981) (163,1.2369) (164,0.5252) (165,0.5812) (166,0.5819) (167,0.3649) (168,0.9421) (169,0.8038) (170,1.1735) (171,1.4540) (172,1.8920) (173,1.5382) (174,1.5490) (175,1.6890) (176,1.9574) (177,1.1987) (178,1.1848) (179,1.4116) (180,0.9931) (181,1.5912) (182,1.2018) (183,0.3722) (184,0.4134) (185,0.4683) (186,0.4447) (187,0.4191) (188,0.8847) (189,0.8742) (190,1.0704) (191,0.7830) (192,0.9085) (193,1.1111) (194,0.2227) (195,0.3619) (196,0.0002) (197,0.3166) (198,0.2082) (199,0.0005) (200,0.0005) (201,0.8215) (202,0.6277) (203,0.8937) (204,0.9532) (205,1.2079) (206,0.3776) (207,0.4337) (208,0.3995) (209,1.0046) (210,1.6570) (211,0.5975) (212,1.1999) (213,0.5363) (214,0.6310) (215,1.6813) (216,0.5812) (217,1.6492) (218,1.9531) (219,4.2774) (220,2.8537) (221,2.3175) (222,2.6499) (223,4.3216) (224,3.5679) (225,4.2754) (226,3.7527) (227,2.6622) (228,2.7579) (229,1.9199) (230,3.1164) (231,2.4025) (232,2.6950) (233,2.2211) (234,2.2139) (235,1.9063) (236,1.4305) (237,1.6327) (238,1.5261) (239,0.7141) (240,0.7321) (241,0.2478) (242,1.4072) (243,1.1666) (244,0.6759) (245,0.9214) (246,0.7739) (247,0.7110) (248,0.7453) (249,1.2750) (250,0.7933) (251,1.0174) (252,0.9600) (253,1.6311) (254,1.1513) (255,1.0313) (256,0.5516) (257,1.2646) (258,1.7440) (259,1.7247) (260,1.0938) (261,0.9575) (262,0.9642) (263,0.6486) (264,1.7131) (265,0.8818) (266,0.6635) (267,0.6510) (268,1.3969) (269,0.6426) (270,1.2047) (271,1.0854) (272,1.1609) (273,1.2919) (274,0.7271) (275,0.5769) (276,0.1680) (277,0.3133) (278,1.0645) (279,0.3659) (280,1.5845) (281,1.2106) (282,0.8257) (283,1.6514) (284,0.9843) (285,0.7083) (286,0.8646) (287,0.9331) (288,0.6955) (289,0.4291) (290,0.8804) (291,2.0328) (292,1.3252) (293,1.6405) (294,1.8053) (295,1.4315) (296,2.4370) (297,2.8787) (298,3.2196) (299,3.8192) (300,3.8659) (301,2.8525) (302,3.1317) (303,3.2097) (304,2.7821) (305,2.1208) (306,1.9716)
\psline[linecolor=gray] (0,2.3580) (1,2.8199) (2,3.0038) (3,2.8059) (4,2.5416) (5,2.7584) (6,2.5719) (7,2.3696) (8,2.5157) (9,2.6019) (10,3.0086) (11,3.2421) (12,2.6566) (13,2.1667) (14,1.4851) (15,2.7069) (16,2.0821) (17,2.8683) (18,1.2521) (19,0.7212) (20,0.9474) (21,0.8663) (22,2.0217) (23,1.5760) (24,2.4940) (25,2.4356) (26,2.4064) (27,1.1842) (28,1.0525) (29,0.7111) (30,1.1808) (31,1.3523) (32,2.3524) (33,1.3436) (34,0.6923) (35,1.5538) (36,1.1524) (37,0.8828) (38,0.5815) (39,0.6164) (40,0.7936) (41,0.6081) (42,0.9860) (43,2.0510) (44,1.7528) (45,2.8415) (46,1.7879) (47,3.2511) (48,1.0359) (49,0.5857) (50,0.5522) (51,0.6164) (52,0.6795) (53,0.6590) (54,0.9023) (55,1.3927) (56,1.2624) (57,1.2028) (58,1.7875) (59,2.3100) (60,1.2075) (61,1.1167) (62,1.5115) (63,1.6525) (64,1.3175) (65,1.4082) (66,1.0770) (67,1.0095) (68,1.9407) (69,1.7816) (70,1.6825) (71,1.4004) (72,1.5961) (73,1.8630) (74,1.6426) (75,1.8914) (76,1.3945) (77,1.3720) (78,2.3451) (79,1.7104) (80,1.4479) (81,1.6928) (82,1.1614) (83,1.3773) (84,1.6846) (85,1.5349) (86,1.0540) (87,0.5813) (88,0.9298) (89,0.8698) (90,1.0437) (91,0.8611) (92,1.3455) (93,1.5875) (94,1.4680) (95,1.3810) (96,1.8010) (97,2.9672) (98,2.9193) (99,1.8872) (100,1.1439) (101,1.2304) (102,1.2047) (103,1.1307) (104,0.9924) (105,0.8149) (106,1.5887) (107,1.5386) (108,1.7585) (109,2.1235) (110,1.2106) (111,1.4248) (112,2.1757) (113,1.9063) (114,2.0471) (115,3.0828) (116,2.2012) (117,1.5267) (118,1.3157) (119,1.3817) (120,1.3718) (121,1.4430) (122,2.1000) (123,2.0762) (124,1.8488) (125,1.6203) (126,1.2774) (127,2.5266) (128,3.5132) (129,3.5885) (130,2.1757) (131,2.0973) (132,1.9023) (133,2.4452) (134,2.1444) (135,2.5964) (136,3.4662) (137,1.0783) (138,1.6574) (139,1.4519) (140,1.4136) (141,1.0456) (142,0.8547) (143,0.4602) (144,0.4619) (145,0.7514) (146,1.0452) (147,0.7677) (148,0.9297) (149,0.4284) (150,1.1757) (151,1.1286) (152,1.4170) (153,1.6702) (154,2.3941) (155,2.8091) (156,2.4906) (157,3.3891) (158,2.7313) (159,2.0636) (160,2.2729) (161,1.5453) (162,1.8897) (163,1.7203) (164,1.0910) (165,1.0443) (166,1.2478) (167,0.9029) (168,1.1863) (169,1.4558) (170,1.5556) (171,1.9709) (172,2.2265) (173,1.8961) (174,1.5976) (175,1.6696) (176,1.6760) (177,1.0005) (178,0.7064) (179,0.6769) (180,0.4052) (181,1.7754) (182,1.5663) (183,0.7097) (184,0.8314) (185,0.6582) (186,0.6426) (187,0.6057) (188,1.1635) (189,1.2393) (190,1.0861) (191,1.1496) (192,0.9118) (193,1.0819) (194,0.2327) (195,0.3770) (196,0.0012) (197,0.4564) (198,0.2141) (199,0.0006) (200,0.0005) (201,0.2556) (202,0.3129) (203,0.3322) (204,0.2071) (205,1.1095) (206,0.5440) (207,1.5592) (208,0.4513) (209,0.9937) (210,0.9347) (211,0.7341) (212,0.8280) (213,0.8171) (214,0.2474) (215,1.0295) (216,0.4139) (217,1.6390) (218,1.6551) (219,3.4994) (220,1.5998) (221,1.8418) (222,2.3371) (223,2.4612) (224,2.9170) (225,3.1900) (226,3.1723) (227,2.1781) (228,2.4194) (229,1.5305) (230,2.5558) (231,2.1587) (232,2.4000) (233,1.9457) (234,1.5476) (235,1.2859) (236,1.4500) (237,1.3465) (238,1.0063) (239,0.6974) (240,0.8302) (241,0.8644) (242,1.3204) (243,0.9069) (244,0.9640) (245,1.5721) (246,1.1913) (247,0.9002) (248,0.7081) (249,0.8725) (250,0.7563) (251,1.0638) (252,1.1610) (253,1.3797) (254,0.9279) (255,1.0959) (256,0.7865) (257,1.3888) (258,2.0217) (259,2.0528) (260,1.5549) (261,1.0463) (262,1.2930) (263,0.8130) (264,2.0365) (265,1.7158) (266,1.4381) (267,1.5344) (268,1.4058) (269,0.6411) (270,1.0967) (271,1.1627) (272,1.2881) (273,1.1142) (274,0.9555) (275,1.3180) (276,0.6970) (277,0.3013) (278,0.5446) (279,0.8136) (280,1.1736) (281,1.0216) (282,0.8737) (283,1.1994) (284,1.2479) (285,1.1923) (286,1.3868) (287,1.9590) (288,1.1225) (289,1.0495) (290,0.7678) (291,0.5947) (292,0.6530) (293,1.2330) (294,1.4377) (295,1.4860) (296,2.3766) (297,2.2142) (298,3.1714) (299,3.5081) (300,3.4159) (301,2.5264) (302,2.4289) (303,2.4559) (304,2.2260) (305,1.5862) (306,1.3722)
\end{pspicture}
\begin{pspicture}(-1,-0.6)(4,5.6)
\psset{xunit=0.02,yunit=1}
\psline{->}(0,0)(320,0) \rput(330,0){t} \psline{->}(0,0)(0,5) \rput(15,5){$\theta $} \scriptsize  \rput(-20,-0.2){0} \psline(100,-0.1)(100,0.1) \rput(100,-0.3){100} \psline(200,-0.1)(200,0.1) \rput(200,-0.3){200} \psline(300,-0.1)(300,0.1) \rput(300,-0.3){300} \psline(-5,1.667)(5,1.667) \rput(-20,1.667){$5$} \psline(-5,3.333)(5,3.333) \rput(-20,3.333){$10$}
\psline (0,0.9157) (1,1.5207) (2,1.5146) (3,2.9793) (4,2.4942) (5,1.9010) (6,2.2878) (7,2.6853) (8,1.3207) (9,1.0186) (10,1.0602) (11,1.1810) (12,1.6974) (13,2.5538) (14,1.4100) (15,1.3779) (16,1.2990) (17,2.0276) (18,1.3818) (19,1.2610) (20,1.3655) (21,1.5872) (22,1.2150) (23,1.7796) (24,1.9439) (25,1.9841) (26,1.4405) (27,1.1599) (28,0.8672) (29,1.1464) (30,1.7530) (31,1.1490) (32,2.1000) (33,1.4025) (34,0.7463) (35,1.3116) (36,2.8199) (37,1.4100) (38,1.0648) (39,1.0484) (40,0.6115) (41,0.9282) (42,0.2855) (43,1.7195) (44,0.6473) (45,1.2785) (46,1.6843) (47,2.4894) (48,1.4457) (49,1.1301) (50,0.6592) (51,0.9612) (52,0.7470) (53,0.7144) (54,1.2298) (55,0.9828) (56,1.0499) (57,0.5375) (58,1.7582) (59,1.1777) (60,0.7486) (61,2.6792) (62,0.4932) (63,1.0205) (64,0.6365) (65,1.7221) (66,2.3582) (67,0.4950) (68,0.4633) (69,1.5927) (70,1.2465) (71,1.1726) (72,0.8487) (73,0.8041) (74,2.9691) (75,1.7413) (76,1.7917) (77,1.1993) (78,1.0817) (79,1.2551) (80,1.1308) (81,0.6820) (82,0.5945) (83,1.2555) (84,1.3834) (85,3.0927) (86,3.2017) (87,2.0736) (88,1.1229) (89,1.1972) (90,2.7703) (91,2.8288) (92,1.5519) (93,1.0516) (94,1.0623) (95,1.3066) (96,1.1820) (97,1.5467) (98,1.9189) (99,2.4570) (100,1.3851) (101,1.0416) (102,2.0667) (103,1.5973) (104,1.1170) (105,1.5564) (106,0.8671) (107,1.1727) (108,1.6198) (109,1.3636) (110,0.9356) (111,1.7377) (112,1.1812) (113,1.3148) (114,0.8308) (115,1.4894) (116,1.3547) (117,0.5697) (118,0.4440) (119,0.9238) (120,1.2687) (121,1.0101) (122,1.1038) (123,0.8734) (124,0.3439) (125,0.3620) (126,0.5316) (127,0.9835) (128,0.2088) (129,0.8436) (130,1.5116) (131,0.9470) (132,1.3175) (133,2.0155) (134,1.1474) (135,0.8068) (136,0.7388) (137,1.0374) (138,1.2584) (139,2.1174) (140,0.6949) (141,1.5062) (142,1.2796) (143,0.9473) (144,0.5403) (145,0.6649) (146,1.0197) (147,0.9731) (148,1.0306) (149,0.9333) (150,0.7954) (151,0.6302) (152,1.7660) (153,1.4852) (154,1.2678) (155,1.2276) (156,1.4896) (157,0.8015) (158,1.3254) (159,0.8998) (160,1.1767) (161,1.7192) (162,1.3587) (163,1.1738) (164,1.0237) (165,0.6237) (166,0.6165) (167,1.1815) (168,1.5196) (169,0.6373) (170,0.7770) (171,0.6566) (172,1.0092) (173,0.8330) (174,0.8178) (175,0.6494) (176,0.5660) (177,0.7808) (178,1.7131) (179,1.3660) (180,0.8157) (181,0.6082) (182,0.4626) (183,0.7847) (184,0.6318) (185,0.6934) (186,1.6235) (187,1.6855) (188,1.5955) (189,1.5962) (190,1.4439) (191,0.8677) (192,1.2226) (193,1.5971) (194,0.1681) (195,0.1280) (196,0.0277) (197,0.9807) (198,0.9155) (199,0.4735) (200,0.2248) (201,0.8524) (202,0.8585) (203,3.1949) (204,1.3477) (205,1.2921) (206,0.7030) (207,0.5184) (208,0.8313) (209,0.8723) (210,1.2897) (211,1.6614) (212,0.8470) (213,1.3533) (214,1.7816) (215,2.2514) (216,0.5175) (217,2.3311) (218,0.9801) (219,1.9576) (220,2.0923) (221,0.8596) (222,0.7226) (223,1.6925) (224,0.5735) (225,1.6935) (226,0.9983) (227,1.0433) (228,1.5155) (229,0.7221) (230,1.0608) (231,0.6444) (232,0.6122) (233,1.3654) (234,0.6664) (235,1.2876) (236,0.8528) (237,1.3068) (238,1.6490) (239,2.5171) (240,1.0030) (241,0.8749) (242,2.1147) (243,1.0047) (244,1.1240) (245,1.6702) (246,1.2432) (247,0.8002) (248,1.3620) (249,1.7217) (250,2.5089) (251,1.6508) (252,1.2891) (253,0.7341) (254,0.7318) (255,1.9470) (256,0.9261) (257,0.8741) (258,1.7260) (259,1.6900) (260,1.7566) (261,1.4397) (262,3.0319) (263,2.2945) (264,1.2540) (265,0.9633) (266,1.0059) (267,1.6689) (268,1.2936) (269,1.7090) (270,1.5774) (271,1.5893) (272,0.5089) (273,0.9129) (274,0.4431) (275,1.5732) (276,1.0714) (277,2.8474) (278,0.7400) (279,1.2794) (280,2.5387) (281,1.0158) (282,1.4614) (283,1.1264) (284,2.9164) (285,0.6415) (286,1.0257) (287,0.8486) (288,1.0227) (289,1.9488) (290,0.8881) (291,1.1276) (292,0.7643) (293,0.5299) (294,1.2259) (295,2.1411) (296,1.4883) (297,1.1141) (298,0.8896) (299,1.9665) (300,0.4558) (301,1.4337) (302,0.6235) (303,1.1397) (304,1.6217) (305,1.4527) (306,1.2730)
\psline[linecolor=gray] (0,1.3144) (1,1.8452) (2,1.8914) (3,2.2111) (4,1.3065) (5,1.5219) (6,1.7996) (7,2.8579) (8,0.5454) (9,0.7633) (10,1.1092) (11,1.2717) (12,2.1244) (13,2.7330) (14,2.4117) (15,1.8660) (16,1.7796) (17,1.9375) (18,2.2482) (19,1.2216) (20,1.7391) (21,1.7116) (22,0.9607) (23,1.6440) (24,1.5329) (25,1.5520) (26,1.0536) (27,0.6465) (28,1.2053) (29,1.1242) (30,0.7800) (31,0.4055) (32,2.1252) (33,2.2653) (34,0.8081) (35,0.9919) (36,2.9102) (37,1.5496) (38,1.3124) (39,1.4414) (40,0.6893) (41,1.3682) (42,1.3251) (43,0.5553) (44,1.2930) (45,1.1853) (46,0.9991) (47,2.6151) (48,0.6274) (49,0.9401) (50,0.7114) (51,1.2870) (52,0.6851) (53,0.5100) (54,0.6450) (55,0.8135) (56,0.7262) (57,0.5773) (58,0.8017) (59,0.8443) (60,0.4575) (61,1.1817) (62,0.8309) (63,0.6732) (64,0.7274) (65,1.9021) (66,1.4103) (67,1.2711) (68,1.7243) (69,1.2764) (70,2.5799) (71,2.4126) (72,1.2554) (73,1.8144) (74,2.2786) (75,2.0286) (76,1.6342) (77,0.7905) (78,1.6417) (79,1.8603) (80,0.5780) (81,1.1591) (82,1.5864) (83,1.0888) (84,0.6855) (85,3.3337) (86,1.0201) (87,1.4649) (88,1.2807) (89,1.3627) (90,2.3317) (91,2.0938) (92,0.4500) (93,0.4054) (94,0.6476) (95,0.6929) (96,1.0536) (97,0.8018) (98,0.9746) (99,0.8889) (100,0.6727) (101,0.9822) (102,1.1991) (103,2.3183) (104,0.8233) (105,1.6951) (106,0.7027) (107,0.7553) (108,1.2522) (109,0.7682) (110,1.6510) (111,0.7056) (112,0.8735) (113,0.8661) (114,0.7053) (115,1.4568) (116,1.3099) (117,0.6345) (118,0.5619) (119,0.4015) (120,0.9781) (121,0.7087) (122,1.4792) (123,1.2351) (124,0.8074) (125,1.0664) (126,0.4928) (127,1.1302) (128,0.5730) (129,0.8039) (130,1.4453) (131,0.4033) (132,1.1299) (133,1.8650) (134,0.8801) (135,1.4722) (136,0.4005) (137,1.3334) (138,0.4903) (139,1.4564) (140,1.4293) (141,1.0779) (142,1.0222) (143,0.9011) (144,1.7381) (145,1.5913) (146,1.8059) (147,1.1735) (148,1.3095) (149,1.1486) (150,1.2593) (151,0.7135) (152,2.6053) (153,2.3738) (154,2.1928) (155,1.3365) (156,1.9231) (157,1.3357) (158,1.1990) (159,1.4208) (160,1.2316) (161,1.4937) (162,2.2699) (163,1.0845) (164,1.4346) (165,1.6809) (166,1.0176) (167,1.4086) (168,0.7395) (169,0.4172) (170,0.9118) (171,0.6641) (172,1.5955) (173,1.3815) (174,1.2618) (175,0.9869) (176,0.6606) (177,0.7941) (178,1.2116) (179,0.8243) (180,1.1838) (181,2.1558) (182,0.8033) (183,1.5133) (184,1.1420) (185,1.1870) (186,1.9044) (187,0.6881) (188,1.6622) (189,1.8700) (190,1.5310) (191,1.2790) (192,0.5040) (193,0.1614) (194,0.0443) (195,0.6544) (196,0.0496) (197,0.3904) (198,0.3673) (199,0.4414) (200,1.5567) (201,1.6083) (202,0.6133) (203,1.3218) (204,0.1547) (205,2.5073) (206,2.7077) (207,0.8202) (208,0.1566) (209,0.3754) (210,0.4006) (211,0.7003) (212,0.7721) (213,1.6612) (214,0.6342) (215,1.7112) (216,0.5544) (217,2.3439) (218,1.2687) (219,1.6517) (220,1.5706) (221,1.5225) (222,0.6482) (223,1.5190) (224,0.4965) (225,1.8925) (226,0.7653) (227,1.4712) (228,1.3429) (229,0.7854) (230,1.1634) (231,0.9976) (232,2.2094) (233,1.3348) (234,0.7649) (235,2.4078) (236,1.6360) (237,1.2904) (238,1.3588) (239,2.2243) (240,1.2875) (241,1.2811) (242,1.8242) (243,0.9169) (244,1.8987) (245,1.6241) (246,2.7638) (247,0.9484) (248,1.3812) (249,1.9387) (250,1.4764) (251,1.0011) (252,1.9175) (253,1.2481) (254,0.4545) (255,1.1352) (256,0.5430) (257,0.6541) (258,0.7069) (259,1.4878) (260,0.9779) (261,0.9318) (262,1.0427) (263,1.4007) (264,2.4230) (265,2.7026) (266,0.9007) (267,2.3697) (268,2.0911) (269,2.2949) (270,1.4941) (271,1.5274) (272,0.7388) (273,0.6923) (274,0.6666) (275,0.3990) (276,0.8013) (277,1.1168) (278,0.5521) (279,1.0604) (280,1.1246) (281,0.8006) (282,2.0144) (283,0.5687) (284,1.0658) (285,0.3291) (286,0.5112) (287,0.3238) (288,0.7884) (289,0.3192) (290,0.3798) (291,0.7918) (292,1.2040) (293,0.6102) (294,1.9474) (295,2.6109) (296,1.8603) (297,1.4689) (298,0.8976) (299,1.3264) (300,1.7700) (301,1.4259) (302,0.6017) (303,0.5317) (304,0.7476) (305,1.0431) (306,0.9069)
\end{pspicture}

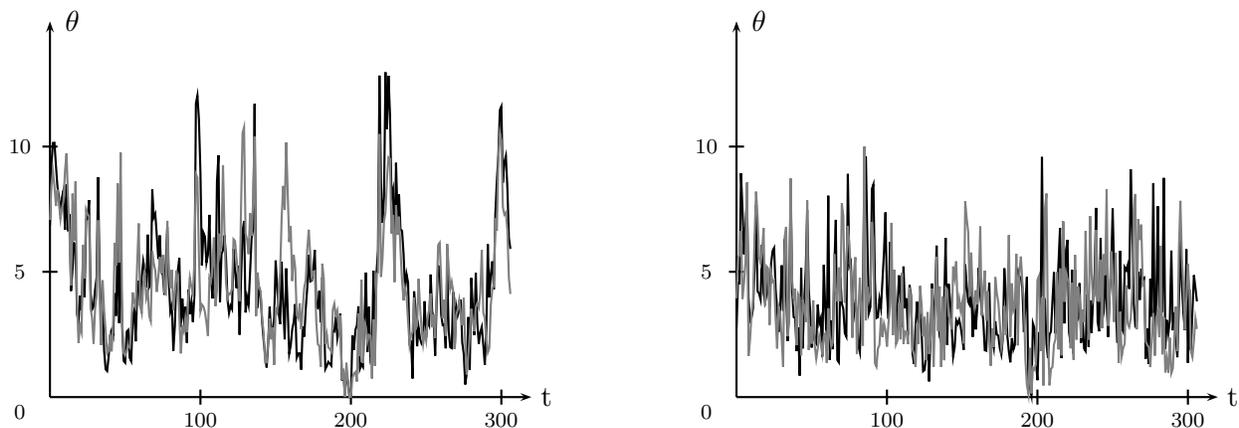
\noindent Fig. 12. Angles (in degrees) between vectors $RI^{pred}$ and $RI^{real}$ for windows of 100 days with sliding windows of 5 days, for the original data (left) and for the residues of the regression (right), both without short-selling. The results without cleaning are in black lines, and the results with cleaning are in gray lines.

\vskip 0.3 cm

\subsection{Distance and Kullback-Leibler Distance}

Here we place the graphs of the Distance as measured by (\ref{distance}) and the Kullback-Leibler distance (\ref{KL}). As we wrote before, they are measures of the differences between the correlation matrices and not of the efficient frontiers of the portfolios. So, they are more fundamental, in terms of not depending on the building of portfolios or, for example, if short selling is allowed or not.

Figure 13 shows the evolution of the distance ($Dist$) in time, and Figure 13 shows the evolution of the Kullback-Leibler distance ($D_{KL}$) in time. The graph for the simple distance shows a peak around the time of high volatility of the BM\&F-Bovespa (the Subprime Crisis of 2008), and the results for cleaning and no cleaning are very similar for the original data (without the regression). Distances are about the same for the correlation matrices built from the residues of the regression and also more stable in time. The cleaned result is consistently smaller than the uncleaned one, except for the period of highest volatility.

The results for the Kullback-Leibler distance are very similar to the ones obtained with the simple distance, also showing stronger peaks during times of high volatility and a similiar behavior for cleaned and uncleaned correlation matrices based on the original data. For correlation matrices based on the residues of the regression, the distances are much smaller, and the interesting result is that the distances for cleaned correlation matrices are much larger than the ones for the uncleaned correlation matrices.

\begin{pspicture}(-1,-0.6)(8,5.6)
\psset{xunit=0.02,yunit=1}
\psline{->}(0,0)(320,0) \rput(330,0){t} \psline{->}(0,0)(0,5) \rput(25,5){$Dist$} \scriptsize  \rput(-20,-0.2){0} \psline(100,-0.1)(100,0.1) \rput(100,-0.3){100} \psline(200,-0.1)(200,0.1) \rput(200,-0.3){200} \psline(300,-0.1)(300,0.1) \rput(300,-0.3){300} \psline(-5,1.25)(5,1.25) \rput(-25,1.25){$50$} \psline(-5,2.5)(5,2.5) \rput(-25,2.5){$100$} \psline(-5,3.75)(5,3.75) \rput(-25,3.75){$150$}
\psline (0,1.7484) (1,1.8458) (2,2.3023) (3,2.3394) (4,2.2079) (5,2.4560) (6,2.6385) (7,2.5561) (8,2.5376) (9,2.7747) (10,2.8151) (11,2.8113) (12,2.9889) (13,3.0024) (14,2.7637) (15,2.5814) (16,2.3822) (17,2.1275) (18,1.5590) (19,1.5956) (20,1.5071) (21,1.3268) (22,1.2019) (23,1.1161) (24,1.0644) (25,1.1383) (26,1.2545) (27,1.2898) (28,1.3757) (29,1.4334) (30,1.6443) (31,1.8335) (32,1.9677) (33,1.8063) (34,1.6371) (35,1.6555) (36,1.7182) (37,1.6557) (38,1.7681) (39,1.9491) (40,1.8721) (41,1.4200) (42,1.3560) (43,1.2718) (44,1.0271) (45,0.9758) (46,1.1032) (47,1.1233) (48,1.0065) (49,0.9971) (50,1.0302) (51,1.0532) (52,1.0984) (53,1.0706) (54,1.1329) (55,1.1901) (56,1.2513) (57,1.2918) (58,1.5465) (59,1.8521) (60,1.7364) (61,1.7228) (62,2.0140) (63,2.1177) (64,1.8956) (65,1.8655) (66,1.7798) (67,1.7646) (68,1.9873) (69,1.8909) (70,2.0678) (71,1.6087) (72,1.6282) (73,1.6769) (74,1.5790) (75,1.5112) (76,1.5523) (77,1.5393) (78,1.2912) (79,1.2380) (80,1.3050) (81,1.2527) (82,1.4106) (83,1.3467) (84,1.4425) (85,1.5764) (86,1.4902) (87,1.8278) (88,2.2148) (89,2.0148) (90,2.3162) (91,2.2200) (92,2.4792) (93,2.6029) (94,2.5106) (95,2.5985) (96,2.6833) (97,2.8528) (98,2.5345) (99,1.6900) (100,1.1647) (101,1.1216) (102,1.3339) (103,1.1400) (104,1.3424) (105,1.4730) (106,1.5324) (107,2.4470) (108,2.5854) (109,2.4288) (110,1.7556) (111,1.7104) (112,2.0159) (113,2.1361) (114,2.2720) (115,2.0970) (116,2.1396) (117,1.3403) (118,1.2502) (119,1.0453) (120,1.1402) (121,1.2807) (122,1.5334) (123,1.5291) (124,1.7531) (125,1.6238) (126,1.6990) (127,2.5532) (128,2.6239) (129,2.7433) (130,1.7708) (131,1.8691) (132,2.1582) (133,2.1285) (134,2.2336) (135,1.9008) (136,1.9578) (137,1.1311) (138,1.3065) (139,1.3496) (140,1.2990) (141,1.3104) (142,1.1624) (143,0.9474) (144,0.9427) (145,0.9606) (146,1.0539) (147,0.9206) (148,0.9028) (149,0.8420) (150,0.8565) (151,0.8839) (152,1.0231) (153,1.0264) (154,1.1199) (155,1.1616) (156,1.2469) (157,1.5653) (158,1.6412) (159,1.6425) (160,1.3555) (161,1.5497) (162,1.2665) (163,1.1612) (164,1.1156) (165,1.1162) (166,1.1253) (167,1.1371) (168,1.1093) (169,1.1252) (170,1.1959) (171,1.1840) (172,1.1366) (173,1.0707) (174,1.1508) (175,1.1451) (176,1.1593) (177,1.0835) (178,1.0350) (179,0.9181) (180,0.8646) (181,1.4486) (182,1.2347) (183,1.0650) (184,1.0307) (185,1.0162) (186,1.0327) (187,1.2666) (188,1.4023) (189,1.6056) (190,1.7435) (191,1.7451) (192,1.6024) (193,1.3747) (194,1.4247) (195,1.7205) (196,1.6550) (197,2.5374) (198,2.4409) (199,2.9488) (200,3.0270) (201,4.2073) (202,3.8759) (203,3.3897) (204,3.3383) (205,3.1513) (206,3.0168) (207,3.3041) (208,3.3664) (209,3.6101) (210,3.5673) (211,3.5505) (212,3.0145) (213,2.5869) (214,1.1832) (215,1.0850) (216,0.9674) (217,1.9526) (218,2.1880) (219,4.4084) (220,3.4658) (221,3.3520) (222,3.8434) (223,3.9900) (224,3.9147) (225,3.3235) (226,3.0880) (227,2.9671) (228,3.0868) (229,2.9688) (230,3.0391) (231,2.8178) (232,2.6716) (233,2.5969) (234,2.1950) (235,1.7923) (236,2.1932) (237,1.8561) (238,1.6853) (239,1.0748) (240,1.2370) (241,1.0479) (242,0.9331) (243,0.8807) (244,1.0993) (245,1.0748) (246,1.2123) (247,1.3522) (248,1.1003) (249,0.9973) (250,0.9881) (251,1.0440) (252,1.0245) (253,0.9758) (254,1.1149) (255,1.0862) (256,1.1878) (257,1.1912) (258,1.2370) (259,1.2173) (260,1.2330) (261,1.1451) (262,1.1829) (263,1.2716) (264,1.2894) (265,1.2745) (266,1.5895) (267,1.6702) (268,1.3142) (269,1.2068) (270,1.3341) (271,1.7038) (272,1.5437) (273,1.6511) (274,1.4157) (275,1.4056) (276,1.2909) (277,1.2905) (278,1.3619) (279,1.2485) (280,1.2666) (281,1.1581) (282,1.5666) (283,1.6456) (284,1.5308) (285,1.7078) (286,1.8066) (287,1.8764) (288,1.7597) (289,1.3314) (290,1.3108) (291,1.4486) (292,1.3940) (293,1.3206) (294,1.2713) (295,1.2856) (296,1.6370) (297,2.0626) (298,2.0509) (299,2.2923) (300,2.0542) (301,1.9864) (302,1.4836) (303,1.4246) (304,1.6389) (305,1.7224) (306,1.6160)
\psline[linecolor=gray] (0,1.2138) (1,1.2891) (2,1.7138) (3,1.7362) (4,1.5847) (5,1.8425) (6,2.1897) (7,1.9098) (8,1.8679) (9,2.2492) (10,2.3123) (11,2.2987) (12,2.4872) (13,2.4787) (14,2.2810) (15,2.0451) (16,2.0293) (17,1.6356) (18,1.2110) (19,1.2975) (20,1.2255) (21,0.9167) (22,0.8412) (23,0.7774) (24,0.8032) (25,0.9064) (26,0.9810) (27,1.0620) (28,1.1490) (29,1.0916) (30,1.5022) (31,1.6090) (32,1.7470) (33,1.5512) (34,1.3861) (35,1.4817) (36,1.5073) (37,1.5722) (38,1.5890) (39,1.7557) (40,1.7223) (41,1.1293) (42,1.0911) (43,1.0323) (44,0.7520) (45,0.7055) (46,0.8290) (47,0.8195) (48,0.7612) (49,0.7515) (50,0.7654) (51,0.7187) (52,0.7601) (53,0.7356) (54,0.7871) (55,0.8190) (56,0.8847) (57,0.9086) (58,1.1288) (59,1.3953) (60,1.2758) (61,1.2517) (62,1.5166) (63,1.5606) (64,1.3478) (65,1.3367) (66,1.3111) (67,1.2953) (68,1.5021) (69,1.4169) (70,1.5276) (71,1.1310) (72,1.1672) (73,1.2333) (74,1.1186) (75,1.0373) (76,0.9762) (77,0.9511) (78,0.9134) (79,0.8839) (80,0.9682) (81,0.9744) (82,1.0113) (83,1.0507) (84,1.0480) (85,1.1658) (86,1.1104) (87,1.5014) (88,1.8918) (89,1.6664) (90,1.9442) (91,1.8257) (92,2.0779) (93,2.2440) (94,2.1623) (95,2.2686) (96,2.4040) (97,2.5759) (98,2.3361) (99,1.4780) (100,0.8969) (101,0.8488) (102,0.8347) (103,0.7751) (104,0.7776) (105,0.8819) (106,0.9465) (107,1.9433) (108,2.0660) (109,1.9199) (110,1.3027) (111,1.2668) (112,1.5693) (113,1.6739) (114,1.8247) (115,1.6385) (116,1.5075) (117,0.8490) (118,0.8080) (119,0.6309) (120,0.7252) (121,0.8959) (122,1.1709) (123,1.1264) (124,1.3434) (125,1.2168) (126,1.2716) (127,2.3101) (128,2.3648) (129,2.5100) (130,1.4883) (131,1.5852) (132,1.8727) (133,1.8460) (134,1.9585) (135,1.6065) (136,1.5200) (137,0.6676) (138,0.8265) (139,0.8568) (140,0.8204) (141,0.8447) (142,0.7372) (143,0.5599) (144,0.5674) (145,0.6283) (146,0.6975) (147,0.5482) (148,0.5360) (149,0.4647) (150,0.4666) (151,0.4800) (152,0.6150) (153,0.6557) (154,0.7522) (155,0.7697) (156,0.8843) (157,1.1910) (158,1.2573) (159,1.2323) (160,0.9175) (161,1.1625) (162,0.8412) (163,0.7364) (164,0.6846) (165,0.6892) (166,0.6907) (167,0.7186) (168,0.7139) (169,0.7419) (170,0.8487) (171,0.8261) (172,0.7234) (173,0.6491) (174,0.7070) (175,0.7039) (176,0.7105) (177,0.6122) (178,0.5726) (179,0.4704) (180,0.3875) (181,1.0425) (182,0.6921) (183,0.5348) (184,0.6833) (185,0.6722) (186,0.6828) (187,0.9261) (188,1.0524) (189,1.2519) (190,1.4146) (191,1.4171) (192,1.2671) (193,1.0510) (194,1.1794) (195,1.4038) (196,1.3461) (197,2.2815) (198,2.1756) (199,2.6998) (200,2.7137) (201,4.0331) (202,3.5913) (203,3.0941) (204,3.2164) (205,3.0451) (206,2.9225) (207,3.2159) (208,3.2798) (209,3.5518) (210,3.5002) (211,3.4945) (212,2.9594) (213,2.5036) (214,1.0520) (215,0.9398) (216,0.8282) (217,1.7901) (218,2.0429) (219,4.1732) (220,3.2078) (221,3.1265) (222,3.5987) (223,3.7269) (224,3.6673) (225,2.8707) (226,2.6709) (227,2.5790) (228,2.7304) (229,2.6233) (230,2.6734) (231,2.4704) (232,2.3476) (233,2.2824) (234,1.8812) (235,1.5036) (236,1.8977) (237,1.5723) (238,1.3843) (239,0.8111) (240,0.9648) (241,0.7499) (242,0.7852) (243,0.5372) (244,0.8994) (245,0.7718) (246,0.8181) (247,0.9515) (248,0.5500) (249,0.5516) (250,0.5235) (251,0.5897) (252,0.5481) (253,0.5301) (254,0.6699) (255,0.6475) (256,0.7445) (257,0.7753) (258,0.8422) (259,0.8200) (260,0.8537) (261,0.8867) (262,0.8314) (263,1.0389) (264,0.9532) (265,0.9240) (266,1.2617) (267,1.3431) (268,0.9384) (269,0.8621) (270,0.9404) (271,1.2744) (272,1.0908) (273,1.1714) (274,0.9448) (275,0.9765) (276,0.9050) (277,0.8654) (278,0.9464) (279,0.6847) (280,0.9036) (281,0.8066) (282,1.1359) (283,1.2117) (284,1.1130) (285,1.2965) (286,1.4127) (287,1.4821) (288,1.3666) (289,1.0963) (290,1.1025) (291,1.1935) (292,1.1748) (293,1.0270) (294,0.9150) (295,0.9447) (296,1.2964) (297,1.6361) (298,1.5926) (299,1.9427) (300,1.7736) (301,1.4974) (302,0.9557) (303,1.0373) (304,1.1947) (305,1.3154) (306,1.2689)
\end{pspicture}
\begin{pspicture}(-1,-0.6)(4,5.6)
\psset{xunit=0.02,yunit=1}
\psline{->}(0,0)(320,0) \rput(330,0){t} \psline{->}(0,0)(0,5) \rput(25,5){$Dist$} \scriptsize  \rput(-20,-0.2){0} \psline(100,-0.1)(100,0.1) \rput(100,-0.3){100} \psline(200,-0.1)(200,0.1) \rput(200,-0.3){200} \psline(300,-0.1)(300,0.1) \rput(300,-0.3){300} \psline(-5,1.25)(5,1.25) \rput(-25,1.25){$25$} \psline(-5,2.5)(5,2.5) \rput(-25,2.5){$50$} \psline(-5,3.75)(5,3.75) \rput(-25,3.75){$75$}
\psline (0,2.8258) (1,2.9274) (2,2.9181) (3,2.7924) (4,2.8407) (5,2.6611) (6,2.5820) (7,2.7642) (8,2.7872) (9,2.8636) (10,2.6592) (11,2.6946) (12,2.6259) (13,2.6961) (14,2.6576) (15,2.8781) (16,2.9019) (17,2.9253) (18,3.0504) (19,2.7331) (20,2.7401) (21,2.9964) (22,2.9392) (23,2.7923) (24,2.7571) (25,2.6012) (26,2.5638) (27,2.7499) (28,2.7771) (29,2.7788) (30,2.7780) (31,2.8641) (32,2.8107) (33,2.7834) (34,2.9004) (35,2.9604) (36,2.8700) (37,2.7500) (38,2.8643) (39,2.8961) (40,2.7957) (41,2.8731) (42,2.7633) (43,2.7031) (44,2.7201) (45,2.6928) (46,2.6366) (47,2.7667) (48,2.6393) (49,2.4805) (50,2.6081) (51,2.6480) (52,2.7229) (53,2.7176) (54,2.7639) (55,2.8003) (56,2.6383) (57,2.6671) (58,2.7296) (59,2.8255) (60,2.9791) (61,2.9466) (62,2.8516) (63,2.8259) (64,2.6788) (65,2.6733) (66,2.4912) (67,2.4960) (68,2.5160) (69,2.4893) (70,2.5824) (71,2.6524) (72,2.6681) (73,2.6627) (74,2.7002) (75,2.6506) (76,2.5616) (77,2.5609) (78,2.4842) (79,2.5434) (80,2.6793) (81,2.7781) (82,2.7533) (83,2.7992) (84,2.7158) (85,2.7212) (86,2.7952) (87,2.7845) (88,2.8724) (89,2.8652) (90,2.9978) (91,2.9786) (92,2.9152) (93,2.9637) (94,2.8853) (95,2.8318) (96,2.6589) (97,2.7164) (98,2.6774) (99,2.7718) (100,2.6471) (101,2.6263) (102,2.7010) (103,2.7772) (104,2.8274) (105,2.8684) (106,2.9681) (107,3.0225) (108,3.1141) (109,3.1023) (110,3.0863) (111,3.0732) (112,3.1094) (113,3.1926) (114,3.0566) (115,3.1632) (116,2.9229) (117,2.8456) (118,2.8342) (119,2.8593) (120,2.9415) (121,2.8504) (122,2.7319) (123,2.9077) (124,2.9357) (125,2.9793) (126,3.0741) (127,3.0190) (128,2.9728) (129,3.0150) (130,2.9883) (131,3.1179) (132,3.3658) (133,3.2889) (134,3.2032) (135,3.3496) (136,3.0508) (137,3.0122) (138,2.9305) (139,2.8859) (140,2.9856) (141,2.9453) (142,2.9167) (143,2.8671) (144,2.7939) (145,2.7303) (146,2.7653) (147,2.8540) (148,2.8462) (149,2.8561) (150,2.8849) (151,2.8902) (152,3.0341) (153,2.9593) (154,2.9524) (155,2.9824) (156,2.8568) (157,2.9603) (158,2.9702) (159,2.8617) (160,3.0292) (161,3.1876) (162,3.1700) (163,3.1271) (164,3.0522) (165,2.9292) (166,2.8712) (167,2.9602) (168,2.9391) (169,2.8267) (170,2.7610) (171,2.8819) (172,3.1237) (173,3.0502) (174,3.0471) (175,2.9713) (176,3.0104) (177,3.0078) (178,3.0788) (179,3.1627) (180,3.1081) (181,3.1167) (182,3.0839) (183,3.0521) (184,3.0248) (185,3.1464) (186,3.3059) (187,3.3549) (188,3.3530) (189,3.4422) (190,3.4359) (191,3.5492) (192,3.8512) (193,3.6259) (194,3.7477) (195,3.6055) (196,3.5176) (197,3.8793) (198,4.0657) (199,4.1134) (200,4.4519) (201,4.2935) (202,4.1046) (203,4.1110) (204,3.5958) (205,3.8358) (206,3.9220) (207,3.5474) (208,3.6210) (209,3.5989) (210,3.4883) (211,3.7336) (212,3.9376) (213,4.0276) (214,3.9021) (215,3.8786) (216,3.9098) (217,4.4986) (218,3.8322) (219,3.7247) (220,4.1132) (221,3.9237) (222,3.6481) (223,3.7184) (224,3.6011) (225,3.5276) (226,3.4387) (227,3.4169) (228,3.5741) (229,3.5794) (230,3.4367) (231,3.3664) (232,3.3687) (233,3.4944) (234,3.4313) (235,3.4745) (236,3.5603) (237,3.7414) (238,3.6107) (239,3.2918) (240,3.4381) (241,3.2755) (242,2.9336) (243,2.9667) (244,3.2953) (245,2.8515) (246,2.6887) (247,2.6823) (248,2.6918) (249,2.8022) (250,2.8685) (251,2.9451) (252,2.9884) (253,2.8374) (254,2.8145) (255,2.7112) (256,2.7192) (257,2.6664) (258,2.5964) (259,2.6529) (260,2.6372) (261,2.5828) (262,2.5592) (263,2.5851) (264,2.5826) (265,2.6273) (266,2.6099) (267,2.5530) (268,2.6630) (269,2.5329) (270,2.6595) (271,2.6915) (272,2.9741) (273,2.9260) (274,2.9629) (275,3.0748) (276,3.0997) (277,3.2623) (278,3.2364) (279,3.1740) (280,2.9427) (281,2.9487) (282,2.9794) (283,3.0050) (284,2.9936) (285,3.1039) (286,2.8155) (287,2.8228) (288,2.8149) (289,2.8696) (290,2.8555) (291,2.9272) (292,2.9494) (293,2.8865) (294,2.8337) (295,2.7611) (296,2.7474) (297,2.9197) (298,2.9438) (299,2.9319) (300,2.9464) (301,3.0441) (302,3.1969) (303,3.0742) (304,3.1277) (305,3.2675) (306,2.9934)
\psline[linecolor=gray] (0,2.0078) (1,2.0852) (2,2.0526) (3,1.9988) (4,1.9221) (5,1.6974) (6,1.7205) (7,2.1326) (8,2.0230) (9,2.3068) (10,2.1217) (11,2.1858) (12,2.2088) (13,2.0561) (14,1.9350) (15,2.4116) (16,2.1647) (17,2.3292) (18,2.4690) (19,1.8945) (20,2.2765) (21,2.5982) (22,2.4789) (23,2.2718) (24,2.0488) (25,1.8476) (26,1.6686) (27,1.8645) (28,1.9799) (29,1.8498) (30,1.8766) (31,1.8086) (32,1.9684) (33,1.7169) (34,2.1445) (35,2.2426) (36,2.0901) (37,1.9368) (38,2.1200) (39,2.0523) (40,2.0569) (41,1.9464) (42,1.9479) (43,1.9243) (44,1.9711) (45,1.9997) (46,1.7510) (47,1.9375) (48,1.9438) (49,1.8430) (50,1.9110) (51,1.8806) (52,2.0558) (53,1.9456) (54,1.9787) (55,1.9737) (56,1.5853) (57,1.5387) (58,1.6090) (59,1.7891) (60,1.9401) (61,2.0941) (62,2.0881) (63,1.9475) (64,1.7319) (65,1.6062) (66,1.6276) (67,1.7222) (68,1.6106) (69,1.6446) (70,1.7091) (71,1.9162) (72,1.9958) (73,1.9629) (74,2.1374) (75,2.0331) (76,1.8197) (77,1.5985) (78,1.7222) (79,1.8690) (80,1.9367) (81,2.0651) (82,1.9878) (83,2.3010) (84,1.7963) (85,1.7643) (86,1.6633) (87,1.7753) (88,1.8940) (89,1.8496) (90,2.0207) (91,2.0034) (92,2.0400) (93,2.0932) (94,1.9254) (95,1.8958) (96,1.7655) (97,1.5740) (98,1.5732) (99,1.6390) (100,1.7075) (101,1.7567) (102,2.0118) (103,1.9266) (104,2.1170) (105,2.1670) (106,2.1700) (107,2.1678) (108,2.1105) (109,2.1025) (110,2.0902) (111,1.9686) (112,2.0889) (113,2.2601) (114,1.9094) (115,1.9816) (116,1.6591) (117,1.9103) (118,1.9695) (119,1.9870) (120,2.1601) (121,1.8280) (122,2.1230) (123,2.3864) (124,2.3942) (125,2.4911) (126,2.6665) (127,2.4656) (128,2.3044) (129,2.4394) (130,2.6115) (131,2.7401) (132,2.8207) (133,2.8395) (134,2.9267) (135,2.8480) (136,2.4072) (137,2.2758) (138,2.2220) (139,2.0518) (140,2.3798) (141,2.2279) (142,2.1562) (143,2.0965) (144,1.9506) (145,2.0561) (146,1.9784) (147,1.9107) (148,2.0339) (149,2.0796) (150,1.8473) (151,2.0422) (152,2.0069) (153,2.3726) (154,2.4219) (155,2.5520) (156,2.1516) (157,2.1665) (158,2.3518) (159,1.9651) (160,2.1500) (161,2.2814) (162,2.1687) (163,2.0021) (164,1.7582) (165,1.5878) (166,1.4046) (167,1.7052) (168,1.6119) (169,1.7318) (170,1.5363) (171,1.7164) (172,1.9750) (173,2.1166) (174,2.3990) (175,2.4188) (176,2.0507) (177,2.1445) (178,2.4388) (179,2.3665) (180,2.3896) (181,2.5675) (182,2.3495) (183,2.2050) (184,2.3238) (185,2.3569) (186,2.6613) (187,2.6861) (188,2.4710) (189,2.8262) (190,2.8700) (191,2.9970) (192,3.3839) (193,2.8241) (194,2.9994) (195,2.8100) (196,2.4148) (197,3.0546) (198,3.3063) (199,3.5094) (200,3.7767) (201,4.1071) (202,3.4118) (203,3.4441) (204,2.9584) (205,3.2255) (206,3.2153) (207,2.7924) (208,2.9119) (209,2.9511) (210,2.8788) (211,3.1757) (212,3.3004) (213,3.4100) (214,3.3361) (215,3.2422) (216,3.0201) (217,3.7563) (218,2.9291) (219,2.9942) (220,3.2411) (221,3.2110) (222,3.1277) (223,2.8163) (224,2.9243) (225,2.8245) (226,2.8557) (227,2.8246) (228,3.0251) (229,3.0773) (230,2.9247) (231,2.9624) (232,2.9140) (233,2.8439) (234,2.6983) (235,2.6853) (236,2.7957) (237,3.0828) (238,2.9231) (239,2.7052) (240,2.8595) (241,2.6770) (242,2.2509) (243,2.3997) (244,2.5463) (245,2.1732) (246,2.2150) (247,1.9108) (248,1.9005) (249,2.0104) (250,2.1607) (251,2.1874) (252,2.1880) (253,1.8888) (254,1.8424) (255,1.6856) (256,1.5511) (257,1.5809) (258,1.7043) (259,1.7809) (260,1.7805) (261,1.7030) (262,1.7773) (263,1.9262) (264,1.9088) (265,2.0679) (266,2.2529) (267,1.9368) (268,2.1332) (269,1.9659) (270,2.1919) (271,2.1321) (272,2.3086) (273,2.0228) (274,2.3440) (275,2.4046) (276,2.4793) (277,2.6964) (278,2.3957) (279,2.4869) (280,2.4334) (281,2.5059) (282,2.5862) (283,2.4556) (284,2.5948) (285,2.7541) (286,2.3913) (287,2.4294) (288,2.4504) (289,2.3390) (290,2.3476) (291,2.4491) (292,2.4754) (293,2.4063) (294,2.3416) (295,2.2361) (296,2.3455) (297,2.6920) (298,2.8393) (299,2.7177) (300,2.6405) (301,2.7923) (302,2.7772) (303,2.8060) (304,2.7235) (305,2.9867) (306,2.6187)
\end{pspicture}

\noindent Fig. 13. Distance ($Dist$) between the predicted and realized correlation matrices for windows of 100 days with sliding windows of 5 days, for the original data (left) and for the residues of the regression (right). The results without cleaning are in black lines, and the results with cleaning are in gray lines.

\vskip 0.3 cm

\begin{pspicture}(-1,-0.6)(8,5.6)
\psset{xunit=0.02,yunit=1}
\psline{->}(0,0)(320,0) \rput(330,0){t} \psline{->}(0,0)(0,4.8) \rput(23,4.8){$D_{KL}$} \scriptsize  \rput(-20,-0.2){0} \psline(100,-0.1)(100,0.1) \rput(100,-0.3){100} \psline(200,-0.1)(200,0.1) \rput(200,-0.3){200} \psline(300,-0.1)(300,0.1) \rput(300,-0.3){300} \psline(-5,1.2)(5,1.2) \rput(-25,1.2){$0.5$} \psline(-5,2.4)(5,2.4) \rput(-25,2.4){$1$} \psline(-5,3.6)(5,3.6) \rput(-25,3.6){$1.5$}
\psline (0,0.9434) (1,1.0240) (2,1.5079) (3,1.6109) (4,1.4335) (5,1.6492) (6,1.9057) (7,1.9490) (8,2.0494) (9,2.4196) (10,2.4806) (11,2.5579) (12,2.6792) (13,2.6264) (14,2.3316) (15,2.0072) (16,1.6809) (17,1.2994) (18,0.4624) (19,0.5822) (20,0.5672) (21,0.2482) (22,0.0877) (23,0.0730) (24,0.0624) (25,0.1641) (26,0.1091) (27,0.3215) (28,0.4858) (29,0.5973) (30,0.9202) (31,1.1161) (32,1.4192) (33,1.2070) (34,0.8700) (35,0.7769) (36,1.0325) (37,0.8594) (38,0.8212) (39,1.4277) (40,1.4632) (41,0.7985) (42,0.6251) (43,0.5744) (44,0.1847) (45,0.0810) (46,0.0797) (47,0.0273) (48,0.0277) (49,0.0091) (50,0.0198) (51,0.0463) (52,0.1032) (53,0.0715) (54,0.0968) (55,0.1332) (56,0.2201) (57,0.2052) (58,0.5495) (59,0.7707) (60,0.6675) (61,0.5452) (62,0.6636) (63,0.6028) (64,0.3279) (65,0.2869) (66,0.2035) (67,0.2158) (68,0.6420) (69,0.4704) (70,0.9344) (71,0.6914) (72,0.6470) (73,0.6349) (74,0.5104) (75,0.4574) (76,0.4431) (77,0.4702) (78,0.1507) (79,0.0979) (80,0.1339) (81,0.1519) (82,0.3733) (83,0.2451) (84,0.3047) (85,0.4279) (86,0.4386) (87,0.7267) (88,1.0422) (89,0.8597) (90,1.1937) (91,1.1652) (92,1.2224) (93,1.2844) (94,1.4361) (95,1.5563) (96,1.5703) (97,1.4948) (98,1.0861) (99,0.4949) (100,0.1945) (101,0.1211) (102,0.1936) (103,0.1318) (104,0.3387) (105,0.4597) (106,0.5105) (107,1.3586) (108,1.3843) (109,1.3533) (110,0.7131) (111,0.6983) (112,0.9620) (113,1.0780) (114,1.0498) (115,0.9061) (116,1.1036) (117,0.4065) (118,0.2266) (119,0.0651) (120,0.0737) (121,0.2270) (122,0.6784) (123,0.6866) (124,0.9056) (125,0.8192) (126,0.6998) (127,1.7224) (128,1.6981) (129,2.0503) (130,0.7275) (131,0.8387) (132,1.1182) (133,1.0960) (134,1.2404) (135,0.9516) (136,1.1327) (137,0.0752) (138,0.1938) (139,0.2248) (140,0.1411) (141,0.1030) (142,0.0417) (143,0.0376) (144,0.0521) (145,0.1191) (146,0.2062) (147,0.0474) (148,0.0430) (149,0.0289) (150,0.0124) (151,0.0539) (152,0.0389) (153,0.0730) (154,0.1522) (155,0.2089) (156,0.1975) (157,0.5126) (158,0.5883) (159,0.4705) (160,0.2715) (161,0.6813) (162,0.2615) (163,0.1809) (164,0.1604) (165,0.2249) (166,0.2681) (167,0.2274) (168,0.2229) (169,0.2741) (170,0.4054) (171,0.3717) (172,0.2501) (173,0.2109) (174,0.1684) (175,0.1810) (176,0.1544) (177,0.2813) (178,0.1255) (179,0.0386) (180,0.0466) (181,1.1267) (182,0.6441) (183,0.5153) (184,0.4446) (185,0.4735) (186,0.3553) (187,0.9081) (188,1.0661) (189,1.1453) (190,1.3936) (191,1.3638) (192,0.9700) (193,0.4293) (194,0.1537) (195,0.7071) (196,0.7846) (197,2.9253) (198,2.4913) (199,3.5553) (200,2.8075) (201,3.9486) (202,3.9290) (203,3.2429) (204,3.2746) (205,2.8704) (206,2.6758) (207,2.8927) (208,2.9113) (209,2.9159) (210,2.9775) (211,2.8739) (212,2.3922) (213,2.0193) (214,0.3472) (215,0.1705) (216,0.1338) (217,1.0963) (218,1.2068) (219,3.6880) (220,3.2017) (221,3.0394) (222,3.4153) (223,3.4545) (224,3.2490) (225,2.7984) (226,2.6806) (227,2.4010) (228,2.5214) (229,2.4100) (230,2.2915) (231,2.1223) (232,1.9240) (233,1.8001) (234,1.3916) (235,0.9714) (236,1.2778) (237,0.8740) (238,0.7149) (239,0.1733) (240,0.3106) (241,0.1780) (242,0.0733) (243,0.0876) (244,0.0895) (245,0.1728) (246,0.3891) (247,0.4861) (248,0.2602) (249,0.1370) (250,0.1328) (251,0.1785) (252,-0.0222) (253,0.1194) (254,0.0741) (255,0.0437) (256,0.0769) (257,0.1732) (258,0.1584) (259,0.2119) (260,0.2684) (261,0.1199) (262,0.2138) (263,0.2805) (264,0.4433) (265,0.4675) (266,0.6849) (267,0.7563) (268,0.3923) (269,0.3349) (270,0.4884) (271,0.9986) (272,0.6644) (273,0.7082) (274,0.5265) (275,0.3914) (276,0.3519) (277,0.2633) (278,0.2251) (279,0.1501) (280,0.1562) (281,0.0515) (282,0.5139) (283,0.6374) (284,0.5505) (285,0.6110) (286,0.6743) (287,0.6604) (288,0.5621) (289,0.0935) (290,0.0827) (291,0.1337) (292,0.1019) (293,0.0691) (294,0.1588) (295,0.2075) (296,0.6803) (297,0.9769) (298,0.9853) (299,1.2082) (300,0.9747) (301,1.1228) (302,0.3418) (303,0.3095) (304,0.4865) (305,0.3101) (306,0.2083)
\psline[linecolor=gray] (0,1.1103) (1,1.0730) (2,1.7112) (3,1.8820) (4,1.7536) (5,1.9710) (6,1.7894) (7,2.9461) (8,1.5928) (9,2.5865) (10,2.4130) (11,2.5448) (12,2.9645) (13,2.7677) (14,2.5765) (15,2.2477) (16,1.6318) (17,1.4631) (18,0.4193) (19,0.4367) (20,0.4537) (21,0.2280) (22,0.0558) (23,0.0319) (24,0.0779) (25,0.2527) (26,0.3774) (27,0.5158) (28,0.6679) (29,0.8450) (30,1.2522) (31,1.5541) (32,1.8141) (33,1.5321) (34,1.1944) (35,1.1240) (36,1.2630) (37,1.3204) (38,1.5989) (39,2.1982) (40,2.1734) (41,1.3675) (42,1.2244) (43,0.5432) (44,0.1788) (45,0.0862) (46,0.1790) (47,0.0962) (48,0.1990) (49,0.0776) (50,0.1199) (51,0.0343) (52,0.0829) (53,0.0640) (54,0.0787) (55,0.1256) (56,0.2052) (57,0.1414) (58,0.5447) (59,0.8276) (60,0.6570) (61,0.4910) (62,0.6051) (63,0.6369) (64,0.3961) (65,0.3498) (66,0.1853) (67,0.1849) (68,0.7518) (69,0.4890) (70,1.1033) (71,0.7844) (72,0.7313) (73,0.7755) (74,0.5114) (75,0.4659) (76,0.4423) (77,0.4361) (78,0.2269) (79,0.2927) (80,0.3896) (81,0.3787) (82,0.5728) (83,0.4270) (84,0.5383) (85,0.6387) (86,0.7456) (87,1.0643) (88,1.2629) (89,1.2238) (90,1.6833) (91,1.6317) (92,1.4611) (93,1.6002) (94,1.5255) (95,1.6585) (96,1.7505) (97,1.9234) (98,1.3626) (99,0.5139) (100,0.0913) (101,0.0285) (102,0.2707) (103,0.1029) (104,0.2694) (105,0.2974) (106,0.3967) (107,1.2292) (108,1.3423) (109,1.2853) (110,0.6091) (111,0.5852) (112,0.8521) (113,0.9741) (114,0.9468) (115,0.8144) (116,1.4061) (117,0.0388) (118,-0.0109) (119,0.1239) (120,0.4161) (121,0.6779) (122,1.6969) (123,0.2121) (124,1.7194) (125,1.6734) (126,1.4956) (127,2.8527) (128,2.7919) (129,2.9038) (130,1.0654) (131,1.1978) (132,1.6109) (133,1.4396) (134,0.5671) (135,2.2034) (136,0.6917) (137,0.0828) (138,0.1031) (139,0.1691) (140,0.0694) (141,0.0958) (142,0.0264) (143,0.0847) (144,0.1242) (145,0.2576) (146,0.3272) (147,0.1079) (148,0.0945) (149,0.0028) (150,-0.0672) (151,-0.0472) (152,0.0857) (153,0.0617) (154,0.2832) (155,0.3436) (156,0.3403) (157,0.8506) (158,0.9609) (159,0.8435) (160,0.5597) (161,1.5411) (162,0.7712) (163,0.7778) (164,0.6570) (165,0.6954) (166,0.4099) (167,0.7267) (168,1.1498) (169,1.4165) (170,2.2124) (171,2.0495) (172,0.9332) (173,0.7816) (174,0.3157) (175,0.4071) (176,0.4126) (177,0.2195) (178,0.0969) (179,0.1027) (180,0.0529) (181,1.3176) (182,0.8329) (183,0.4872) (184,0.6101) (185,0.6544) (186,0.7808) (187,1.1861) (188,1.1740) (189,1.3734) (190,1.5774) (191,1.6020) (192,1.1555) (193,0.6493) (194,0.3729) (195,1.4890) (196,1.6117) (197,3.1238) (198,3.2245) (199,4.3911) (200,3.7059) (201,4.8080) (202,4.4373) (203,4.4662) (204,4.1888) (205,3.4940) (206,3.3986) (207,3.7495) (208,3.6177) (209,3.8024) (210,3.8505) (211,3.7021) (212,3.0975) (213,2.5552) (214,0.4983) (215,0.2072) (216,0.1912) (217,1.2710) (218,1.3341) (219,4.0667) (220,3.2366) (221,3.0852) (222,3.7569) (223,3.6329) (224,3.6758) (225,3.5435) (226,3.3308) (227,2.7639) (228,2.5424) (229,2.4116) (230,2.1631) (231,1.9173) (232,2.0737) (233,2.0010) (234,1.2691) (235,0.8499) (236,1.3316) (237,0.9437) (238,0.8079) (239,0.1210) (240,0.2460) (241,0.2312) (242,0.0649) (243,0.1332) (244,0.1650) (245,0.2273) (246,0.6430) (247,0.9054) (248,0.5802) (249,-0.1000) (250,-0.1163) (251,-0.0848) (252,-0.0757) (253,-0.1023) (254,0.2628) (255,0.2402) (256,0.3578) (257,0.4832) (258,0.3618) (259,0.3638) (260,0.3918) (261,0.2463) (262,0.3845) (263,0.6057) (264,0.8960) (265,0.9516) (266,1.4022) (267,1.3249) (268,0.6389) (269,0.4426) (270,0.4914) (271,0.6834) (272,0.6020) (273,0.7881) (274,0.4639) (275,0.3468) (276,0.2425) (277,0.1935) (278,0.1538) (279,0.0846) (280,0.0963) (281,0.0323) (282,0.4461) (283,0.6409) (284,0.5455) (285,0.5951) (286,0.6205) (287,0.6265) (288,0.5396) (289,0.1545) (290,0.1320) (291,0.1626) (292,0.1299) (293,0.1949) (294,0.1936) (295,0.2290) (296,0.7222) (297,1.0870) (298,1.1531) (299,1.3149) (300,1.0536) (301,1.0611) (302,0.5094) (303,0.3998) (304,0.5249) (305,0.3168) (306,0.1364)
\end{pspicture}
\begin{pspicture}(-1,-0.6)(4,5.6)
\psset{xunit=0.02,yunit=1}
\psline{->}(0,0)(320,0) \rput(330,0){t} \psline{->}(0,0)(0,5.2) \rput(23,5.2){$D_{KL}$} \scriptsize  \rput(-20,-0.2){0} \psline(100,-0.1)(100,0.1) \rput(100,-0.3){100} \psline(200,-0.1)(200,0.1) \rput(200,-0.3){200} \psline(300,-0.1)(300,0.1) \rput(300,-0.3){300} \psline(-5,1.3)(5,1.3) \rput(-25,1.3){$0.1$} \psline(-5,2.6)(5,2.6) \rput(-25,2.6){$0.2$} \psline(-5,3.9)(5,3.9) \rput(-25,3.9){$0.3$}
\psline (0,0.1468) (1,0.1154) (2,0.2808) (3,0.2364) (4,0.2699) (5,0.2275) (6,0.1907) (7,0.1855) (8,0.2312) (9,0.1850) (10,0.2189) (11,0.1821) (12,0.2203) (13,0.2261) (14,0.1755) (15,0.1020) (16,0.0468) (17,0.1072) (18,0.0521) (19,0.1057) (20,0.0866) (21,0.1442) (22,0.2945) (23,0.1868) (24,0.3717) (25,0.4265) (26,0.4871) (27,0.4639) (28,0.3926) (29,0.2986) (30,0.4932) (31,0.2867) (32,0.3124) (33,0.3384) (34,0.5629) (35,0.3151) (36,0.3077) (37,0.5517) (38,0.3766) (39,0.2667) (40,0.2661) (41,0.1562) (42,0.1840) (43,0.0874) (44,0.2462) (45,0.2875) (46,0.4909) (47,0.4243) (48,0.5940) (49,0.4966) (50,0.3917) (51,0.3243) (52,0.2884) (53,0.3121) (54,0.3165) (55,0.4475) (56,0.5157) (57,0.2967) (58,0.5920) (59,0.5625) (60,0.4642) (61,0.7362) (62,0.4071) (63,0.3027) (64,0.1326) (65,0.2303) (66,0.1697) (67,0.0643) (68,0.1146) (69,0.1893) (70,0.4310) (71,0.1176) (72,0.1298) (73,0.1062) (74,0.1173) (75,0.1496) (76,0.1054) (77,0.1825) (78,0.1304) (79,0.1985) (80,0.2731) (81,0.2044) (82,0.1953) (83,0.2559) (84,0.2683) (85,0.2610) (86,0.2755) (87,0.2960) (88,0.3143) (89,0.2609) (90,0.2857) (91,0.2126) (92,0.2662) (93,0.3626) (94,0.2436) (95,0.3081) (96,0.3708) (97,0.3885) (98,0.3003) (99,0.2299) (100,0.2172) (101,0.1757) (102,0.1429) (103,0.1571) (104,0.1215) (105,0.1102) (106,0.1691) (107,0.2142) (108,0.1780) (109,0.2288) (110,0.2650) (111,0.1468) (112,0.1053) (113,0.1456) (114,0.2263) (115,0.0033) (116,0.1344) (117,0.2541) (118,0.1365) (119,0.2642) (120,0.2510) (121,0.1626) (122,0.3338) (123,0.3156) (124,0.4475) (125,0.2788) (126,0.2050) (127,0.2308) (128,0.2651) (129,0.3037) (130,0.3028) (131,0.0232) (132,0.1984) (133,0.1656) (134,0.1905) (135,0.1870) (136,0.1145) (137,0.0893) (138,0.0254) (139,0.0584) (140,0.1309) (141,0.1461) (142,0.0624) (143,-0.0110) (144,0.2014) (145,0.1599) (146,-0.0043) (147,0.1104) (148,0.1634) (149,0.1303) (150,0.2429) (151,0.1218) (152,0.1960) (153,0.1458) (154,0.1317) (155,0.0635) (156,0.1021) (157,0.1181) (158,0.0951) (159,0.2384) (160,0.0786) (161,0.1444) (162,0.1879) (163,0.2224) (164,0.3064) (165,0.1113) (166,0.0812) (167,0.0966) (168,0.1205) (169,0.1004) (170,0.2535) (171,0.2782) (172,0.2864) (173,0.1472) (174,0.2473) (175,0.1728) (176,0.2700) (177,0.2277) (178,0.1471) (179,0.1919) (180,0.1765) (181,0.1444) (182,0.1437) (183,0.2074) (184,0.2537) (185,0.2782) (186,0.2539) (187,0.4955) (188,0.4700) (189,0.6288) (190,0.9538) (191,1.1303) (192,1.0142) (193,1.0016) (194,0.8748) (195,1.0335) (196,0.9250) (197,1.1963) (198,1.5745) (199,1.4311) (200,1.7104) (201,1.3947) (202,1.4057) (203,1.3638) (204,1.0572) (205,1.0416) (206,0.7848) (207,0.3975) (208,0.6163) (209,0.3803) (210,0.2001) (211,0.1743) (212,0.1950) (213,0.2517) (214,0.2595) (215,0.2042) (216,0.2867) (217,0.3566) (218,0.2141) (219,0.2482) (220,0.3020) (221,0.2830) (222,0.4003) (223,0.4696) (224,0.3050) (225,0.5952) (226,0.5688) (227,0.6249) (228,1.4648) (229,1.1475) (230,0.9607) (231,0.6572) (232,0.8703) (233,1.1920) (234,1.6630) (235,1.6751) (236,2.1473) (237,2.0429) (238,2.2409) (239,0.3102) (240,0.8845) (241,1.2406) (242,0.6763) (243,1.1755) (244,1.2866) (245,0.6652) (246,0.5562) (247,0.3971) (248,0.0834) (249,0.0790) (250,0.1896) (251,0.1532) (252,0.0710) (253,0.1521) (254,0.1273) (255,0.1290) (256,0.1505) (257,0.2245) (258,0.2105) (259,0.4929) (260,0.1800) (261,0.5470) (262,0.2865) (263,0.4368) (264,0.3076) (265,0.2938) (266,0.2791) (267,0.3374) (268,0.4809) (269,0.2157) (270,0.1972) (271,0.1030) (272,0.1117) (273,0.2899) (274,0.5244) (275,0.6752) (276,0.3133) (277,0.2229) (278,0.1728) (279,0.2343) (280,0.1042) (281,0.2670) (282,0.2168) (283,0.2596) (284,0.2792) (285,0.3830) (286,0.1989) (287,0.3501) (288,0.3504) (289,0.1666) (290,0.1581) (291,0.2360) (292,0.1471) (293,0.5295) (294,0.2929) (295,0.4352) (296,0.2340) (297,0.1678) (298,0.2199) (299,0.1001) (300,0.0807) (301,0.1651) (302,0.1214) (303,0.1250) (304,0.2442) (305,0.3510) (306,0.2299)
\psline[linecolor=gray] (0,1.1366) (1,0.7951) (2,0.5708) (3,0.7557) (4,0.4614) (5,0.2381) (6,0.3357) (7,0.4448) (8,0.1990) (9,0.3473) (10,0.3168) (11,0.4595) (12,0.4212) (13,0.4365) (14,0.5404) (15,0.8950) (16,0.2984) (17,0.2350) (18,0.2417) (19,0.5381) (20,0.3665) (21,0.5608) (22,0.6299) (23,0.7986) (24,1.1604) (25,0.9392) (26,0.9789) (27,0.6176) (28,0.9342) (29,0.4627) (30,0.7192) (31,0.3357) (32,0.3227) (33,0.5173) (34,1.1004) (35,0.5633) (36,0.5484) (37,1.0472) (38,0.6626) (39,0.6948) (40,0.5745) (41,0.2985) (42,0.3511) (43,0.2285) (44,0.3966) (45,0.3499) (46,0.1962) (47,0.1389) (48,0.2918) (49,0.3152) (50,0.3291) (51,0.4284) (52,0.3699) (53,0.4175) (54,0.1346) (55,0.1740) (56,0.3403) (57,0.2037) (58,0.0727) (59,0.0498) (60,0.1085) (61,0.5204) (62,0.6630) (63,0.5084) (64,0.5639) (65,0.3396) (66,0.3453) (67,0.5466) (68,0.3724) (69,0.6009) (70,0.3686) (71,0.5056) (72,0.4807) (73,0.3450) (74,0.4162) (75,0.4149) (76,0.4292) (77,0.1968) (78,0.3114) (79,0.3209) (80,0.7250) (81,0.7354) (82,0.7046) (83,0.8886) (84,0.4614) (85,0.6492) (86,0.3186) (87,0.3905) (88,0.4372) (89,0.4013) (90,0.6927) (91,0.5509) (92,0.6698) (93,0.9423) (94,0.6596) (95,0.6238) (96,0.6670) (97,1.0365) (98,1.2173) (99,1.3359) (100,1.5189) (101,0.5116) (102,2.1503) (103,0.7286) (104,0.7641) (105,0.7207) (106,1.5599) (107,0.5873) (108,0.4113) (109,0.8618) (110,0.2675) (111,0.2265) (112,0.3059) (113,0.2401) (114,0.2325) (115,0.3861) (116,0.4297) (117,0.2254) (118,0.0255) (119,0.4200) (120,0.2354) (121,0.2366) (122,0.3027) (123,0.2180) (124,0.0272) (125,0.0810) (126,0.3240) (127,0.1549) (128,0.2159) (129,0.1723) (130,0.7190) (131,0.6851) (132,0.8569) (133,0.4298) (134,0.3119) (135,0.7073) (136,0.7732) (137,0.3712) (138,0.4047) (139,1.2970) (140,1.2752) (141,0.9226) (142,0.5479) (143,0.7668) (144,0.3510) (145,0.5483) (146,0.3936) (147,0.5768) (148,0.7014) (149,0.5278) (150,0.7881) (151,0.8383) (152,1.0098) (153,0.8346) (154,0.8238) (155,0.6883) (156,0.4709) (157,0.7178) (158,1.2113) (159,0.6356) (160,0.4267) (161,0.3744) (162,0.7271) (163,0.5991) (164,0.6003) (165,0.7034) (166,0.3113) (167,0.5322) (168,0.4912) (169,0.3856) (170,0.5229) (171,0.8034) (172,0.7733) (173,0.8230) (174,0.4998) (175,0.6112) (176,0.7120) (177,0.8661) (178,0.4901) (179,1.0454) (180,0.7937) (181,0.6960) (182,0.2454) (183,0.4511) (184,1.0561) (185,1.2638) (186,1.7607) (187,2.3933) (188,2.2914) (189,2.5054) (190,4.5431) (191,3.7110) (192,3.7965) (193,3.4377) (194,2.1738) (195,2.0677) (196,1.6535) (197,2.3949) (198,2.6977) (199,2.8592) (200,3.2216) (201,3.3759) (202,3.4634) (203,3.9882) (204,2.6587) (205,2.6173) (206,2.2076) (207,1.3858) (208,1.5442) (209,1.1028) (210,0.8675) (211,0.7401) (212,0.8462) (213,0.7502) (214,0.8002) (215,0.8762) (216,1.4155) (217,1.0874) (218,0.5052) (219,0.1795) (220,0.3852) (221,0.5632) (222,0.4426) (223,0.4116) (224,0.4933) (225,0.7911) (226,0.7801) (227,0.8845) (228,1.4809) (229,1.6921) (230,1.4292) (231,0.9745) (232,0.9173) (233,1.0656) (234,1.5624) (235,1.2672) (236,1.4027) (237,1.3228) (238,1.7237) (239,0.5692) (240,0.1887) (241,0.2419) (242,0.6749) (243,0.3145) (244,0.7615) (245,0.4779) (246,0.4396) (247,0.1735) (248,0.1559) (249,0.1794) (250,0.1398) (251,0.3595) (252,0.0591) (253,0.1657) (254,0.1978) (255,0.3332) (256,0.3356) (257,0.2783) (258,0.3335) (259,0.3848) (260,0.2117) (261,0.3743) (262,0.2443) (263,0.2783) (264,0.2990) (265,0.2207) (266,0.5735) (267,0.4145) (268,0.6369) (269,1.0904) (270,0.8747) (271,1.2539) (272,0.8000) (273,0.8319) (274,0.8469) (275,0.8957) (276,0.6746) (277,1.0619) (278,0.7201) (279,0.7367) (280,0.4092) (281,0.6231) (282,0.9418) (283,1.1225) (284,1.1609) (285,1.7170) (286,0.8833) (287,1.2409) (288,1.0351) (289,0.3607) (290,0.3328) (291,0.4368) (292,0.5315) (293,0.3803) (294,0.5313) (295,0.5664) (296,0.6256) (297,0.9543) (298,0.6608) (299,0.6871) (300,0.8065) (301,0.9136) (302,0.4685) (303,0.4183) (304,0.6405) (305,0.5963) (306,0.6414)
\end{pspicture}

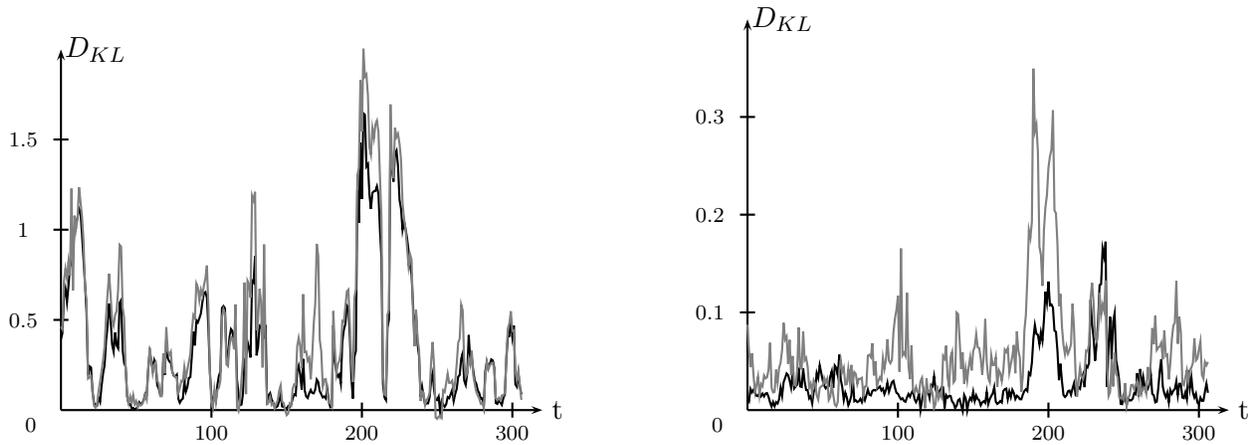
\noindent Fig. 14. Kullback-Leibler distance ($D_{KL}$) between the predicted and realized correlation matrices for windows of 100 days with sliding windows of 5 days, for the original data (left) and for the residues of the regression (right). The results without cleaning are in black lines, and the results with cleaning are in gray lines.

\vskip 0.3 cm

\subsection{Global Riskiness}

One final analysis must be made. A porfolio obtained from a cleaning procedure may produce risk predictions that are closer to the realized risk, but at the cost of augmenting the global riskiness of the portfolio (Tola, Lillo, \& Mantegna (2008), Tumminello, Lillo, \& Mantegna (2010)). In order to analyze this, we calculated the miminum and maximum realized risks for each window of 100 days with a step of 5 days for the diversity of procedures we are using. Figure 15 shows the minimum and maximum expected and realized risks for the original data, no short selling, and no cleaning (left) and also for the residues of the regression with the single index, no short selling, and no cleaning (right). The graphs with the cleaning procedure are not represented, because they are nearly indistinguishable from their uncleaned counterparts.

One can notice few differences between the two graphs. The minimum predicted and realized risks are lower for the residues of the regression model, but they also present a steeper peak than in the case of original data. So, the procedures of making a regression or of cleaning do not affect much the minimum and realized risks of the portfolios.

\begin{pspicture}(-0.6,-0.6)(8,6.6)
\psset{xunit=0.02,yunit=1}
\psline{->}(0,0)(320,0) \rput(330,0){t} \psline{->}(0,0)(0,6) \rput(35,6){Risk} \scriptsize  \rput(-20,-0.2){0} \psline(100,-0.1)(100,0.1) \rput(100,-0.3){100} \psline(200,-0.1)(200,0.1) \rput(200,-0.3){200} \psline(300,-0.1)(300,0.1) \rput(300,-0.3){300} \psline(-5,1.2)(5,1.2) \rput(-25,1.2){$0.001$} \psline(-5,2.4)(5,2.4) \rput(-25,2.4){$0.002$} \psline(-5,3.6)(5,3.6) \rput(-25,3.6){$0.003$} \psline(-5,4.8)(5,4.8) \rput(-25,4.8){$0.004$}
% Minimum predicted risk
\psline[linecolor=gray] (0,0.0826) (1,0.0826) (2,0.0841) (3,0.0849) (4,0.0791) (5,0.0795) (6,0.0707) (7,0.0751) (8,0.0745) (9,0.0794) (10,0.0744) (11,0.0707) (12,0.0705) (13,0.0667) (14,0.0615) (15,0.0539) (16,0.0576) (17,0.0620) (18,0.0478) (19,0.0459) (20,0.0465) (21,0.0505) (22,0.0485) (23,0.0490) (24,0.0515) (25,0.0538) (26,0.0552) (27,0.0590) (28,0.0636) (29,0.0740) (30,0.0609) (31,0.0620) (32,0.0608) (33,0.0620) (34,0.0670) (35,0.0773) (36,0.0744) (37,0.0781) (38,0.1043) (39,0.0726) (40,0.0609) (41,0.0584) (42,0.0568) (43,0.0573) (44,0.0613) (45,0.0609) (46,0.0591) (47,0.0577) (48,0.0612) (49,0.0561) (50,0.0596) (51,0.0642) (52,0.0617) (53,0.0593) (54,0.0578) (55,0.0587) (56,0.0616) (57,0.0625) (58,0.0679) (59,0.0691) (60,0.0703) (61,0.0692) (62,0.0689) (63,0.0712) (64,0.0683) (65,0.0692) (66,0.0696) (67,0.0722) (68,0.0733) (69,0.0756) (70,0.0762) (71,0.0672) (72,0.0668) (73,0.0685) (74,0.0706) (75,0.0633) (76,0.0632) (77,0.0661) (78,0.0651) (79,0.0642) (80,0.0660) (81,0.0697) (82,0.0787) (83,0.0825) (84,0.0741) (85,0.0754) (86,0.0752) (87,0.0692) (88,0.0585) (89,0.0579) (90,0.0498) (91,0.0535) (92,0.0497) (93,0.0478) (94,0.0488) (95,0.0494) (96,0.0493) (97,0.0458) (98,0.0434) (99,0.0456) (100,0.0468) (101,0.0470) (102,0.0559) (103,0.0492) (104,0.0518) (105,0.0514) (106,0.0549) (107,0.0553) (108,0.0573) (109,0.0558) (110,0.0586) (111,0.0584) (112,0.0572) (113,0.0601) (114,0.0605) (115,0.0619) (116,0.0628) (117,0.0723) (118,0.0758) (119,0.0737) (120,0.0734) (121,0.0701) (122,0.0558) (123,0.0578) (124,0.0579) (125,0.0596) (126,0.0559) (127,0.0503) (128,0.0491) (129,0.0510) (130,0.0582) (131,0.0586) (132,0.0546) (133,0.0544) (134,0.0467) (135,0.0481) (136,0.0464) (137,0.0609) (138,0.0690) (139,0.0801) (140,0.0960) (141,0.0988) (142,0.0992) (143,0.1050) (144,0.0972) (145,0.0949) (146,0.0945) (147,0.0967) (148,0.0955) (149,0.0983) (150,0.0900) (151,0.0927) (152,0.0994) (153,0.1026) (154,0.1156) (155,0.1143) (156,0.1178) (157,0.0968) (158,0.0837) (159,0.0852) (160,0.0889) (161,0.1029) (162,0.1170) (163,0.1268) (164,0.1260) (165,0.1282) (166,0.1314) (167,0.1362) (168,0.1404) (169,0.1391) (170,0.1339) (171,0.1367) (172,0.1355) (173,0.1383) (174,0.1356) (175,0.1319) (176,0.1310) (177,0.1384) (178,0.1379) (179,0.1318) (180,0.1335) (181,0.1440) (182,0.1384) (183,0.1328) (184,0.1410) (185,0.1451) (186,0.1430) (187,0.1320) (188,0.1371) (189,0.1300) (190,0.1351) (191,0.1302) (192,0.1510) (193,0.1743) (194,0.2980) (195,0.2304) (196,0.6714) (197,0.4831) (198,0.3542) (199,0.8206) (200,0.8721) (201,0.5433) (202,0.4600) (203,0.4059) (204,0.4294) (205,0.4185) (206,0.5886) (207,0.6299) (208,0.6195) (209,0.5378) (210,0.5499) (211,0.5356) (212,0.5042) (213,0.4804) (214,0.5030) (215,0.5409) (216,0.4932) (217,0.4859) (218,0.4518) (219,0.3910) (220,0.3312) (221,0.3076) (222,0.2679) (223,0.2434) (224,0.2331) (225,0.2103) (226,0.1798) (227,0.1349) (228,0.1264) (229,0.1208) (230,0.1226) (231,0.1227) (232,0.1226) (233,0.1305) (234,0.1264) (235,0.1207) (236,0.1257) (237,0.0996) (238,0.0930) (239,0.0759) (240,0.0734) (241,0.0659) (242,0.0603) (243,0.0580) (244,0.0505) (245,0.0523) (246,0.0509) (247,0.0470) (248,0.0473) (249,0.0492) (250,0.0492) (251,0.0522) (252,0.0500) (253,0.0516) (254,0.0567) (255,0.0603) (256,0.0538) (257,0.0545) (258,0.0527) (259,0.0541) (260,0.0571) (261,0.0570) (262,0.0572) (263,0.0548) (264,0.0499) (265,0.0492) (266,0.0495) (267,0.0528) (268,0.0536) (269,0.0555) (270,0.0577) (271,0.0561) (272,0.0562) (273,0.0593) (274,0.0608) (275,0.0756) (276,0.0864) (277,0.1569) (278,0.0601) (279,0.0523) (280,0.0524) (281,0.0535) (282,0.0540) (283,0.0554) (284,0.0566) (285,0.0574) (286,0.0539) (287,0.0544) (288,0.0521) (289,0.0492) (290,0.0484) (291,0.0435) (292,0.0479) (293,0.0443) (294,0.0512) (295,0.0494) (296,0.0536) (297,0.0483) (298,0.0459) (299,0.0413) (300,0.0423) (301,0.0445) (302,0.0424) (303,0.0429) (304,0.0476) (305,0.0489) (306,0.0451)
% Minimum realized risk
\psline (0,0.0507) (1,0.0506) (2,0.0468) (3,0.0482) (4,0.0476) (5,0.0461) (6,0.0468) (7,0.0487) (8,0.0483) (9,0.0496) (10,0.0445) (11,0.0404) (12,0.0424) (13,0.0414) (14,0.0409) (15,0.0362) (16,0.0365) (17,0.0364) (18,0.0375) (19,0.0377) (20,0.0404) (21,0.0491) (22,0.0541) (23,0.0555) (24,0.0704) (25,0.0787) (26,0.0784) (27,0.0734) (28,0.0840) (29,0.0920) (30,0.0815) (31,0.0808) (32,0.0793) (33,0.0792) (34,0.0730) (35,0.0841) (36,0.0863) (37,0.0901) (38,0.1172) (39,0.0839) (40,0.0728) (41,0.0692) (42,0.0615) (43,0.0635) (44,0.0588) (45,0.0563) (46,0.0548) (47,0.0534) (48,0.0608) (49,0.0616) (50,0.0641) (51,0.0619) (52,0.0550) (53,0.0527) (54,0.0530) (55,0.0489) (56,0.0526) (57,0.0581) (58,0.0579) (59,0.0537) (60,0.0555) (61,0.0557) (62,0.0563) (63,0.0561) (64,0.0574) (65,0.0574) (66,0.0585) (67,0.0571) (68,0.0446) (69,0.0434) (70,0.0387) (71,0.0422) (72,0.0407) (73,0.0397) (74,0.0406) (75,0.0398) (76,0.0382) (77,0.0367) (78,0.0456) (79,0.0498) (80,0.0516) (81,0.0544) (82,0.0826) (83,0.0786) (84,0.0741) (85,0.0730) (86,0.0757) (87,0.0708) (88,0.0725) (89,0.0711) (90,0.0689) (91,0.0688) (92,0.0662) (93,0.0683) (94,0.0711) (95,0.0713) (96,0.0750) (97,0.0762) (98,0.0731) (99,0.0668) (100,0.0631) (101,0.0597) (102,0.0453) (103,0.0464) (104,0.0498) (105,0.0505) (106,0.0482) (107,0.0439) (108,0.0429) (109,0.0436) (110,0.0525) (111,0.0507) (112,0.0449) (113,0.0452) (114,0.0434) (115,0.0443) (116,0.0408) (117,0.0649) (118,0.0744) (119,0.0759) (120,0.0763) (121,0.0755) (122,0.0775) (123,0.0794) (124,0.0756) (125,0.0749) (126,0.0696) (127,0.0717) (128,0.0710) (129,0.0752) (130,0.0660) (131,0.0727) (132,0.0729) (133,0.0713) (134,0.0647) (135,0.0670) (136,0.0721) (137,0.0658) (138,0.0621) (139,0.0676) (140,0.0848) (141,0.0878) (142,0.0935) (143,0.1060) (144,0.1045) (145,0.1041) (146,0.1042) (147,0.1055) (148,0.1060) (149,0.1008) (150,0.1011) (151,0.0971) (152,0.1017) (153,0.1115) (154,0.1290) (155,0.1300) (156,0.1304) (157,0.1311) (158,0.1127) (159,0.0971) (160,0.0949) (161,0.1210) (162,0.1402) (163,0.1305) (164,0.1299) (165,0.1305) (166,0.1307) (167,0.1416) (168,0.1542) (169,0.1650) (170,0.1643) (171,0.1725) (172,0.1720) (173,0.1643) (174,0.1634) (175,0.1651) (176,0.1728) (177,0.1648) (178,0.1662) (179,0.1660) (180,0.1593) (181,0.1209) (182,0.1233) (183,0.1258) (184,0.1300) (185,0.1349) (186,0.1405) (187,0.1189) (188,0.1147) (189,0.0979) (190,0.1037) (191,0.1028) (192,0.1307) (193,0.1648) (194,0.2962) (195,0.2367) (196,0.6714) (197,0.5202) (198,0.3931) (199,0.8206) (200,0.8721) (201,0.6347) (202,0.5393) (203,0.4820) (204,0.5018) (205,0.5057) (206,0.6333) (207,0.6700) (208,0.6731) (209,0.6365) (210,0.6401) (211,0.6182) (212,0.5794) (213,0.5690) (214,0.4989) (215,0.4783) (216,0.4658) (217,0.3373) (218,0.2787) (219,0.1765) (220,0.1731) (221,0.1550) (222,0.1270) (223,0.1145) (224,0.0972) (225,0.0959) (226,0.0874) (227,0.0648) (228,0.0606) (229,0.0598) (230,0.0601) (231,0.0589) (232,0.0586) (233,0.0624) (234,0.0680) (235,0.0705) (236,0.0689) (237,0.0673) (238,0.0684) (239,0.0656) (240,0.0637) (241,0.0605) (242,0.0583) (243,0.0561) (244,0.0500) (245,0.0489) (246,0.0463) (247,0.0468) (248,0.0540) (249,0.0629) (250,0.0617) (251,0.0708) (252,0.0684) (253,0.0688) (254,0.0651) (255,0.0677) (256,0.0630) (257,0.0610) (258,0.0523) (259,0.0546) (260,0.0574) (261,0.0593) (262,0.0628) (263,0.0653) (264,0.0649) (265,0.0650) (266,0.0664) (267,0.0693) (268,0.0588) (269,0.0478) (270,0.0479) (271,0.0391) (272,0.0439) (273,0.0419) (274,0.0519) (275,0.0656) (276,0.0888) (277,0.1706) (278,0.0726) (279,0.0615) (280,0.0561) (281,0.0580) (282,0.0518) (283,0.0500) (284,0.0507) (285,0.0503) (286,0.0470) (287,0.0485) (288,0.0487) (289,0.0469) (290,0.0490) (291,0.0489) (292,0.0460) (293,0.0455) (294,0.0402) (295,0.0378) (296,0.0318) (297,0.0245) (298,0.0245) (299,0.0225) (300,0.0236) (301,0.0258) (302,0.0267) (303,0.0285) (304,0.0298) (305,0.0313) (306,0.0321)
% Maximum predicted risk
\psline[linecolor=gray] (0,0.6477) (1,0.6387) (2,0.6416) (3,0.5739) (4,0.5626) (5,0.5554) (6,0.5344) (7,0.5548) (8,0.5089) (9,0.4921) (10,0.3708) (11,0.4922) (12,0.3231) (13,0.6210) (14,0.2983) (15,0.4895) (16,0.5261) (17,0.5436) (18,0.5355) (19,0.5463) (20,0.4102) (21,0.4717) (22,0.5141) (23,0.5183) (24,0.4606) (25,0.6653) (26,0.7672) (27,0.9414) (28,0.9660) (29,0.9995) (30,0.8106) (31,0.4985) (32,0.9067) (33,0.8093) (34,1.0376) (35,0.6186) (36,0.9383) (37,0.9751) (38,1.0532) (39,0.7289) (40,0.7468) (41,0.6992) (42,0.5632) (43,0.5301) (44,0.3983) (45,0.4548) (46,0.2327) (47,0.4698) (48,0.8790) (49,0.9315) (50,0.8437) (51,0.9783) (52,0.8755) (53,0.4967) (54,0.3907) (55,0.4494) (56,0.4845) (57,1.0114) (58,0.4121) (59,0.4311) (60,1.1204) (61,0.3992) (62,0.5389) (63,0.5202) (64,0.4421) (65,0.4926) (66,0.5304) (67,0.5345) (68,0.4067) (69,0.4853) (70,0.3783) (71,0.5802) (72,0.5795) (73,0.5238) (74,0.5296) (75,0.4658) (76,0.6062) (77,0.5728) (78,0.6697) (79,0.7482) (80,0.7752) (81,0.9728) (82,1.1758) (83,0.7935) (84,0.3681) (85,0.5136) (86,0.3868) (87,0.5171) (88,0.4832) (89,0.5155) (90,0.4836) (91,1.9178) (92,0.4158) (93,0.4008) (94,0.4083) (95,0.4078) (96,0.4115) (97,0.2278) (98,0.2128) (99,0.2081) (100,0.1776) (101,0.3739) (102,0.5682) (103,0.7377) (104,0.8285) (105,0.2937) (106,0.2797) (107,0.3065) (108,0.2276) (109,0.3784) (110,0.3071) (111,0.1667) (112,0.2529) (113,0.3004) (114,0.4916) (115,0.2274) (116,0.2698) (117,0.2744) (118,0.2936) (119,0.3039) (120,0.3085) (121,0.3174) (122,0.2971) (123,0.2952) (124,0.3404) (125,0.2569) (126,0.2409) (127,0.2921) (128,0.3840) (129,0.5504) (130,0.5839) (131,0.6231) (132,0.9016) (133,0.4131) (134,0.2217) (135,0.2260) (136,0.3726) (137,0.2372) (138,0.3125) (139,0.3406) (140,0.3377) (141,0.3390) (142,0.3413) (143,0.3503) (144,0.3606) (145,0.3692) (146,0.3670) (147,0.3533) (148,0.3391) (149,0.3055) (150,0.2968) (151,0.3334) (152,0.7081) (153,0.7389) (154,0.6548) (155,0.8064) (156,0.8497) (157,0.8252) (158,0.7881) (159,0.8480) (160,0.8283) (161,1.1424) (162,0.9423) (163,1.1024) (164,1.0610) (165,0.9471) (166,0.7460) (167,0.6585) (168,0.5949) (169,0.5895) (170,0.6099) (171,0.9714) (172,0.8099) (173,0.7920) (174,0.5084) (175,0.5424) (176,0.6253) (177,0.7057) (178,0.7537) (179,0.7578) (180,0.7477) (181,0.7727) (182,0.7364) (183,0.7994) (184,0.8058) (185,0.8358) (186,0.8478) (187,0.8552) (188,0.8666) (189,0.9109) (190,0.9556) (191,0.9358) (192,0.9919) (193,0.5110) (194,0.6965) (195,0.6983) (196,0.6732) (197,0.7583) (198,0.7026) (199,0.8210) (200,0.8727) (201,0.8788) (202,0.8721) (203,3.3707) (204,0.9171) (205,3.0822) (206,1.5001) (207,2.5044) (208,1.8762) (209,1.8731) (210,1.6799) (211,4.5476) (212,1.8725) (213,1.8532) (214,1.5809) (215,2.6564) (216,1.4564) (217,2.3092) (218,2.2464) (219,1.6519) (220,1.2902) (221,2.1952) (222,2.0009) (223,1.5803) (224,1.4179) (225,1.2555) (226,1.2925) (227,1.2873) (228,1.6050) (229,0.8915) (230,0.8557) (231,1.1009) (232,1.0539) (233,1.0775) (234,1.0130) (235,1.0211) (236,2.2501) (237,1.9866) (238,1.8242) (239,1.6581) (240,1.7552) (241,1.0334) (242,0.6606) (243,0.7862) (244,0.7667) (245,0.3369) (246,0.6982) (247,0.5229) (248,0.6818) (249,0.6699) (250,0.7984) (251,0.7995) (252,0.7647) (253,0.6079) (254,0.7420) (255,0.6785) (256,0.8647) (257,0.6416) (258,0.5940) (259,0.4417) (260,0.2640) (261,0.3727) (262,0.4595) (263,0.4921) (264,0.4727) (265,0.4213) (266,0.3943) (267,0.4741) (268,0.8130) (269,0.6950) (270,0.6198) (271,0.7544) (272,0.3790) (273,0.4482) (274,0.5255) (275,0.4855) (276,0.2935) (277,0.5819) (278,0.5472) (279,0.2118) (280,0.4835) (281,0.3813) (282,0.4524) (283,0.3473) (284,0.3090) (285,0.1556) (286,0.1678) (287,0.3930) (288,0.1600) (289,0.1793) (290,0.3016) (291,0.2030) (292,0.3187) (293,0.5018) (294,0.3424) (295,0.3468) (296,0.2842) (297,0.2049) (298,0.2313) (299,0.2728) (300,0.2870) (301,0.2254) (302,0.2603) (303,0.2537) (304,0.2884) (305,0.3396) (306,0.3145)
% Maximum realized risk
\psline (0,0.5685) (1,0.6208) (2,0.6314) (3,0.5692) (4,0.5497) (5,0.5417) (6,0.5311) (7,0.5465) (8,0.4996) (9,0.4813) (10,0.3314) (11,0.4825) (12,0.2998) (13,0.6157) (14,0.2744) (15,0.4786) (16,0.5230) (17,0.5367) (18,0.5340) (19,0.5440) (20,0.4167) (21,0.4580) (22,0.4564) (23,0.5089) (24,0.4068) (25,0.6379) (26,0.6556) (27,0.9069) (28,0.9606) (29,1.0056) (30,0.8383) (31,0.5089) (32,0.9166) (33,0.8910) (34,1.0494) (35,0.5824) (36,0.9513) (37,0.9758) (38,1.0571) (39,0.7511) (40,0.7442) (41,0.7016) (42,0.5659) (43,0.5579) (44,0.3983) (45,0.5146) (46,0.2138) (47,0.4874) (48,0.8917) (49,0.9350) (50,0.8615) (51,0.9835) (52,0.8911) (53,0.4823) (54,0.4149) (55,0.4524) (56,0.4852) (57,1.0258) (58,0.4088) (59,0.4203) (60,1.1002) (61,0.3711) (62,0.5343) (63,0.5273) (64,0.4696) (65,0.5433) (66,0.5784) (67,0.5543) (68,0.4269) (69,0.4927) (70,0.3767) (71,0.5781) (72,0.5590) (73,0.4833) (74,0.5246) (75,0.4248) (76,0.5644) (77,0.5493) (78,0.6148) (79,0.7441) (80,0.7774) (81,0.9739) (82,1.1770) (83,0.8117) (84,0.4256) (85,0.5130) (86,0.3797) (87,0.5178) (88,0.4958) (89,0.5224) (90,0.5062) (91,1.9275) (92,0.4163) (93,0.4095) (94,0.4130) (95,0.4132) (96,0.4191) (97,0.2074) (98,0.2067) (99,0.1780) (100,0.1987) (101,0.3762) (102,0.5921) (103,0.7149) (104,0.8186) (105,0.2677) (106,0.2527) (107,0.2565) (108,0.1748) (109,0.3407) (110,0.2719) (111,0.1919) (112,0.2143) (113,0.2626) (114,0.4340) (115,0.1990) (116,0.2178) (117,0.2766) (118,0.2962) (119,0.3062) (120,0.3114) (121,0.3199) (122,0.3169) (123,0.3023) (124,0.3493) (125,0.2709) (126,0.2743) (127,0.3319) (128,0.3962) (129,0.5605) (130,0.5906) (131,0.6274) (132,0.9017) (133,0.4136) (134,0.2259) (135,0.2308) (136,0.3564) (137,0.2366) (138,0.3124) (139,0.3413) (140,0.3337) (141,0.3314) (142,0.3380) (143,0.3454) (144,0.3615) (145,0.3687) (146,0.3585) (147,0.3495) (148,0.3368) (149,0.2907) (150,0.3008) (151,0.3273) (152,0.6986) (153,0.7339) (154,0.6510) (155,0.8025) (156,0.8479) (157,0.8228) (158,0.7832) (159,0.8447) (160,0.8123) (161,1.1399) (162,0.9193) (163,1.0901) (164,1.0543) (165,0.9345) (166,0.7096) (167,0.6465) (168,0.5534) (169,0.5977) (170,0.6034) (171,0.9727) (172,0.8017) (173,0.7852) (174,0.5073) (175,0.5386) (176,0.6212) (177,0.7027) (178,0.7510) (179,0.7560) (180,0.7455) (181,0.7683) (182,0.7359) (183,0.8006) (184,0.8061) (185,0.8369) (186,0.8472) (187,0.8542) (188,0.8715) (189,0.9130) (190,0.9609) (191,0.9345) (192,0.9961) (193,0.5467) (194,0.6960) (195,0.7160) (196,0.6732) (197,0.7629) (198,0.7613) (199,0.8210) (200,0.8727) (201,0.8839) (202,0.8831) (203,3.3455) (204,0.9146) (205,3.0609) (206,1.5196) (207,2.5102) (208,1.8751) (209,1.8662) (210,1.6388) (211,4.5512) (212,1.8571) (213,1.8719) (214,1.5906) (215,2.6860) (216,1.4813) (217,2.3146) (218,2.2420) (219,1.6031) (220,1.1889) (221,2.1487) (222,1.9904) (223,1.5657) (224,1.3648) (225,1.1448) (226,1.2766) (227,1.2895) (228,1.5023) (229,0.6347) (230,0.8356) (231,1.0868) (232,1.0241) (233,1.0605) (234,0.9873) (235,1.0193) (236,2.2564) (237,1.9971) (238,1.8470) (239,1.7021) (240,1.7623) (241,1.0099) (242,0.6805) (243,0.8075) (244,0.7857) (245,0.3446) (246,0.6838) (247,0.4597) (248,0.6597) (249,0.6411) (250,0.7883) (251,0.7937) (252,0.7651) (253,0.5473) (254,0.7295) (255,0.5833) (256,0.8446) (257,0.5692) (258,0.5931) (259,0.4255) (260,0.2449) (261,0.3622) (262,0.4548) (263,0.4934) (264,0.4751) (265,0.4492) (266,0.4263) (267,0.5451) (268,0.8254) (269,0.7318) (270,0.6606) (271,0.7596) (272,0.3866) (273,0.4493) (274,0.5270) (275,0.4860) (276,0.2941) (277,0.5820) (278,0.5498) (279,0.2224) (280,0.5005) (281,0.4216) (282,0.4487) (283,0.3345) (284,0.2909) (285,0.1449) (286,0.1568) (287,0.3893) (288,0.1407) (289,0.1703) (290,0.2904) (291,0.1798) (292,0.2935) (293,0.4921) (294,0.3457) (295,0.3423) (296,0.2692) (297,0.1711) (298,0.2005) (299,0.2661) (300,0.2766) (301,0.1906) (302,0.2575) (303,0.2438) (304,0.2773) (305,0.3345) (306,0.3082)
\end{pspicture}
\begin{pspicture}(-0.6,-0.6)(8,6.6)
\psset{xunit=0.02,yunit=1}
\psline{->}(0,0)(320,0) \rput(330,0){t} \psline{->}(0,0)(0,6) \rput(35,6){Risk} \scriptsize  \rput(-20,-0.2){0} \psline(100,-0.1)(100,0.1) \rput(100,-0.3){100} \psline(200,-0.1)(200,0.1) \rput(200,-0.3){200} \psline(300,-0.1)(300,0.1) \rput(300,-0.3){300} \psline(-5,1.2)(5,1.2) \rput(-25,1.2){$0.001$} \psline(-5,2.4)(5,2.4) \rput(-25,2.4){$0.002$} \psline(-5,3.6)(5,3.6) \rput(-25,3.6){$0.003$} \psline(-5,4.8)(5,4.8) \rput(-25,4.8){$0.004$}
% Minimum predicted risk
\psline[linecolor=gray] (0,0.00000153) (1,0.00000215) (2,0.00000208) (3,0.00000000) (4,0.00000002) (5,0.00000010) (6,0.00000012) (7,0.00000035) (8,0.00000020) (9,0.00000262) (10,0.00000198) (11,0.00000091) (12,0.00000113) (13,0.00000009) (14,0.00000001) (15,0.00000115) (16,0.00000015) (17,0.00000000) (18,0.00000000) (19,0.00000131) (20,0.00000081) (21,0.00000050) (22,0.00000068) (23,0.00000020) (24,0.00000013) (25,0.00000008) (26,0.00056242) (27,0.00020605) (28,0.00135273) (29,0.00161539) (30,0.00000005) (31,0.00022203) (32,0.00000046) (33,0.00000048) (34,0.00015572) (35,0.00694440) (36,0.00342992) (37,0.00922635) (38,0.02626882) (39,0.00167168) (40,0.00052947) (41,0.00000011) (42,0.00000569) (43,0.00000003) (44,0.00000059) (45,0.00000000) (46,0.00000005) (47,0.00000002) (48,0.00000069) (49,0.00000030) (50,0.00000045) (51,0.00000057) (52,0.00000126) (53,0.00000020) (54,0.00000001) (55,0.00000003) (56,0.00000154) (57,0.00000020) (58,0.00000032) (59,0.00000064) (60,0.00000007) (61,0.00000076) (62,0.00000143) (63,0.00000017) (64,0.00000089) (65,0.00000084) (66,0.00000036) (67,0.00000087) (68,0.00000140) (69,0.00000012) (70,0.00000028) (71,0.00000118) (72,0.00000009) (73,0.00000024) (74,0.00000072) (75,0.00000001) (76,0.00000044) (77,0.00000049) (78,0.00000003) (79,0.00000039) (80,0.00000059) (81,0.00037154) (82,0.00190622) (83,0.00420682) (84,0.00015025) (85,0.00077817) (86,0.00320281) (87,0.00257821) (88,0.00088469) (89,0.00007209) (90,0.00099148) (91,0.00037158) (92,0.00183044) (93,0.00029551) (94,0.00215574) (95,0.00266553) (96,0.00616797) (97,0.00148729) (98,0.00000895) (99,0.00000339) (100,0.00000149) (101,0.00000077) (102,0.00000083) (103,0.00000000) (104,0.00000069) (105,0.00000002) (106,0.00000002) (107,0.00000035) (108,0.00000007) (109,0.00000079) (110,0.00000011) (111,0.00000000) (112,0.00000086) (113,0.00000006) (114,0.00000018) (115,0.00000007) (116,0.00000007) (117,0.00000068) (118,0.00000122) (119,0.00000152) (120,0.00000074) (121,0.00000003) (122,0.00000030) (123,0.00000142) (124,0.00000001) (125,0.00000122) (126,0.00000014) (127,0.00000042) (128,0.00000163) (129,0.00000036) (130,0.00000095) (131,0.00000147) (132,0.00000037) (133,0.00000150) (134,0.00000032) (135,0.00000084) (136,0.00000000) (137,0.00000025) (138,0.00000002) (139,0.00000007) (140,0.00000007) (141,0.00000144) (142,0.00000024) (143,0.00000073) (144,0.00000081) (145,0.00000036) (146,0.00000001) (147,0.00000034) (148,0.00000055) (149,0.00000003) (150,0.00000010) (151,0.00000105) (152,0.00000090) (153,0.00000048) (154,0.00000010) (155,0.00000010) (156,0.00003320) (157,0.00000016) (158,0.00000061) (159,0.00000002) (160,0.00000002) (161,0.00077939) (162,0.00028852) (163,0.00000253) (164,0.00000388) (165,0.00000099) (166,0.00048496) (167,0.00567396) (168,0.00068643) (169,0.00319258) (170,0.00241584) (171,0.00079995) (172,0.00030256) (173,0.00000060) (174,0.00000013) (175,0.00000574) (176,0.00000001) (177,0.00000015) (178,0.00000001) (179,0.00000133) (180,0.00000264) (181,0.00000001) (182,0.00000039) (183,0.00027368) (184,0.00006388) (185,0.00068150) (186,0.00120375) (187,0.00066125) (188,0.00179697) (189,0.00424293) (190,0.00629816) (191,0.00539129) (192,0.02430444) (193,0.04748228) (194,0.19251897) (195,0.06581218) (196,0.73132796) (197,0.29798181) (198,0.17157150) (199,0.89489172) (200,0.95093070) (201,0.34384906) (202,0.15701839) (203,0.06137022) (204,0.12104580) (205,0.07961309) (206,0.19752963) (207,0.35272173) (208,0.23277074) (209,0.16091590) (210,0.23492472) (211,0.19176359) (212,0.09430866) (213,0.03553726) (214,0.06353860) (215,0.11564290) (216,0.01780615) (217,0.00893527) (218,0.00067291) (219,0.00000142) (220,0.00016402) (221,0.00000121) (222,0.00000011) (223,0.00000212) (224,0.00000187) (225,0.00000016) (226,0.00000000) (227,0.00000168) (228,0.00000005) (229,0.00000037) (230,0.00000231) (231,0.00000206) (232,0.00000052) (233,0.00000003) (234,0.00000178) (235,0.00000191) (236,0.00000001) (237,0.00000011) (238,0.00000006) (239,0.00000011) (240,0.00000151) (241,0.00000001) (242,0.00000267) (243,0.00000090) (244,0.00000019) (245,0.00000000) (246,0.00000039) (247,0.00000030) (248,0.00000049) (249,0.00000033) (250,0.00000044) (251,0.00000006) (252,0.00000001) (253,0.00000000) (254,0.00000102) (255,0.00000071) (256,0.00000206) (257,0.00000068) (258,0.00000063) (259,0.00000009) (260,0.00000028) (261,0.00000039) (262,0.00000001) (263,0.00000023) (264,0.00000089) (265,0.00000008) (266,0.00000000) (267,0.00000058) (268,0.00000032) (269,0.00000003) (270,0.00000017) (271,0.00000005) (272,0.00031790) (273,0.00116281) (274,0.00230669) (275,0.02368882) (276,0.03324051) (277,0.10867934) (278,0.01692250) (279,0.00478311) (280,0.00015438) (281,0.00322726) (282,0.00023832) (283,0.00111442) (284,0.00000882) (285,0.00008915) (286,0.00001205) (287,0.00038497) (288,0.00006392) (289,0.00107671) (290,0.00027447) (291,0.00001411) (292,0.00000017) (293,0.00000002) (294,0.00000007) (295,0.00000130) (296,0.00000016) (297,0.00000054) (298,0.00000018) (299,0.00000071) (300,0.00000004) (301,0.00000061) (302,0.00000052) (303,0.00000002) (304,0.00000007) (305,0.00000027) (306,0.00000005)
% Minimum realized risk
\psline (0,0.00000037) (1,0.00000003) (2,0.00000023) (3,0.00000168) (4,0.00000107) (5,0.00000040) (6,0.00000019) (7,0.00000005) (8,0.00000144) (9,0.00000020) (10,0.00000023) (11,0.00000061) (12,0.00000043) (13,0.00000006) (14,0.00000054) (15,0.00000234) (16,0.00000088) (17,0.00000003) (18,0.00000008) (19,0.00000156) (20,0.00000074) (21,0.00000015) (22,0.00000070) (23,0.00000015) (24,0.00000007) (25,0.00000071) (26,0.00046767) (27,0.00018966) (28,0.00138523) (29,0.00189066) (30,0.00000055) (31,0.00030356) (32,0.00000002) (33,0.00000009) (34,0.00029335) (35,0.00705129) (36,0.00347918) (37,0.00669413) (38,0.02306617) (39,0.00260398) (40,0.00112684) (41,0.00000026) (42,0.00003894) (43,0.00000009) (44,0.00000041) (45,0.00000098) (46,0.00000076) (47,0.00000056) (48,0.00000033) (49,0.00000016) (50,0.00000068) (51,0.00000001) (52,0.00000171) (53,0.00000062) (54,0.00000100) (55,0.00000082) (56,0.00000041) (57,0.00000022) (58,0.00000000) (59,0.00000022) (60,0.00000117) (61,0.00000043) (62,0.00000066) (63,0.00000000) (64,0.00000000) (65,0.00000001) (66,0.00000032) (67,0.00000085) (68,0.00000207) (69,0.00000239) (70,0.00000224) (71,0.00000122) (72,0.00000189) (73,0.00000270) (74,0.00000250) (75,0.00000110) (76,0.00000384) (77,0.00000726) (78,0.00000779) (79,0.00000407) (80,0.00000236) (81,0.00020399) (82,0.00205000) (83,0.00673180) (84,0.00025495) (85,0.00139504) (86,0.00415711) (87,0.00320068) (88,0.00160387) (89,0.00027050) (90,0.00151806) (91,0.00046969) (92,0.00342858) (93,0.00042290) (94,0.00218530) (95,0.00263968) (96,0.00614341) (97,0.00144605) (98,0.00000008) (99,0.00000023) (100,0.00000025) (101,0.00000034) (102,0.00000045) (103,0.00000077) (104,0.00000006) (105,0.00000056) (106,0.00000000) (107,0.00000004) (108,0.00000107) (109,0.00000001) (110,0.00000082) (111,0.00000039) (112,0.00000145) (113,0.00000034) (114,0.00000000) (115,0.00000081) (116,0.00000018) (117,0.00000010) (118,0.00000072) (119,0.00000034) (120,0.00000011) (121,0.00000047) (122,0.00000115) (123,0.00000111) (124,0.00000047) (125,0.00000005) (126,0.00000098) (127,0.00000078) (128,0.00000103) (129,0.00000146) (130,0.00000000) (131,0.00000314) (132,0.00000012) (133,0.00000150) (134,0.00000122) (135,0.00000001) (136,0.00000010) (137,0.00000063) (138,0.00000071) (139,0.00000009) (140,0.00000089) (141,0.00000095) (142,0.00000222) (143,0.00000001) (144,0.00000009) (145,0.00000003) (146,0.00000003) (147,0.00000123) (148,0.00000114) (149,0.00000010) (150,0.00000129) (151,0.00000029) (152,0.00000072) (153,0.00000001) (154,0.00000002) (155,0.00000029) (156,0.00005945) (157,0.00000002) (158,0.00000046) (159,0.00000101) (160,0.00000022) (161,0.00026948) (162,0.00005747) (163,0.00000148) (164,0.00000073) (165,0.00000085) (166,0.00017356) (167,0.00365594) (168,0.00037444) (169,0.00246629) (170,0.00221607) (171,0.00057588) (172,0.00022263) (173,0.00000000) (174,0.00000122) (175,0.00000007) (176,0.00000343) (177,0.00000462) (178,0.00000299) (179,0.00000034) (180,0.00000278) (181,0.00000190) (182,0.00000121) (183,0.00019917) (184,0.00001195) (185,0.00059542) (186,0.00093523) (187,0.00073175) (188,0.00218443) (189,0.00439376) (190,0.00629324) (191,0.00612918) (192,0.03270963) (193,0.06181729) (194,0.20551309) (195,0.09230526) (196,0.73176961) (197,0.31707185) (198,0.20332180) (199,0.89486242) (200,0.95094644) (201,0.47322013) (202,0.21774919) (203,0.07408230) (204,0.13941476) (205,0.09365658) (206,0.22550739) (207,0.36015801) (208,0.28238712) (209,0.21583508) (210,0.28783850) (211,0.17549066) (212,0.09549060) (213,0.04686389) (214,0.06940379) (215,0.10384942) (216,0.02692333) (217,0.01389688) (218,0.00080051) (219,0.00000101) (220,0.00011760) (221,0.00000134) (222,0.00000243) (223,0.00000002) (224,0.00000403) (225,0.00000078) (226,0.00000293) (227,0.00000126) (228,0.00000006) (229,0.00000057) (230,0.00000008) (231,0.00000001) (232,0.00000009) (233,0.00000338) (234,0.00000001) (235,0.00000216) (236,0.00000265) (237,0.00000120) (238,0.00000102) (239,0.00000031) (240,0.00000321) (241,0.00000114) (242,0.00000127) (243,0.00000034) (244,0.00000144) (245,0.00000008) (246,0.00000131) (247,0.00000016) (248,0.00000001) (249,0.00000070) (250,0.00000017) (251,0.00000197) (252,0.00000143) (253,0.00000003) (254,0.00000018) (255,0.00000009) (256,0.00000010) (257,0.00000108) (258,0.00000087) (259,0.00000132) (260,0.00000057) (261,0.00000055) (262,0.00000061) (263,0.00000026) (264,0.00000009) (265,0.00000089) (266,0.00000006) (267,0.00000028) (268,0.00000074) (269,0.00000011) (270,0.00000032) (271,0.00000003) (272,0.00020827) (273,0.00109742) (274,0.00248004) (275,0.02694352) (276,0.03671738) (277,0.11579895) (278,0.01661712) (279,0.00500536) (280,0.00022149) (281,0.00375028) (282,0.00045456) (283,0.00181433) (284,0.00007760) (285,0.00026066) (286,0.00011702) (287,0.00069863) (288,0.00021094) (289,0.00244083) (290,0.00056959) (291,0.00005297) (292,0.00000095) (293,0.00000104) (294,0.00000118) (295,0.00000077) (296,0.00000125) (297,0.00000016) (298,0.00000061) (299,0.00000073) (300,0.00000081) (301,0.00000008) (302,0.00000022) (303,0.00000034) (304,0.00000001) (305,0.00000008) (306,0.00000035)
% Maximum predicted risk
\psline[linecolor=gray] (0,0.5882) (1,0.6134) (2,0.6652) (3,0.6097) (4,0.5892) (5,0.5779) (6,0.5736) (7,0.5841) (8,0.5336) (9,0.4972) (10,0.3548) (11,0.5212) (12,0.2772) (13,0.6652) (14,0.2894) (15,0.5163) (16,0.5521) (17,0.5753) (18,0.5753) (19,0.5736) (20,0.3122) (21,0.4230) (22,0.4330) (23,0.5175) (24,0.3441) (25,0.6511) (26,0.6328) (27,0.9709) (28,1.0362) (29,1.0691) (30,0.8243) (31,0.5114) (32,0.9784) (33,0.8085) (34,1.1083) (35,0.5985) (36,0.9926) (37,1.0360) (38,1.1221) (39,0.7308) (40,0.8041) (41,0.7257) (42,0.4613) (43,0.4373) (44,0.3116) (45,0.3497) (46,0.1633) (47,0.4306) (48,0.9171) (49,0.9911) (50,0.8544) (51,1.0383) (52,0.9096) (53,0.5228) (54,0.3079) (55,0.4397) (56,0.5141) (57,1.0438) (58,0.3195) (59,0.3550) (60,1.1461) (61,0.2976) (62,0.5608) (63,0.5169) (64,0.3577) (65,0.4703) (66,0.5045) (67,0.5517) (68,0.3632) (69,0.4885) (70,0.2983) (71,0.5857) (72,0.5028) (73,0.3064) (74,0.3682) (75,0.3704) (76,0.4893) (77,0.5066) (78,0.6083) (79,0.7904) (80,0.8287) (81,1.0387) (82,1.2645) (83,0.8058) (84,0.2078) (85,0.5282) (86,0.3845) (87,0.5588) (88,0.5215) (89,0.5459) (90,0.4250) (91,2.0744) (92,0.4519) (93,0.4316) (94,0.4409) (95,0.4400) (96,0.4397) (97,0.2126) (98,0.2215) (99,0.2061) (100,0.1668) (101,0.3981) (102,0.5226) (103,0.7677) (104,0.8820) (105,0.2749) (106,0.2443) (107,0.2987) (108,0.2187) (109,0.3952) (110,0.3196) (111,0.1733) (112,0.1713) (113,0.3099) (114,0.5170) (115,0.2117) (116,0.1419) (117,0.2975) (118,0.3184) (119,0.3304) (120,0.3359) (121,0.3450) (122,0.3180) (123,0.3190) (124,0.3699) (125,0.2783) (126,0.2595) (127,0.3125) (128,0.4060) (129,0.5742) (130,0.6056) (131,0.6360) (132,0.9831) (133,0.4410) (134,0.2128) (135,0.2402) (136,0.3706) (137,0.2530) (138,0.3360) (139,0.3682) (140,0.3626) (141,0.3492) (142,0.3633) (143,0.3707) (144,0.3884) (145,0.3877) (146,0.3729) (147,0.3706) (148,0.3375) (149,0.3136) (150,0.2808) (151,0.3524) (152,0.7323) (153,0.7705) (154,0.3872) (155,0.8086) (156,0.8871) (157,0.8592) (158,0.8176) (159,0.8618) (160,0.7457) (161,1.1658) (162,0.6490) (163,0.9724) (164,1.0701) (165,0.9382) (166,0.6382) (167,0.6477) (168,0.4566) (169,0.6262) (170,0.5741) (171,1.0226) (172,0.8046) (173,0.7799) (174,0.5361) (175,0.5707) (176,0.6665) (177,0.7623) (178,0.8149) (179,0.8215) (180,0.8091) (181,0.8392) (182,0.7981) (183,0.8695) (184,0.8772) (185,0.9089) (186,0.9236) (187,0.9301) (188,0.9413) (189,0.9875) (190,1.0384) (191,1.0157) (192,1.0725) (193,0.5153) (194,0.7518) (195,0.6766) (196,0.7346) (197,0.8120) (198,0.6551) (199,0.8958) (200,0.9523) (201,0.9493) (202,0.9324) (203,3.6252) (204,0.9763) (205,3.2261) (206,1.4519) (207,2.7161) (208,2.0307) (209,2.0116) (210,1.7530) (211,4.9074) (212,2.0164) (213,1.9782) (214,1.6779) (215,2.8268) (216,1.5138) (217,2.4010) (218,2.3971) (219,1.5035) (220,0.9203) (221,2.2739) (222,2.0633) (223,1.5912) (224,1.0366) (225,0.9492) (226,1.3403) (227,1.1888) (228,1.5860) (229,0.3272) (230,0.8727) (231,1.1494) (232,1.0692) (233,1.1235) (234,0.8954) (235,0.9548) (236,2.4024) (237,2.0738) (238,1.8754) (239,1.6665) (240,1.7991) (241,1.0686) (242,0.5687) (243,0.7165) (244,0.7852) (245,0.2228) (246,0.7165) (247,0.4077) (248,0.6651) (249,0.6376) (250,0.8503) (251,0.8317) (252,0.7807) (253,0.4282) (254,0.7744) (255,0.5563) (256,0.9036) (257,0.5307) (258,0.5930) (259,0.4098) (260,0.1994) (261,0.3601) (262,0.4921) (263,0.5252) (264,0.5077) (265,0.3987) (266,0.3584) (267,0.4212) (268,0.8670) (269,0.6808) (270,0.5527) (271,0.7848) (272,0.3719) (273,0.4846) (274,0.5695) (275,0.5234) (276,0.3181) (277,0.6331) (278,0.5924) (279,0.2137) (280,0.5180) (281,0.3879) (282,0.4818) (283,0.3487) (284,0.3100) (285,0.1555) (286,0.1612) (287,0.4192) (288,0.1235) (289,0.1669) (290,0.2998) (291,0.1903) (292,0.2703) (293,0.5312) (294,0.3337) (295,0.3679) (296,0.2785) (297,0.1653) (298,0.2111) (299,0.2867) (300,0.2962) (301,0.1892) (302,0.2749) (303,0.2446) (304,0.2862) (305,0.3550) (306,0.3210)
% Maximum realized risk
\psline (0,0.5197) (1,0.6207) (2,0.6672) (3,0.6067) (4,0.5831) (5,0.5746) (6,0.5738) (7,0.5834) (8,0.5344) (9,0.5096) (10,0.3169) (11,0.5103) (12,0.2734) (13,0.6646) (14,0.2755) (15,0.5069) (16,0.5594) (17,0.5710) (18,0.5684) (19,0.5787) (20,0.3651) (21,0.4150) (22,0.4107) (23,0.5050) (24,0.3293) (25,0.6555) (26,0.5692) (27,0.9481) (28,1.0351) (29,1.0741) (30,0.8312) (31,0.5021) (32,0.9761) (33,0.8726) (34,1.1240) (35,0.5789) (36,1.0058) (37,1.0360) (38,1.1237) (39,0.7561) (40,0.7971) (41,0.7239) (42,0.4229) (43,0.4264) (44,0.3165) (45,0.4127) (46,0.1653) (47,0.4727) (48,0.9295) (49,0.9947) (50,0.8630) (51,1.0497) (52,0.9196) (53,0.4827) (54,0.3084) (55,0.4459) (56,0.4974) (57,1.0582) (58,0.3265) (59,0.3537) (60,1.1589) (61,0.2775) (62,0.5609) (63,0.5159) (64,0.3786) (65,0.5140) (66,0.5412) (67,0.5661) (68,0.3968) (69,0.5017) (70,0.3307) (71,0.5962) (72,0.4856) (73,0.2765) (74,0.4004) (75,0.3513) (76,0.4683) (77,0.4984) (78,0.5448) (79,0.7562) (80,0.8240) (81,1.0034) (82,1.2551) (83,0.8490) (84,0.2406) (85,0.5115) (86,0.3361) (87,0.5516) (88,0.5221) (89,0.5347) (90,0.4071) (91,2.0780) (92,0.4504) (93,0.4360) (94,0.4439) (95,0.4429) (96,0.4432) (97,0.1835) (98,0.2125) (99,0.1698) (100,0.1900) (101,0.4021) (102,0.6063) (103,0.7496) (104,0.8611) (105,0.2673) (106,0.2289) (107,0.2659) (108,0.1696) (109,0.3623) (110,0.2880) (111,0.2009) (112,0.2161) (113,0.2795) (114,0.4665) (115,0.1722) (116,0.1453) (117,0.2991) (118,0.3180) (119,0.3293) (120,0.3350) (121,0.3451) (122,0.3348) (123,0.3202) (124,0.3513) (125,0.2761) (126,0.2749) (127,0.3260) (128,0.3976) (129,0.5715) (130,0.5999) (131,0.6298) (132,0.9829) (133,0.4407) (134,0.1704) (135,0.2314) (136,0.3301) (137,0.2520) (138,0.3387) (139,0.3703) (140,0.3594) (141,0.3404) (142,0.3601) (143,0.3660) (144,0.3862) (145,0.3864) (146,0.3666) (147,0.3616) (148,0.3437) (149,0.2998) (150,0.2935) (151,0.3490) (152,0.7226) (153,0.7841) (154,0.6026) (155,0.8398) (156,0.9068) (157,0.8772) (158,0.8350) (159,0.9023) (160,0.8096) (161,1.1935) (162,0.7806) (163,1.0572) (164,1.0969) (165,0.8998) (166,0.5498) (167,0.6117) (168,0.3437) (169,0.6412) (170,0.5278) (171,1.0256) (172,0.8432) (173,0.8265) (174,0.5448) (175,0.5769) (176,0.6682) (177,0.7611) (178,0.8123) (179,0.8155) (180,0.8045) (181,0.8357) (182,0.8000) (183,0.8679) (184,0.8747) (185,0.9083) (186,0.9185) (187,0.9248) (188,0.9446) (189,0.9878) (190,1.0413) (191,1.0104) (192,1.0637) (193,0.4752) (194,0.7523) (195,0.6987) (196,0.7346) (197,0.8185) (198,0.7436) (199,0.8958) (200,0.9523) (201,0.9543) (202,0.9441) (203,3.5796) (204,0.9760) (205,3.2168) (206,1.4813) (207,2.7175) (208,2.0233) (209,2.0033) (210,1.6636) (211,4.8931) (212,1.9865) (213,1.9997) (214,1.6959) (215,2.8461) (216,1.5398) (217,2.4450) (218,2.4055) (219,1.5716) (220,1.0086) (221,2.2378) (222,2.0974) (223,1.6402) (224,1.1652) (225,0.9650) (226,1.3219) (227,1.2061) (228,1.5011) (229,0.2397) (230,0.8402) (231,1.1397) (232,1.0519) (233,1.1220) (234,0.9560) (235,1.0126) (236,2.4335) (237,2.1156) (238,1.9323) (239,1.7471) (240,1.8451) (241,1.0256) (242,0.5840) (243,0.6705) (244,0.8150) (245,0.2220) (246,0.7000) (247,0.3506) (248,0.6371) (249,0.6007) (250,0.8390) (251,0.8425) (252,0.7947) (253,0.3921) (254,0.7679) (255,0.4513) (256,0.8804) (257,0.4393) (258,0.6319) (259,0.4336) (260,0.1989) (261,0.3677) (262,0.4849) (263,0.5322) (264,0.5047) (265,0.4437) (266,0.3743) (267,0.4338) (268,0.8718) (269,0.7489) (270,0.6353) (271,0.8155) (272,0.3777) (273,0.4858) (274,0.5703) (275,0.5245) (276,0.3191) (277,0.6327) (278,0.5927) (279,0.2151) (280,0.5324) (281,0.4320) (282,0.4846) (283,0.3602) (284,0.3026) (285,0.1504) (286,0.1626) (287,0.4216) (288,0.1382) (289,0.1774) (290,0.3045) (291,0.1772) (292,0.2766) (293,0.5326) (294,0.3546) (295,0.3641) (296,0.2853) (297,0.1804) (298,0.2166) (299,0.2894) (300,0.2998) (301,0.1939) (302,0.2779) (303,0.2588) (304,0.2846) (305,0.3558) (306,0.3237)
\end{pspicture}

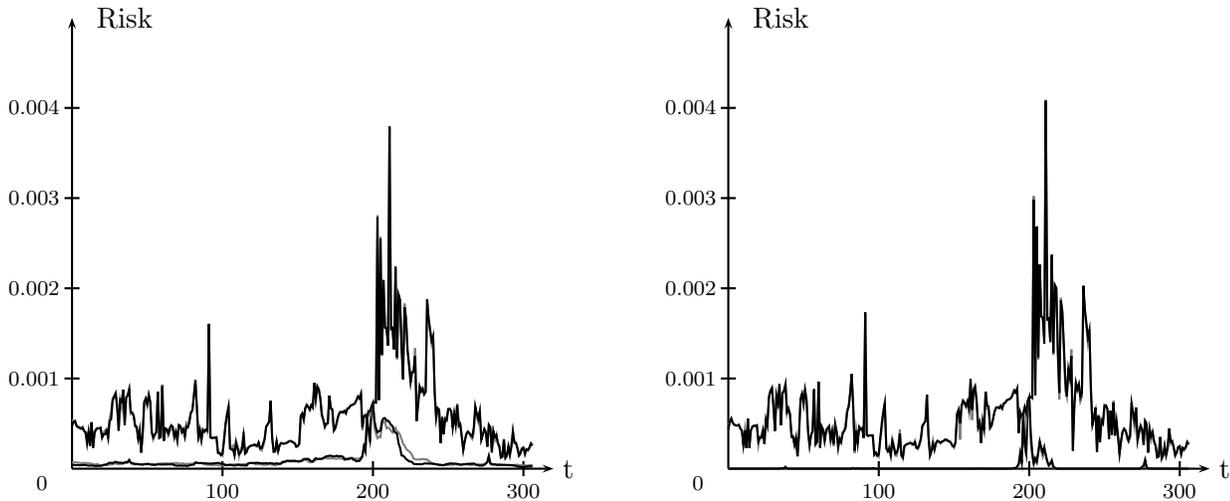
\noindent Fig. 15. Minimum and maximum expected and realized risks for the original data (left) and for the residues of the regression (right), both for no short selling and no cleaning. The predicted risks are in gray and the realized risks are in black.

\vskip 0.3 cm

\section{Final remarks}

In this article, we used two techniques in order to clean the correlation matrix in the building of portfolios using Markowitz's theory. The first technique is the use of Random Matrix Theory in order to clean the correlation matrix built from the time series data of stocks in the year prior to that for which the portfolio is to be built. The second technique is to use a regression model in the removal of the market effect due to the common movement of all stocks. These are used in order to forecast the risk of a portfolio in a particular year using data from its previous year with better precision. The data were the time series returns of the 100\% liquid assets of the BM\&F-Bovespa covering the years from 2004 to 2010. The aim was to combine these two methods in different configurations, and to compare the results in order to obtain the best risk forecasts for portfolios.

Based on a diversity of measures of the aggreement between the forecasted and the realized risks - Agreement ($AG$), Mean Squared Error ($MSE$), and the angle between the risk vectors ($Angle$) - and also of the forecasted and realized correlation matrices - Simple Distance ($Dist$) and Kullback-Leibler distance ($D_{KL}$) - we conclude that the forecasted risk is closer to the realized risk, depending on the volatility of the forecasted year being smaller or larger than the volatility of the year used for the forecast.

In general, the cleaning of the correlation matrix did not produce better results than using the original correlation matrix (without cleaning) for all the measures we used ($AG$, $MSE$, and $Angle$). The use of the regression for the removal of the market produced better results than without the use of the regression in at least 65\% of the cases, according to the measures $MSE$ and $Angle$.

Eterovic and Eterovic (2013) obtained better results for the cleaning procedure of the correlation matrices for the Chilean stock market. Only for data based on 2007 forecasting the results for 2008, and only then for portfolios with the restriction of no short-selling, they obtained a worse result with the cleaning procedure.

The combination of the regression with the cleaning of the correlation matrix lead to a significant improvement in the forecast of the risk of the assets for about 65\% of the cases with the measure $AG$, for 83\% of the cases with the measure $MSE$, and for 83\% of the cases for the measure $Angle$. The use of regression was also paramount for the improvement of the forecasts for the case of short selling allowed, with better results in at least 65\% of the cases. The combination of the regression with the cleaning of the correlation matrix had an excellent performance for the forecasts with short selling allowed, with 65\% of success for the $AG$, 83\% for the $MSE$, and 50\% for the $Angle$.

In the temporal analysis, we could see that the difference between forecasted and realized risks is larger for times of high volatility of the stock market when analyzed by the measures $MSE$ and $Angle$, and that the difference between the predicted and the realized correlation matrices is also larger for times of high volatility when analysed by $Dist$ and by $D_{KL}$. Only the $AG$ gives ambiguous results for the difference between forecasted and realized risks, and one must remind oneself that it is expected from the definition of the $MSE$ that it will be larger in times of high volatility. In general, the regression leads to better results in all measures, but there is no significant difference between results obtained with cleaned or not cleaned correlation matrices. In results that are not posted here, we verified that the use of a model of regression using the eigenvectors corresponding to the first and the second largest eigenvalues of the correlation matrix did not lead to better results.

The cleaning of correlation matrices, especially when dealing with data based on the residues of a regression, usually lead to computational problems due to numerical errors, and one often has to deal with non positive definite covariance matrices, eigenvalues with very small imaginary parts, and non-symmetric correlation matrices, all of them slightly so, but leading to further problems like non invertible covariance matrices, which are the basis of some solving algorithms for portfolios. In order to diminish or control some of these problems, we used a different algorithm, based on quadratic programming, for calculating the optimal portfolios. This process still leads to some errors in the calculations, mainly for the cleaning procedure for the correlation matrices based on residues of the regression, and those errors may be responsible for some of the poor results associated with this procedure.

So, the use of a regression method with a single index in the removal of market effects is usually advisable, but the use of Random Matrix Theory in the removal of noise from the correlation matrices tends to fail in the forecasting for years of high volatility, which are precisely the occasions in which a reliable risk forecast is most needed.

\vskip 0.6 cm

\newpage

\noindent{\bf Acknowledgements}

\vskip 0.4 cm

L. Sandoval Jr. and M.K. Venezuela thank for the support of this work by a grant from Insper, Instituto de Ensino e Pesquisa. We are also grateful to Gustavo Curi Amarante, who collected the data and to Nicolas Eterovic, for useful discussions. Special thanks go to the two anonymous reviewers, who made a large list of suggestions that made this article, without doubt, a much better one. This article was written using \LaTeX, all figures were made using PSTricks, and the calculations were made using Matlab, Ucinet and Excel. All data are freely available upon request on leonidassj@insper.edu.br. Suplementary Material containing more complete tables and graphs for all the pairs of years can be obtained with the authors.

\vskip 0.4 cm

\noindent {\bf References}

\vskip 0.2 cm

\noindent Anton, H. and Rorres, C. (2005). {\sl Elementary Linear Algebra with Applications}, Ninth Edition, Wiley.

\vskip 0.2 cm

\noindent Barunika, J., Astec, T. , Di Matteo, T., and Liuf R. (2012). Understanding the source of multifractality in financial markets. {\sl Phys. A.} 391:4234-4251.

\vskip 0.2 cm

\noindent Biroli, G., Bouchaud, J-P, and Potters, M. (2007). The Student Ensemble of Correlation Matrices: Eigenvalue Spectrum and Kullback–Leibler Entropy. {\sl Acta Physica Polonica B}, 13:4009-4026.

\vskip 0.2 cm

\noindent Blanco, S. (2006). Detecting long and short memory via spectral methods. ArXiv:cond-mat/0610022v1.

\vskip 0.2 cm

\noindent Bodie, Z., Kane, A., \& Marcus, A.J. (2009). {\sl Investments}, Eigth Edition , McGraww-Hill/Irwin.

\vskip 0.2 cm

\noindent Bouchaud, J-P, \&  M. Potters (2011). Financial applications of random matrix theory: a short review. In Akemann, G., Baik, J., Di Francesco, P. (editors), {\sl The Oxford handbook of random matrix theory}, Oxford University Press.

\vskip 0.2 cm

\noindent Chopra, V., \& Ziemba, W.T. (1993). The Effect of Errors in Mean and Co-Variance Estimates on Optimal Portfolio Choice. {\sl J. Port. Management}, 6-11.

\vskip 0.2 cm

\noindent Conlon, T., Ruskin, H.J., \& Crane, M. (2007). Random Matrix Theory and Fund of Funds Portfolio Optmizisation, {\sl Physica A}, 382:565-578.

\vskip 0.2 cm

\noindent DeMiguel, V., Garlappi, L., Nogales, F.J., \& Uppal, R. (2009). A Generalized Approach to Portfolio Optimization: Improving Performance by Constraining Portfolio Norms. {\sl Management Science}, 55:782-812.

\vskip 0.2 cm

\noindent Damião, J.E.F. \& Valls Pereira, P.L. (2009). Comparação de carteiras otimizadas segundo o critério média-variância formadas através de estimativas robustas de risco e retorno. Textos para discussão No 180, Escola de Economia de São Paulo, Fundação Getulio Vargas.

\vskip 0.2 cm

\noindent Dickinson, J.P. (1974). The Reliability of Estimation Procedures in Portfolio Analysis. {\sl J. Fin. Quant. Anal.}, 9:447-462.

\vskip 0.2 cm

\noindent Elton, E.J., Gruber, M.J., Brown, S.J., \& Goetzmann, W. (2009). {\sl Modern Portfolio Theory and Investment Analysis}, Eigth Edition , Wiley.

\vskip 0.2 cm 

\noindent Eterovic, N.A. and Eterovic, D.S. (2013). Separating the wheat from the chaff: understanding portfolio returns in an emerging market. Preprint submitted to Elsevier.

\vskip 0.2 cm

\noindent Frankfurter, G.M., Phillips, H.E., \& Seagle, J.P. (1971). Portfolio Selection: The Effects of Uncertain Means, Variances, and Covariances. {\sl J. Fin. Quant. Anal.}, 6:1251-1262.

\vskip 0.2 cm

\noindent Frankfurter, G.M., Phillips, H.E., \& Seagle, J.P. (1972). Estimation Risk in the Portfolio Selection Model: A Comment. {\sl J. Fin. Quant. Anal.}, 7:1423-1424.

\vskip 0.2 cm 

\noindent Górski, A.Z., Dro\.{z}d\.{z}, S., Kwapie\'{n}, J., and O\'{s}wi\c{e}cimka, P. (2006). Complexity characteristics of currency networks. ArXiv:physics/0606020v1.

\vskip 0.2 cm

\noindent Jiang, Z-Q and Zhou, W-X (2008). Multifractality in stock indexes: Fact or fiction? {\sl Physica A} 387:3605-3614.

\vskip 0.2 cm

\noindent Jobson, J.D., \&  Korkie, B.M. (1980). Estimation for Markowitz Efficient Portfolios. {\sl J. Am. Stat. Assoc.}, 75:544-554.

\vskip 0.2 cm

\noindent Jorion, P.(1986). Bayes-Stein Estimation for Portfolio Analysis. {\sl J. Fin. Quant. Anal.}, 21:279-292.

\vskip 0.2 cm

\noindent Kullback, S., Leibler, R. A. (1951). On information and suficiency. {\sl The Annals of Mathematical Statistics} 22:79-86.

\vskip 0.2 cm

\noindent La Spada, G., Farmer, J.D. and Lillo, F. (2008). The Non-Random Walk of Stock Prices: The Long-Term Correlation between Signs and Sizes. {\sl The European Physical Journal B} 64:607-614.

\vskip 0.2 cm

\noindent Laloux, L., Cizeau, P., Bouchaud, J.-P., \&  Potters, M. (1999) Noise dressing of financial correlation matrices. {Phys. Rev. Lett.} 83:1467-1470.

\vskip 0.2 cm

\noindent Laloux, L., Cizeau, P., Bouchaud, J.-P., \&  Potters, M. (2000). Random Matrix Theory and Financial Correlations. {\sl Int. J. Theor. Appl. Finance.}, 3:391-397.

\vskip 0.2 cm

\noindent Ledoit, O. \& Wolf, M. (2003). Improved estimation of the covariance matrix of stock returns with an application to portfolio selection. {\sl Journal of Empirical Finance} 10:603:621.

\vskip 0.2 cm

\noindent Mar\v{e}nko, V.A., \&  Pastur, L.A. (1967). {\sl USSR-Sb}, 1:457-483.

\vskip 0.2 cm

\noindent Markowitz, H.M. (1952). Portfolio Selection. {\sl The Journal of Finance}, 7:77-91.

\vskip 0.2 cm 

\noindent Medina, L.M.H. and Mansilla, R.C. (2008). Teoria de matrices aleatorias y correlacion de series financieras. El caso de la Bolsa mexicana de Valores. {\sl Revista de Administracion, Finanzas y Economia} 2:125-135.

\vskip 0.2 cm

\noindent Mehta, M. L. (2004). {\sl Random Matrices}, Academic Press.

\vskip 0.2 cm

\noindent Mendes, B.V.M. \& Leal, R.P.C. (2005). Matriz Robusta de Covariâncias e Seleção de Ações Brasileiras. Quinto Encontro Brasileiro de Finanças, São Paulo. {\sl Anais do Quinto Encontro Brasileiro de Finanças, 2005}.

\vskip 0.2 cm

\noindent Michaud, R.O. (1989). The Markowitz Optimization Enigma: Is `Optimized' Optimal? {\sl Fin. Anal. Journal}, 45:31-42.

\vskip 0.2 cm

\noindent Nascimento Jr. H.B., Fulco, U.L., Lyra, M.L., Serva, M, and Viswanathan, G.M. (2007). Why stock markets crash: the origin of fat tailed distributions of returns. {\sl Rev. Bras. Ensino Fís.} 29. http://dx.doi.org/10.1590/ S0102-47442007000300005

\vskip 0.2 cm 

\noindent Nilantha, K.G.D.R., Ranasinghe, and Malmini, P.K.C. (2007). Eigenvalue density of cross-correlations in Sri Lankan financial market. {\sl Physica A} 378:345-356.

\vskip 0.2 cm

\noindent Onnela, J-P., Chakraborti, A., \&  Kaski, K. (2003). Dynamics of market correlations: taxonomy and portfolio analysis. {\sl Phys. Rev. E}, 68:056110.

\vskip 0.2 cm

\noindent Pafka, S., \&  Kondor, I. (2002). Noisy covariance matrices and portfolio optimization. {\sl Eur. Phys. J. B}, 27:277-280.

\vskip 0.2 cm 

\noindent Pan, R.K. and Sinha, S. (2007). Collective behavior of stock price movements in an emerging market. {\sl Physical Review E} 76:046116.

\vskip 0.2 cm

\noindent Pantaleo, E., Tumminello, M., Lillo, F., \&  Mantegna, R.S. (2011). When do improved covariance matrix estimators enhance portfolio optimization? An empirical comparative study of nine estimators. {\sl Quantitative Finance}, 11:1067-1080.

\vskip 0.2 cm

\noindent Plerou, V., Gopikrishnan, P., Rosenow, B., Amaral, L.A.N., Guhr, T., \&  Stanley, H.E. (2002). A Random Matrix Theory Approach to Cross-Correlations in Financial Data. {\sl Phys. Rev. E}, 65:066126.

\vskip 0.2 cm

\noindent Qian, M-C, Jiang, Z-Q, and Zhou, W-X (2010). {J. Phys. A: Math. Theor.} 43:335002. doi:10.1088/1751-8113/43/33/335002.

\vskip 0.2 cm

\noindent Reyna, F., Duarte Jr., A.M.D., Mendes, B.V.M., \& Porto, O. (2005) Optimal Portfolio Structuring in
Emerging Stock Markets Using Robust Statistics, {\sl Brazilian Review of Econometrics}, 25:139–157.

\vskip 0.2 cm

\noindent Rosenow, B., Plerou, V., Gopikrishnan, P., \&  Stanley, H.E. (2002). Portfolio optimization and the random magnet problem. {\sl Europhys. Lett.}, 59:500.

\vskip 0.2 cm

\noindent Ross, S. (1976). The Arbitrage Theory of Capital Asset Pricing. {\sl Journal of Economic Theory}, 50:30-49.

\vskip 0.2 cm

\noindent Sharifi, S., Crane, M., Shamaie, A., \&  Ruskin, H. (2004). Random Matrix Theory for Portfolio Optimization: A Stability Approach. {\sl Physica A}, 335:629-643.

\vskip 0.2 cm

\noindent Schor, A., Bonomo, M., \& Valls Pereira, P.L. (2002). Arbitrage Pricing Theory (APT) e Variáveis Macroeconômicas: Um Estudo Empírico Sobre O Mercado Acionário Brasileiro. {\sl Revista de Economia e Administração} 1:34-46.

\vskip 0.2 cm

\noindent Tola, V., Lillo, F., Gallegati, M., \&  Mantegna, R.N. (2008). Cluster analysis for portfolio optimization. {\sl Journal of Economic Dynamics and Control}, 32:235-258.

\vskip 0.2 cm

\noindent Tumminello, M., Lillo, F., \& Mantegna, R.N. (2007a). Kullback-Leibler distance as a measure of the information filtered from multivariate data. {Phys. Rev. E} 76:031123.

\vskip 0.2 cm

\noindent Tumminello, M., Lillo, F., \& Mantegna, R.N. (2007b). Shrinkage and spectral filtering of correlation matrices: A comparison via the Kullback-Leibler distance. {Acta Physica Polonica B} 38:4079-4088.

\vskip 0.2 cm

\noindent Tumminello, M., Lillo, F., \& Mantegna, R.N. (2010). Correlation, hierarchies, and networks in financial markets. {\sl J. of Economic Behavior and Organization} 75:40-58.

\vskip 0.2 cm

\noindent Vahabi, M., Jafari, G.R. and Movahed, M.S. (2011). Analysis of fractional Gaussian noises using level crossing method. {\sl Journal of Statistical Mechanics: Theory and Experiment}, P11021 doi:10.1088/1742-5468/2011/11/ P11021.

\vskip 0.2 cm

\noindent Vinha, L. G. \& Chiann, C. (2007). Modelos fatoriais para retornos de ativos. {\sl Revista Brasileira de Estatística}, 68:1.

\vskip 0.2 cm

\noindent Wigner, E. P. (1955). Characteristic vectors of bordered matrices with infinite dimensions. {\sl Ann. Math.}, 62:548-564.

\vskip 0.2 cm

\noindent Wigner, E. P.(1958). On the distribution of the roots of certain symmetric matrices. {\sl Ann. Math.}, 67:325-327.

\vskip 0.2 cm

\noindent Wilcox, E. and Gebbie, T. (2004). On the analysis of cross-correlations in South African market data. {\sl Physica A} 344:294-298.

\vskip 0.2 cm 

\noindent Wilcox, E. and Gebbie, T. (2007). An analysis of cross-correlations in an emerging market. {\sl Physica A} 375:584-598.

\vskip 0.2 cm

\noindent Zhou, W-X (2009). The components of empirical multifractality in financial returns. {\sl Europhysics letters} 88:28004, doi:10.1209/0295-5075/88/28004.

\end{document}